\documentclass[%
 reprint,
unsortedaddress,
 amsmath,amssymb,
 aps,
]{revtex4-2}

\usepackage{graphicx}% Include figure files
\graphicspath{ {images/} }
\usepackage{dcolumn}% Align table columns on decimal point
\usepackage{bm}% bold math
\usepackage{xcolor}
\usepackage{url}
\usepackage{multirow}
\usepackage{tabularx}
\usepackage{ragged2e}
\newcommand{\vecTh}{\vec{\Theta}}
\newcommand{\Like}{\mathcal{L}}
\newcommand{\phenomhm}{\texttt{PhenomHM\;}} % changed this to teletype as per convention, but wordspacing after the command is wonky. added an en space here as a hack; see if you can find a better solution, e.g. not using newcommand

\newcommand{\aap}{Astronomy \& Astrophysics}
\newcommand{\apjl}{Astro}
\newcommand{\mnras}{MNRAS}

\usepackage{array}
\newcolumntype{L}[1]{>{\raggedright\let\newline\\\arraybackslash\hspace{0pt}}m{#1}}
\newcolumntype{C}[1]{>{\centering\let\newline\\\arraybackslash\hspace{0pt}}m{#1}}
\newcolumntype{R}[1]{>{\raggedleft\let\newline\\\arraybackslash\hspace{0pt}}m{#1}}

\begin{document}

\preprint{APS/123-QED}

\title{GPU-accelerated massive black hole binary parameter estimation with LISA}% Force line breaks with \\

\author{Michael L. Katz}
 \email{mikekatz04@gmail.com}
 \affiliation{%
 Department of Physics and Astronomy, Northwestern University, Evanston, IL 60201, U.S.A
}%
 \affiliation{%
 Center for Interdisciplinary Exploration and Research in Astrophysics, Evanston, IL 60201, U.S.A
}%
 
\author{Sylvain Marsat}%
\affiliation{%
 Laboratoire Astroparticule et Cosmologie, 10 rue Alice Domon et L\'{e}onie Duquet, 75013 Paris, France
}%

\author{Alvin J. K. Chua}%
\affiliation{%
 Jet Propulsion Laboratory, California Institute of Technology, Pasadena, CA 91109, U.S.A.
}%

\author{Stanislav Babak}%
\affiliation{%
 Laboratoire Astroparticule et Cosmologie, 10 rue Alice Domon et L\'{e}onie Duquet, 75013 Paris, France
}%
\affiliation{%
Moscow Institute of Physics and Technology, Dolgoprudny, Moscow region, Russia
}%

\author{Shane L. Larson}
 \affiliation{%
 Department of Physics and Astronomy, Northwestern University, Evanston, IL 60201, U.S.A
}%
 \affiliation{%
 Center for Interdisciplinary Exploration and Research in Astrophysics, Evanston, IL 60201, U.S.A
}%



\begin{abstract}
The Laser Interferometer Space Antenna (LISA) is slated for launch in the early 2030s. A main target of the mission is massive black hole binaries that have an expected detection rate of $\sim20$ yr$^{-1}$. We present a parameter estimation analysis for a variety of massive black hole binaries. This analysis is performed with a graphics processing unit (GPU) implementation comprising the \phenomhm waveform with higher-order harmonic modes and aligned spins; a fast frequency-domain LISA detector response function; and a GPU-native likelihood computation. The computational performance achieved with the GPU is shown to be 500 times greater than with a similar CPU implementation, which allows us to analyze full noise-infused injections at a realistic Fourier bin width for the LISA mission in a tractable and efficient amount of time. With these fast likelihood computations, we study the effect of adding aligned spins to an analysis with higher-order modes by testing different configurations of spins in the injection, as well as the effect of varied and fixed spins during sampling. Within these tests, we examine three different binaries with varying mass ratios, redshifts, sky locations, and detector-frame total masses ranging over three orders of magnitude. We discuss varied correlations between the total masses and mass ratios; unique spin posteriors for the larger mass binaries; and the constraints on parameters when fixing spins during sampling, allowing us to compare to previous analyses that did not include aligned spins.
\end{abstract}

\maketitle

%\tableofcontents
\section{Introduction}\label{sec:introduction}

The Laser Interferometer Space Antenna (LISA), an ESA-led mission, is officially slated for launch in the early 2030s \citep{LISAMissionProposal_2}. LISA will add the mHz regime of the gravitational wave spectrum to the high-frequency observations by the LIGO--Virgo--KAGRA network \citep{O2_summary_2, LVK2018LivingReview}. A primary source of interest for LISA is the inspiral and merger of massive black hole (MBH) binaries. Recent work has shown that LISA is expected to detect $\sim1$--$20$ MBH binaries per year \citep{Klein2016, Berti2016, Salcido2016, Katz2019, Bonetti2019}. These sources are expected to occur at high signal-to-noise ratios (SNRs) of $\sim100$--$1000$ out to high redshifts ($z\sim10$--$20$) \citep{GravitationalUniverse_2, Barausse2015, LISAMissionProposal_2}. LISA will be sensitive to binaries of $\sim10^3$--$10^9M_\odot$ if they exist \citep{LISAMissionProposal_2, Klein2016, Katz2018}.

The high SNR of MBH signals will help astronomers understand the formation and evolution of these objects over cosmic time \citep{GravitationalUniverse_2, Barausse2015, LISAMissionProposal_2}. Theories of MBH formation channels usually fall into three categories: high-mass seeds of $\sim10^4$--$10^6M_\odot$ from pre-galactic halo collapse at $z\sim10$--$20$ \citep{Loeb1994, Begelman2006, Latif2013, Habouzit2016, Ardaneh2018, Dunn2018}; intermediate seeds of $\sim10^3$--$10^4M_\odot$ from runaway cluster collapse at $z\sim10$--$20$ \citep{Omukai2008, Devecchi2009, Davies2011, Katz2015}; and smaller seeds of $\sim10^2M_\odot$ from the collapse of Population III stars at $z\sim20$--$50$ \citep{Haiman2000, Fryer2001, Heger2003, Volonteri2003, Tanaka2009, Alvarez2009}. The loud signals from MBH binaries can also be used for tests of fundamental physics and independent measurements of cosmological parameters \citep{Petiteau2011, Gair2013, Barausse2015}. In addition, determining MBH binary information like the sky location and masses is crucial in the search for electromagnetic counterparts. This search requires efficient and reliable measurement of such parameters prior to the actual merger of the two MBHs. With electromagnetic counterparts to MBH mergers, further questions about cosmology can be answered \citep{Schutz1986, Holz2005}, as well as questions related to accretion processes and active galactic nuclei \citep{Armitage2002, Burke-Spolaor2013, Bogdanovic2015, DalCanton2019}.

Traditionally, Bayesian inference techniques have been suggested for MBH binary searches and statistical analysis \citep[e.g.][]{Cornish2006, Cornish2007, Porter2014, Porter2015, Marsat2020LISAPE}. Prior to the use of modern Bayesian sampling techniques \citep{ChristensenMeyer1998MCMCOrig}, the Fisher information matrix (FIM) approach was an accepted method for determining the uncertainties of measurable quantities; however, it has been shown that FIM analyses do not encompass the necessary information to make strong statements about parameter estimation with MBH binaries \citep[e.g.][]{Porter2008, Vallisneri2008, Porter2015}. Similarly, it is important to include more advanced waveform models that better describe the physics of binary black hole coalescence. Aspects like higher-order harmonic modes help to further constrain parameters, remove systematic bias, and to further understand the characteristics of actual data analysis when the LISA mission is launched. Any difference between the waveform models used to create our templates and the true signals that will be observed by LISA will create systematic error \citep{Cutler2007ParameterExtracionErrors}. However, in this work, we will inject the same waveform model as the model used for the templates, therefore bypassing this issue.

In this paper, we examine MBH parameter estimation under the new LISA sensitivity \citep{LISAMissionProposal_2, SciRD1} by using modern, computationally enhanced inspiral--merger--ringdown waveforms that include higher-order harmonics and aligned spins. A variety of LISA-related parameter estimation studies have previously been performed for MBH binaries. The work in \citep{Cutler1994, Vecchio2004, Arun2006, Berti2005, Lang2006} addressed this problem with FIM estimates, the low-frequency approximation to the LISA response function, and inspiral-only Newtonian waveforms. The Mock LISA Data Challenges \citep{Babak2010} expanded on these studies by developing MCMC methods towards the source search problem. Improving upon these analyses further, Bayesian inference techniques were developed in \citep{Brown2007, Cornish2006, Crowder2006, Wickham2006, Rover2007, Feroz2009, Gair2009, Petiteau2009, Porter2014, Porter2015}. However, these works still employed inspiral-only waveforms.

The effect of using higher-order harmonic corrections in studies of MBH binaries was examined in \citep{Arun2007, Trias2008, Porter2008, McWilliams2010}. Analyses of waveforms including merger and ringdown were examined with the FIM method in \citep{McWilliams2010, Thorpe2009, McWilliams2010b, McWilliams2011} and with a Bayesian approach in \citet{Babak2008}. \citet{Babak2014} used the FIM method to analyse parameter constraints with a variety of LISA designs. This study used inspiral waveforms with a reweighting scheme to account for the contribution of the merger and ringdown. \citet{Baibhav2020} recently studied the ability to constrain parameters using only ringdown signals with higher-order harmonics. This study used the FIM method, no noise, and an analytical description of the ringdown.

\citet{Marsat2020LISAPE} was the first study to use a full Bayesian analysis on waveforms that include inspiral, merger, and ringdown. This analysis examined Schwarzschild MBHs, and featured a new prescription for the LISA response function by \citet{Marsat2018}. They used MCMC methods to characterize only extrinsic parameters, by employing a noiseless injection that included higher-order modes. In this work, we extend the findings of \citet{Marsat2020LISAPE} to include MBH signals with aligned spins, injected into a data stream that includes realistic LISA noise. We also analyze the full parameter space of MBH binaries, including both intrinsic and extrinsic quantities.
 
Many of the analyses mentioned above were performed using the original LISA sensitivity curve \citep{Larson2000}. The changes to the LISA sensitivity curve in the most recent ESA-approved proposal \citep{LISAMissionProposal_2} are expected to severely affect our ability to measure and characterize MBH binaries \citep{Katz2018}. First, the SNR for a general MBH binary source will be slightly lower due to higher noise at the low-frequency end and a smaller arm length (2.5 Gm versus 5 Gm).
%Additionally, preliminary investigations are showing the general plan for the entire analysis must change. Previously, it was expected that we would observe MBH sources for weeks or months with the original sensitivity \MLK{REFERENCE}. With this large amount of time, a source could be detected and a Bayesian analysis that would take days could run and finish in a sufficient amount of time prior to merger to alert electromagnetic partners.
Additionally, it can be shown we will observe MBH mergers for less time with the new LISA sensitivity curve \citep{LISAMissionProposal_2}, compared to previous predictions under the classic LISA sensitivity curve \citep{Larson2000}. For some mergers, we will observe their signal for only a few days, instead of the originally estimated weeks or months.

\begin{figure}
\begin{center}
\includegraphics[scale=0.39]{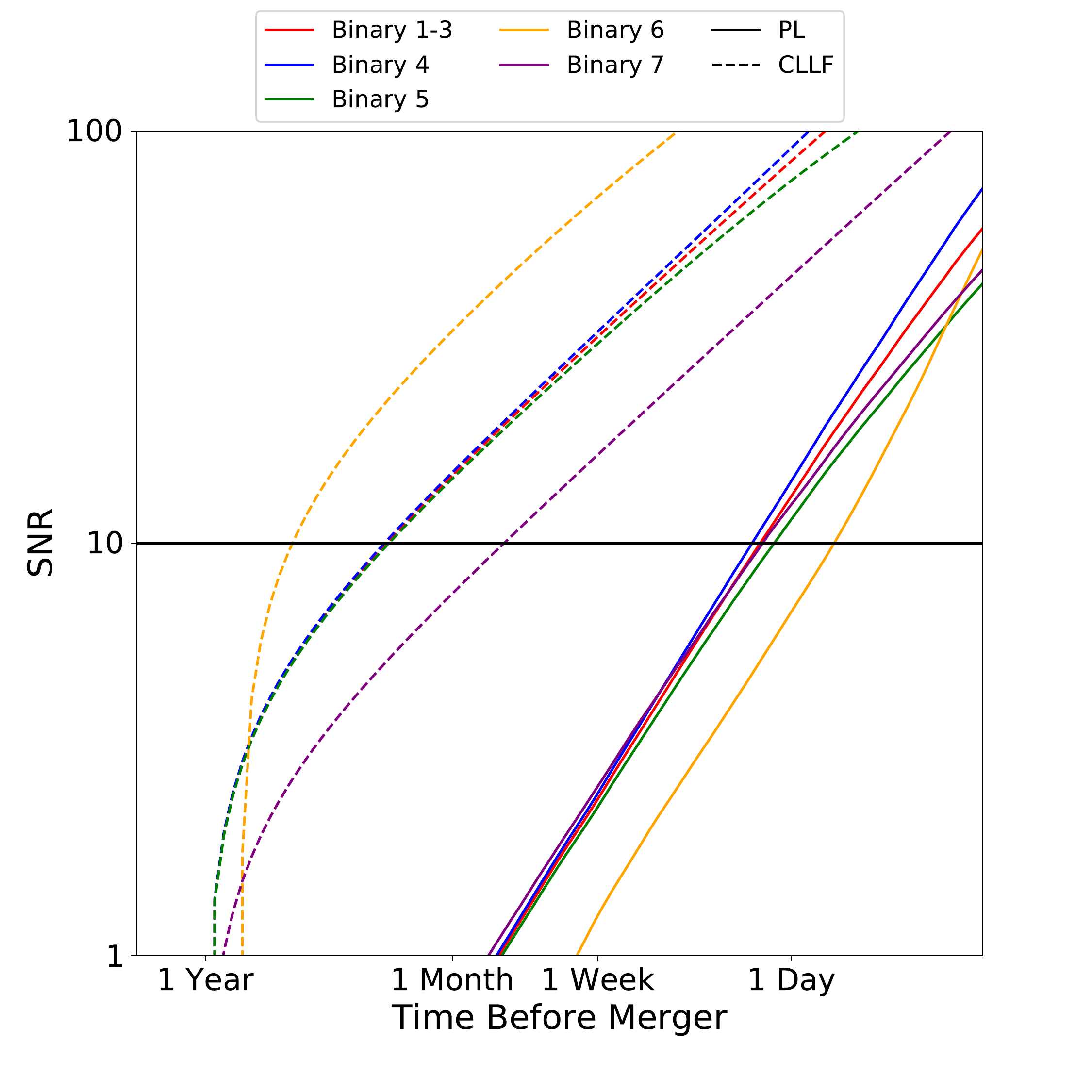}
\end{center}
\caption{SNR versus time before merger is shown above. The focus here is when sources become detectable at an SNR of $\sim10$. This calculation was performed using only the dominant $l=m=2$ mode from \texttt{PhenomD} \citep{Husa2016, Khan2016} as a part of the \texttt{gwsnrcalc} package by \citet{Katz2018}. Solid lines show the SNR computed using the proposed LISA sensitivity curve (PL; \citep{LISAMissionProposal}). Calculations for the classic LISA sensitivity curve (CLLF; \citep{Larson2000}) are shown with dashed lines. CLLF employs a more realistic low-frequency shape compared to the original classic curve by \citet{Larson2000}; for more information on this construction, see \citet{Katz2018}.
%The purpose of this plot is to show the difference in observation time of each binary as the sensitivity has evolved.
The binaries shown are those we analyze in this work; see \mbox{Section \ref{sec:results_discussion}} for a detailed description of these sources. Binaries 1--3 are grouped into one example because their parameters are identical.}\label{fig:time_before_merger}
\end{figure}

\mbox{Figure \ref{fig:time_before_merger}} shows the SNR over time for the binaries (described in \mbox{Section \ref{sec:results_discussion}}) that are studied in this work. For this plot, we employ a simplified calculation of the sky-\mbox{location-}, \mbox{polarization-}, and inclination-averaged SNR for the $l=m=2$ mode, with the \texttt{gwsnrcalc} package \citep{Katz2018} that uses \texttt{PhenomD} \citep{Husa2016, Khan2016}, \mbox{Equation \ref{eq:innerprod}}, and an averaging factor of 16/5 \citep{Cornish2018}. It can be seen that, for the binaries tested, we will observe them for days, whereas with the classic LISA sensitivity, we can observe them for months. As a result of this difference in observation time, sky localization is not expected to be satisfactory until close to or after the merger. Early, accurate notifications for astronomers are critical for detecting electromagnetic counterparts to MBH binary mergers. Developing low-latency algorithms and computational infrastructure improvements for faster analysis is essential for this process. 

In \mbox{Section \ref{sec:bayesinf}}, we discuss Bayesian inference in the context of MBH signals and the LISA mission. This includes our GPU-accelerated implementations of the waveform, LISA response, and the Bayesian likelihood, which prove to be very beneficial when examining a more complete version of the LISA analysis problem that includes a full noise realization. We detail our sampling methods in \mbox{Section \ref{sec:mcmc}}. In \mbox{Section \ref{sec:results_discussion}}, we discuss the sample data sets we analyzed, our ability to infer the proper parameters, and information gained about LISA parameter estimation through running our tests. For our analysis, we use units with $G=c=1$.

\section{Bayesian Inference with Gravitational Waves from Massive Black Holes}\label{sec:bayesinf}

\begin{figure*}[tbh]
\begin{center}
\includegraphics[scale=0.58]{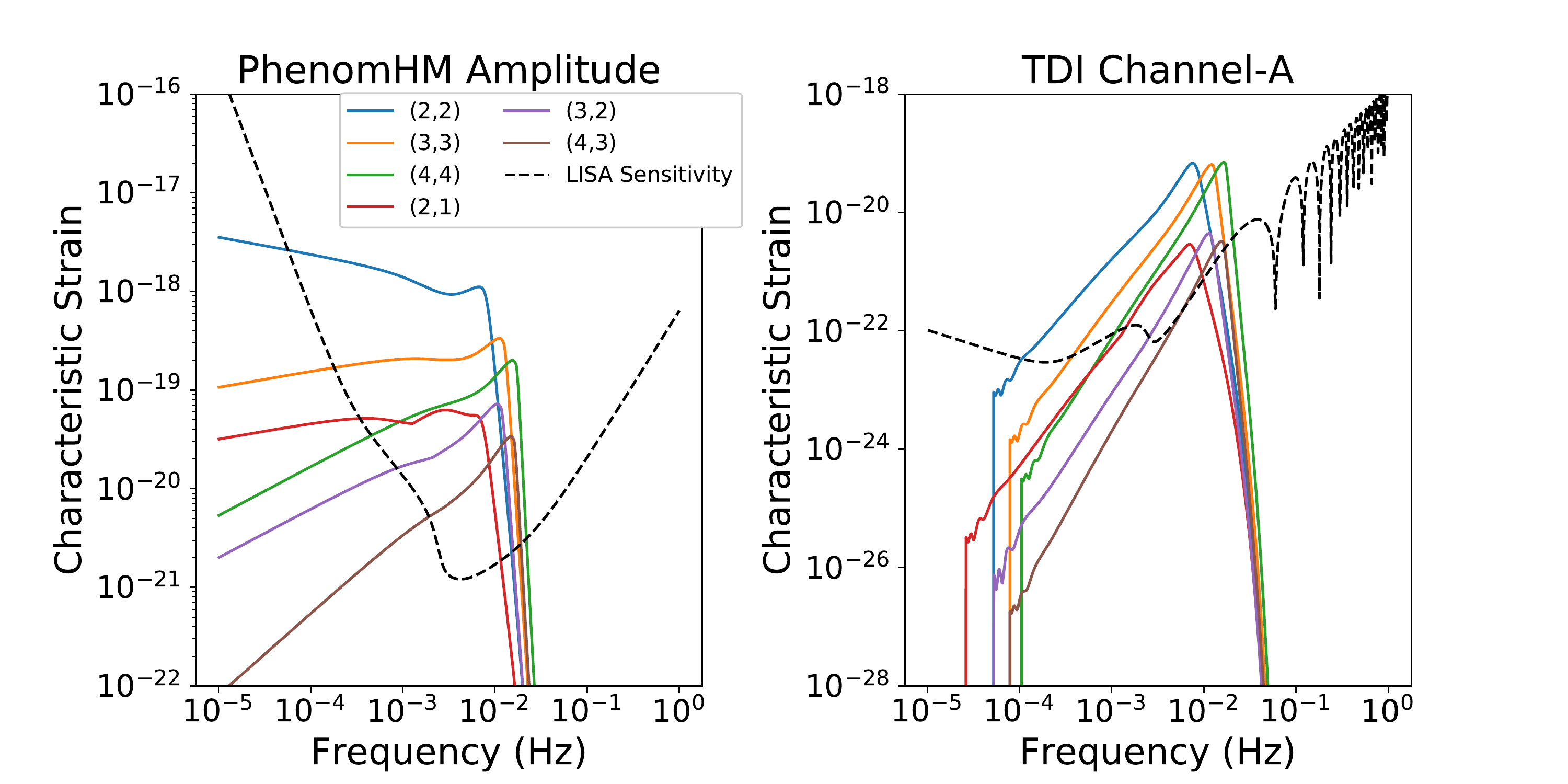}
\caption{The plots above display an example signal for Binary 1 from \mbox{Section \ref{sec:results_discussion}}. Each harmonic mode is shown in a different color in the characteristic strain representation. The characteristic strain, $h_c$ is given as $h_c^2 = f^2|\tilde{h}(f)|^2$ \citep{Finn2000, Moore2015}. The left plot displays the amplitude of each mode determined by \phenomhm \citep{London2018} as well as the LISA sensitivity from \citep{SciRD1} with a contribution from the Galactic background noise after one year of observation described in \citet{Babak2017}. The left plot effectively displays this signal before the LISA response from \citet{Marsat2018} is applied. The right plot shows the same signal put through the LISA response. Here, we display TDI channel A. This is also shown in the characteristic strain representation. The modulations shown at low frequencies are due to the motion of the LISA constellation in its orbit about the Sun. The modes shown in the right plot are the same as those labeled in the left plot.}\label{fig:signal_examples}
\end{center}
\end{figure*}

\begin{figure*}[tbh]
\begin{center}
\includegraphics[scale=0.45]{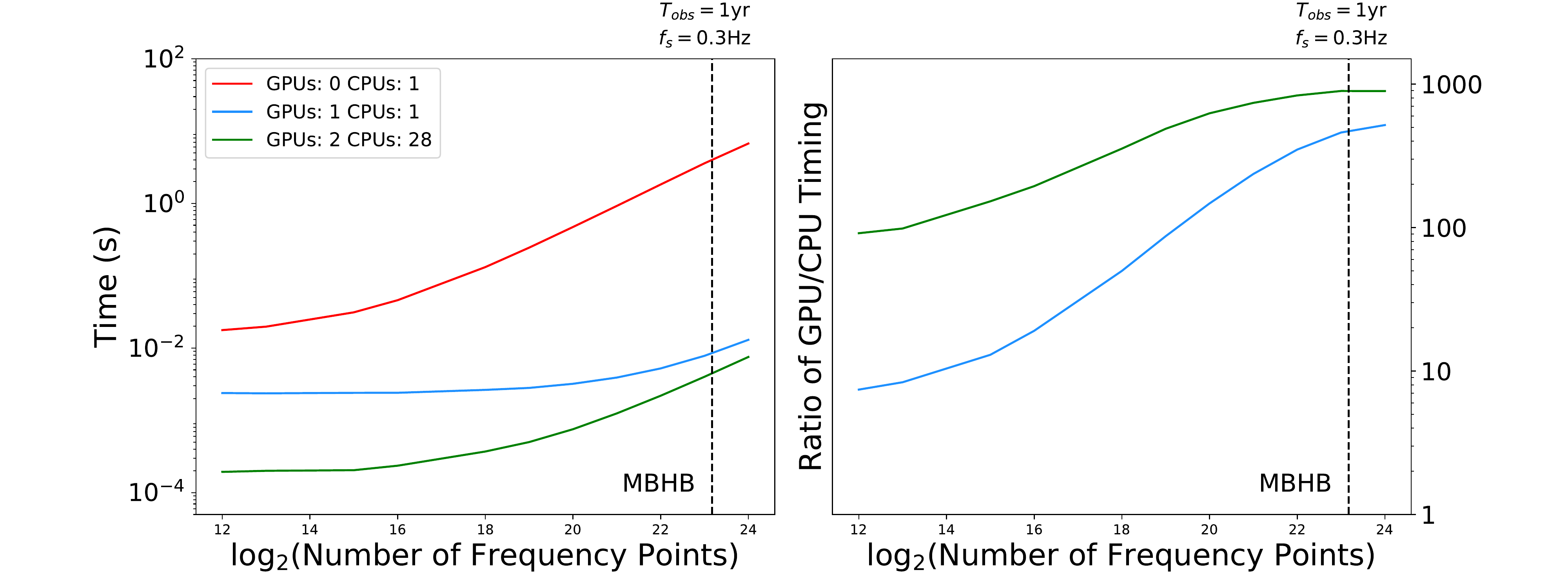}
\caption{Timing results for our GPU-accelerated likelihood computations are shown above. The timing shown represents the total time to generate the waveform, apply the response, and calculate the likelihood in all three channels. The left plot represents the actual time per likelihood calculation. The CPU-only results are shown in red. For these results, no likelihood approximations were included; the calculation is performed in a so-called ``brute-force'' way to compare directly with GPU timings. The blue line shows the usage of one CPU and one GPU. This timing is greatly affected by the time spent on the CPU performing preprocessing for the \phenomhm \citep{London2018} waveform constants. The green line shows the speedup using 28 CPUs and 2 GPUs. This allows us to parallelize the preprocessing and then let the GPUs split the waveform production reducing the calculation time even further. The right plot translates the left plot into an ``effective number of CPUs'' by dividing the CPU-only time by the tests with at least one GPU. The vertical dashed line shows the number of data points in the Fourier transform of a data stream sampled at 0.3 Hz for one year of observation.}\label{fig:timecomp}
\end{center}
\end{figure*}

In the field of gravitational waves, the data stream, $d(t)$, consists of a possible signal, $s(t)$, as well as noise, $n(t)$ ($d(t)=s(t)+n(t))$. In our theoretical study, we will generate the data set by injecting a signal both with and without noise. We will then estimate the parameters of the source in the fabricated data set using Bayesian inference techniques. Bayesian inference methods are based on Bayes rule given by,
\begin{equation}\label{eq:Bayes_rule}
    p(\vecTh| d, \Lambda) = \frac{p(d|\vecTh, \Lambda)\pi(\vecTh|\Lambda)}{p(d|\Lambda)},
\end{equation}
where $\Lambda$ and $\vecTh$ represent the underlying model and model parameters, respectively. The probability on the left-hand side of this equation is the posterior probability, which is the main value we are concerned with determining. The posterior probability represents the probability for the model parameters given the observed data and assumed model. In our study, the parameter set exists in $\mathbb{R}^D$, where $D=11$ is the number of dimensions we test. The parameter set we examine is \{$M_T,\ q,\ a_1,\ a_2,\ D_L,\ \phi_\text{ref},\  \iota,\ \lambda,\ \beta,\ \psi,\ t_\text{ref}$\}, which represents, in order, the total mass, mass ratio, dimensionless spin of the larger black hole, dimensionless spin of the smaller black hole, luminosity distance, reference phase, inclination, ecliptic longitude, ecliptic latitude, polarization angle, and reference time. As discussed in \mbox{Section \ref{sec:GPU}}, the spins of the two MBHs are assumed to be aligned to the orbital angular momentum vector, therefore, reducing our parameter space from $D=15$ to $D=11$ since the spins are represented only as magnitudes without vector angles. For our underlying model, $\Lambda$, we will only examine one waveform model. Therefore, we will drop the $\Lambda$ throughout the rest of this paper. 

The prior probability, $\pi(\vecTh)$, is based on prior knowledge of the potential true parameters representing the injection. For our analysis, we assume limited prior knowledge by employing uniform priors across our parameters throughout an encompassing volume. Information on the priors used can be found in \mbox{Table \ref{tb:priorinfo}}. 

\begin{center}
\begin{table}
\caption{Our priors used in our Bayesian inference are shown below. For each parameter we assume a uniform distribution between the lower and upper bounds. The lower (upper) bound on the distance is based on a redshift of 0.01 (100.0). The inclination ($\iota$) and ecliptic latitude ($\beta$) are listed in terms of the cosine and sine, respectively, of these values because these quantities represent uniform distributions for isotropically distributed angles on the sphere. The upper bound on $t_\text{ref}$ represents 10 years in seconds (the quoted maximum lifetime of the mission).}\label{tb:priorinfo}
\begin{tabular}{| c | c | c |}
 \hline
Parameter & Lower Bound & Upper Bound \\
\hline
$\ln{M_T}$ & $\ln{\left(10^3\right)}$& $\ln{\left(10^9\right)}$ \\
\hline
$q$ & 0.05 & 1.0 \\
\hline
$a_1$ & -0.99 & 0.99 \\
\hline
$a_2$ & -0.99 & 0.99 \\
\hline
$\ln{D_L}$ & $\ln{\left(0.044\right)}$ & $\ln{\left(1298.00\right)}$ \\
\hline
$\phi_\text{ref}$ & 0.0 & $2\pi$ \\
\hline
$\cos{\iota}$ & -1.0 & 1.0 \\
\hline
$\lambda$ & 0.0 & $2\pi$ \\
\hline
$\sin{\beta}$ & -1.0 & 1.0 \\
\hline
$\psi$ & 0.0 & $\pi$ \\
\hline
$\ln{t_\text{ref}}$ & $\ln{\left(1\right)}$ & $\ln{\left(3.15576\times10^8\right)}$ \\
\hline
\end{tabular}
\end{table}
\end{center}

The key element of Bayesian analysis involves the calculation of the Likelihood, $\Like(\vecTh) = p(d|\vecTh)$. In the case of gravitational waves, $\Like$ is the probability for the observed data given a set of parameters and assumed underlying model. The negative log of the Likelihood (NLL) is given by,
\begin{equation}\label{eq:NLL}
    -\log{\Like} = \frac{1}{2}\langle d - h|d - h \rangle = \frac{1}{2}\left(\langle d|d \rangle + \langle h|h \rangle - 2\langle d|h \rangle\right),
\end{equation}
where $h$ is the template waveform, $d$ is the data stream, and $\langle a|b \rangle$ is the noise-weighted inner product of the Fourier transforms of two time series $a(t)$ and $b(t)$. The Fourier  transform of a(t) is represented as $\tilde{a}(f)=\mathcal{F}\{a(t)\}$. The inner product is given by,
\begin{equation}\label{eq:innerprod}
    \langle a|b \rangle = 2\int_0^\infty\frac{\tilde{a}(f)\tilde{b}(f)^* + \tilde{a}(f)^*\tilde{b}(f)}{S_n(f)} df,
\end{equation}
where $S_n(f)$ is the one-sided power spectral density (PSD) of the noise. The template, $h$, is built from the same model as the injected signal, $s$; however, we use $s$ to represent the true signal to avoid confusion. The optimal SNR attainable by a template $h$ is $\langle h|h \rangle^{1/2}$. Here, we treat $S_n(f)$ as constant over the observation duration for simplicity. In reality, the noise is expected to vary slowly with time allowing for noise estimation on the order of $\sim$week when the instantaneous gravitational wave amplitude is well below the noise. For reference, the error in noise estimation is a second-order effect on the statistics illuminated by parameter estimation \citep{EdwardsBayesPSDEstimate, Littenberg2015BayesSpectralEsitmationNoise, Biscoveanu2020PSDEstimation}.  

The denominator on the right-hand side of \mbox{Equation \ref{eq:Bayes_rule}} is referred to as the evidence, $Z=p(d)$. The evidence is the marginalization of the likelihood over the parameter space,
\begin{equation}\label{eq:evidence}
    Z = \int_{\vecTh} \Like(\vecTh) \pi(\vecTh)d\vecTh.
\end{equation}
The evidence, generally, helps estimate the fidelity of a model and to compare models with one another. Computing the evidence directly in the gravitational wave setting is intractable. However, in the Markov Chain Monte Carlo (MCMC) analysis described in \mbox{Section \ref{sec:mcmc}}, the evidence enters only as a multiplicative constant. Therefore, we do not need to compute the evidence for our chosen analysis method. However, as we will describe in \mbox{Section \ref{sec:mcmc}}, we employ a parallel tempering version of MCMC. Within the parallel tempering method, the evidence can be estimated using techniques such as Thermodynamic Integration \citep{Goggans2004ThermoInt, Lartillot2006ThermoInt} or Stepping-Stone Sampling \citep{Maturana-Russel2019SteppingStone}.

To determine the signals and templates, we use the frequency domain waveform model for binary black hole coalescence \phenomhm \citep{Husa2016, Khan2016, London2018}. We refer the interested reader to \citet{Kalaghatgi2019} for questions related to systematic errors in this waveform model. \phenomhm includes aligned spins, higher order spherical harmonic modes, and all three stages of binary black hole coalescence: inspiral, merger, and ringdown. This is the first study of its kind for LISA MBH analysis to include all three of these properties simultaneously. In this work, we analyze the parameter estimation portion of LISA data analysis. We do not perform the initial search for the source; instead, we assume a source has already been identified in the data stream. Additionally, we also ignore any effects from the superposition of signals from other sources simultaneously evolving in the LISA data.

The set of inputs to the waveform model are \{$M_T,\ q,\ a_1,\ a_2,\ D_L$\}. In our implementation of the waveform, we receive the amplitude, A$(f)$, and the phase, $\phi(f)$, of each input spherical harmonic mode $(l,m)$, given by,
\begin{equation}\label{eq:amp_and_phase}
    h_{lm} = A_{lm}(f)e^{i\phi_{lm}(f)}.
\end{equation}
There are six spherical harmonic modes within the \phenomhm model: \mbox{$(l,m)\in \{(2,2), (3,3), (4,4), (2,1), (3,2), (4,3)\}$}. The (2,2) mode is the dominant mode. The other modes are referred to as higher harmonics or higher modes. An example of the \phenomhm amplitudes for each harmonic mode is shown in the left plot of \mbox{Figure \ref{fig:signal_examples}}. 

The frequency bounds of the waveform are determined from the merger time of the signal. We choose a coalescence time, $t_\text{coal}\sim1$ yr into the LISA observing window. In our methodology, for our true signal we set $t_\text{coal}=t_\text{ref}+t_0$. $t_0$ is on the order of a year and defines a temporal marker around which we analyze our signal. $t_\text{ref}$ represents the fine tuning of $t_0$ to the actual coalescence time; therefore, $t_\text{ref}$ is on the scale of seconds to months. $t_\text{coal}$ represents the time at which the waveform reaches $f_\text{max}$ for the (2,2) mode, which is determined internally within \phenomhm to be the frequency at which $f^2A(f)$ reaches a maximum (see \citep{London2018} and \citep{Khan2016} for more information). Due to the complicated nature of the time-frequency correspondence across multiple higher order modes, we determine the time evolution of the system using the derivative of the phase and its relation to $t_\text{coal}$ given by,
\begin{equation}\label{eq:t_f}
    t_{lm}(f) = t_\text{coal} - \frac{1}{2\pi}\frac{d\phi_{lm}(f)}{df}.
\end{equation}
We make a cut in frequency when $t_{lm}(f)$ falls below zero for each individual mode, indicating the initial point of the waveform in the detector when LISA initially begins taking data. 

We then apply the transfer function projecting the waveform onto the detector. The LISA response involves a complex time- and frequency-dependent transfer function \citep{Cutler1994, Larson2000, Cornish2003, Marsat2018}. We apply the fast Fourier domain response from \citet{Marsat2018}. Therefore, the signal in the LISA detector is represented by,
\begin{equation}\label{eq:response}
    h_{lm}^\text{A,E,T}(f, t_{lm}(f)) = \mathcal{T}^\text{A,E,T}(f, t_{lm}(f))h_{lm}(f),
\end{equation}
where $\mathcal{T}(f, t_{lm}(f))$ is the response transfer function and \{A, E, T\} are the time-delay interferometry (TDI) channel indicators. TDI is necessary for LISA to suppress laser noise \citep{Tinto1999, Armstrong1999,Estabrook2000, Dhurandhar2002, Tinto2005}. The A, E, and T channels are transformations of the original X, Y, and Z Michelson TDI observables \citep{Vallisneri2005} given by,
\begin{align}
    A =& \frac{1}{\sqrt{2}}\left(Z-X\right), \\
    E =& \frac{1}{\sqrt{6}}\left(X-2Y+Z\right), \\
    T =&\frac{1}{\sqrt{3}}\left(X+Y+Z\right).
\end{align}
Employing channels A, E, and T, we assume the noise in each channel is uncorrelated, giving a diagonalized noise matrix \citep{Tinto2014, Muratore2020}. $\mathcal{T}(f,t_{lm}(f))$ is determined using the extrinsic parameters, \{$\phi_\text{ref},\  \iota,\ \lambda,\ \beta,\ \psi,\ t_\text{ref}$\} ($D_L$ is factored into $h_{lm}$); the time-frequency correspondance of the waveform; and the orbital and rotational properties of the LISA constellation. \{$\lambda,\ \beta,\ \psi,\ t_\text{ref}$\} are used in the solar system baricenter (SSB) frame when determining $\mathcal{T}(f,t_{lm}(f))$; however, during MCMC runs, these parameters are sampled in the LISA frame and then converted to the SSB frame before waveform generation. For details on the construction and methodology of $\mathcal{T}(f,t_{lm}(f))$, please see \citet{Marsat2018} and \citet{Marsat2020LISAPE}. An example of the \phenomhm waveform amplitudes fed through the response function can be seen in the right plot in \mbox{Figure \ref{fig:signal_examples}}. 

With the waveforms in each TDI channel, we need the PSD of the noise in each channel, $S_n^\text{A,E,T}$. We use the \texttt{tdi} package from the LISA Data Challenges Working Group software collection to generate our PSD in each channel. Within the \texttt{tdi} package, we analyze the ``SciRDv1'' model \citep{SciRD1} for the PSD with a contribution from the Galactic background noise predicted for 1 year into LISA observation. This Galactic background noise is described in \citet{Babak2017}. Figure \ref{fig:signal_examples} shows the PSD$^{1/2}$ shown in the left plot, which includes the contribution from the Galactic background. In the right side plot of the same figure, the PSD$^{1/2}$ is shown in the TDI A channel. In \mbox{Equation \ref{eq:NLL}}, the inner product is really the sum of the inner products over all three channels:
\begin{equation}\label{eq:innerprodsum}
    \langle a | b \rangle = \sum_\text{j=A,E,T} \langle a^j | b^j \rangle.
\end{equation}
We analyze both data sets with noise ($\tilde{n}(f)\neq0$) and without noise ($\tilde{n}(f)=0$). However, the $S_n$ weighting factor in \mbox{Equation \ref{eq:innerprod}} applies in both cases as it defines the SNR. When $\tilde{n}(f)=0$ we are assuming standard LISA noise assumptions while analyzing the unlikely scenario the noise assumes its average value of zero. When we do generate noise as part of the data, we generate it in the frequency domain given by,
\begin{equation}\label{eq:noise_gen}
    \left|\tilde{n}(f)\right| = \mathcal{N}\left(0,\frac{1}{2\sqrt{\Delta f}}\right)\sqrt{S_n(f)},
\end{equation}
where $\mathcal{N}(\mu, \sigma)$ represents a normal distribution with mean $\mu$ and standard deviation $\sigma$, and $\Delta f$ is the Fourier bin width equivalent to 1/$T_\text{obs}$. It is then assigned a random phase from 0 to $2\pi$ based on a uniform distribution.

With the tools to calculate the likelihood, we will now discuss acceleration of the process using graphics processing units (GPUs). 

\subsection{GPU Accelerated Waveforms and Likelihoods}\label{sec:GPU}

A GPU is a special piece of hardware designed for parallel computation. They were originally designed for graphics cards related to video games and other video-related areas; however, as of the mid-2000s, it was realized these units could be re-purposed for general computational and academic use under the description: general purpose computing on graphics processing units (GPGPU). Here, we implement our \phenomhm waveforms, LISA response calculation, and likelihood computation in NVIDIA's proprietary GPU programming language CUDA \citep{CUDA}. GPUs have been used in gravitational wave analysis in \citet{Talbot2019}, where they have implemented \texttt{PhenomPv2} \citep{Hannam2014}, a waveform for the (2,2) mode with precessing spins, directly in CUDA for the LIGO Algorithmic Library \citep[LAL;][]{lalsuite}. We take a different approach to our waveform implementation. Additionally, we have the first GPU implementation for the LISA fast response from \citet{Marsat2018}. We will introduce our implementation and then provide some timing results compared to a similar CPU-based program. For our timing tests, we used a Tesla V100 GPU and a Xeon Gold 6132 2.60 GHz CPU.

The main goal of our implementation is to perform as much of the computation as is feasible and beneficial on the GPU, while reducing the number and size of necessary memory transfers. Before beginning sampling where we will calculate the Likelihood $>10^7$ times, we perform the aspects of this computation that only need to be completed once. These include inputting our data stream, $d$, PSD information, and transferring this information to the GPU. Following these initial steps, we begin the calculation process that occurs each time we call the Likelihood function.

The first part of the \phenomhm creation process, which involves determining fitting constants from input parameters, is implemented in the \texttt{C++} programming language. Structures containing these constants are then copied to the GPU. From here, the actual waveform creation from these fitting constants is implemented in parallel across the GPU because each frequency point is independent of one another. With the waveform constants, the amplitude and phase are generated and stored on the GPU. We then calculate $t(f)$ using \mbox{Equation \ref{eq:t_f}}. $t(f)$ is input into the response function with the extrinsic parameters. $\mathcal{T}(f,t_{lm}(f))$ is calculated on the GPU and stored on the GPU for each frequency point. At this stage, $A(f)$, $\phi(f)$, and $\mathcal{T}(f,t_{lm}(f))$ are represented as smooth functions with $\sim2^{10}$ frequency points. In LISA analysis, the data streams can be represented by upwards of $~10^6-10^7$ points depending on the sampling frequency and duration of observation. Therefore, to maintain speed in data analysis, these smooth functions describing the waveform and detector response are interpolated to the resolution desired. We designed a cubic spline interpolation algorithm, for each of these smooth functions, based on the tridiagonal nature of the cubic spline coefficient solution in the \texttt{Scipy} library \citep{scipy} and the sparse matrix computational tools included in the \texttt{cuSparse} library provided by NVIDIA \citep{cuSparse_2}. This allowed us to avoid memory transfers necessary to perform spline interpolation on the CPU. The final step of waveform creation is to interpolate all input smooth functions to create the signal in each TDI channel. In this step, we weight by the PSD. With the waveform calculated and on the GPU, we use the \texttt{cuBLAS} library \citep{cuBLAS_2} to compute complex dot products for the discretized inner product in \mbox{Equation \ref{eq:NLL}} and \mbox{Equation \ref{eq:innerprodsum}}. The likelihood is then calculated and returned to the sampling program. 

For ease of usage, our program is wrapped into Python using \texttt{Cython} \citep{Cython}. This allows us to run python sampling packages using our GPU-accelerated likelihood calculation. \texttt{Cython} generally does not interface with the CUDA-based compiler \texttt{nvcc}. However, we wrote a special wrapping for it based on \citet{CUDAwrapper}.

Speed comparisons are shown in \mbox{Figure \ref{fig:timecomp}}. Three configurations are shown: 1 CPU; 1 CPU and 1 GPU (host and device); and 28 CPUS and 2 GPUs. The last configuration vectorizes the computation over an array of sources. Therefore, these timings are run in vectorized form and then divided by the number of binaries in the array. As the points in the waveform increase, the difference between the CPU and GPU grows. The time of GPU evaluations barely increases until $\sim10^6$ points. Above this number, the GPU is saturated, running a high percentage of its available threads constantly.

This increase in speed allows us to perform our analysis in a unique way. Traditionally, in a zero-noise representation, each spherical harmonic mode is treated separately in the Likelihood computation (see \citet{Marsat2020LISAPE} for an example of this method). CPUs are not fast enough to efficiently compute the number of points necessary for combining all modes at a high enough resolution to preserve the fine structure created during the mixing of the higher modes. However, this is a balancing act because as more modes are added, the cross-terms between modes in \mbox{Equation \ref{eq:NLL}} become extensive and prohibitive. With GPUs, the time for the calculation allows us to forgo this question and produce templates at a high enough resolution to combine the modes into a one-dimensional data stream. This prevents the calculation of cross-terms and allows us to caclulate one inner product per TDI channel. Additionally, CPU speed prohibits the timely evaluation of Likelihoods with noise infused in the injection. With $T_\text{obs}\sim$yrs, $\Delta f$ in the Fourier transform is $\sim10^{-8}$Hz, leading to data streams with $\sim10^7$ points. Similarly, for a typical source expected for LISA with $M_T=10^6M_\odot$, the waveform must be calculated at $\sim5\times10^6$ points. For lower masses, this is even larger. Therefore, CPUs have large difficulty performing this calculation with noise involved without including approximations and other tricks to speed up the calculation. The GPUs, on the other hand, facilitate the use of the brute-force calculation, without approximations. In the future, we may test different approximations in our GPU implementation to gain even more performance. 

\section{Sampling Methods}\label{sec:mcmc}
There have been a large variety of methods developed to illuminate the posterior distribution, $p(\vecTh|D)$, as well as to calculate the evidence, $Z$. Two commonly used methods include Nested Sampling \citep{Skilling2004, Skilling2006} and Markov Chain Monte Carlo (MCMC) methods, for which the most commonly used type is the Metropolis-Hastings (M-H) algorithm \citep{Metropolis1949, Metropolis1953, Hastings1970}. Nested sampling is generally used to estimate the evidence, $Z$, by effectively performing a numerical integration across the prior volume. MCMC, on the other hand, is used to sample directly from the posterior to produce marginalized posterior distributions across the parameters tested. In this work, we will employ only MCMC techniques to illuminate our posterior distributions for the injected signals we examine. However, we have tested our accelerated likelihood with Nested Sampling techniques and have confirmed its possible usage as a search method. This will be included in future work. For actual sampling, we use variants of the typical MCMC techniques. Below, we will give a brief description of the algorithms used, but refer the interested reader to the cited papers for further details on the specific algorithms and their implementations.

\subsection{Markov Chain Monte Carlo}\label{sec:emcee}

The implementation we use for MCMC methods is based on the Python packages \texttt{emcee} \citep{emcee} and \texttt{ptemcee} \citep{Vousden2016}. Please see their release papers and documentation for details on specific constructions. The most popular form of MCMC is the M-H algorithm. Starting with a position $\vecTh_t$ at step $t$, a new position, $\vecTh_{t+1}$, is sampled from a transition distribution $Q(\vecTh_{t+1}|\vecTh_t)$. This new point is then accepted with a probability given by,
\begin{equation}
    \text{min}\left( 1, \frac{\Like(\vecTh_{t+1})}{\Like(\vecTh_t)} \frac{Q(\vecTh_{t+1}|\vecTh_t)}{Q(\vecTh_t|\vecTh_{t+1})} \right).
\end{equation}
A usual choise for $Q$ is a multivariate Gaussian distribution centered around $\vecTh_t$. This procedure has strong dependence on tuning the distribution for $Q$ (the covariance matrix for a multivariate Gaussian) with order $\sim D^2$ tuning parameters. Additionally, the M-H algorithm can be slow to convergence if tuning is not optimal \citep{emcee}.

\citet{Goodman2010} proposed a different method for MCMC sampling that significantly outperforms M-H in many situations. Specifically, this means the autocorrelation time (see \mbox{Section \ref{sec:autocorrelation}}) of the MCMC chains is much less, indicating less Likelihood evaluations per independent sample. This method is refered to as Affine Invariant MCMC (AIMCMC). In AIMCMC, the sampler evolves an ensemble of $K$ walkers $S=\{\vecTh_k\}$. The algorithm proposes a new position for the $k$th walker based on the other $K-1$ walkers in the ensemble ($S_{[k]} = \{\vecTh_j, \forall j \neq k\}$). First, a walker is chosen at random from $S_{[k]}$, represented by $\vecTh_j$. Then, a new position is proposed given by,
\begin{equation}
    \vecTh_{k,t}\rightarrow \vecTh_{k,t+1} = \vecTh_j + Y\left[\vecTh_{k,t} - \vecTh_j\right],
\end{equation}
where Y is a random variable distributed according to the probability density $g(Y=y)$, given by \citep{Goodman2010},
\begin{equation}\label{eq:g_of_y}
    g(y) = 
  \begin{cases}
    \frac{1}{\sqrt{y}} & \text{if}\ y\in \left[\frac{1}{a},a\right], \\
    0 & \text{otherwise},
  \end{cases}
\end{equation}
where $a$ is a tuning parameter that \citet{Goodman2010} set to 2. This proposal is referred to as the ``stretch'' proposal. The proposed point will be accepted according to the probability, $a_\text{str}$, given by,
\begin{equation}
    a_\text{str} = \text{min}\left( 1, Y^{(D-1)} \frac{\Like(\vecTh_{k,t+1})}{\Like(\vecTh_{k,t})} \right).
\end{equation}

\subsection{Parallel Tempering}\label{sec:paratemp}

In order to efficiently explore the entire prior range, we employ parallel tempering \citep{Swendsen1986, Earl2005}. Specifically, we use the implementation from \citet{Vousden2016} and its associated Python package, \texttt{ptemcee}. Here, we present the main aspects of parallel tempering, but we refer the interested reader to \citet{Vousden2016}, and sources within, for more detailed information. 

In parallel tempering, the Likelihood in \mbox{Equation \ref{eq:Bayes_rule}} becomes $\Like(\vecTh)^{1/T_i}$, where $T_i$ is the temperature of a given chain. Groups of MCMC walkers are assigned to a specific temperature arranged in a ladder with $N$ rungs: $T_i=T_1<T_2<...<T_N$. $T_1$ is 1, indicating chains assigned to that temperature represent the target distribution. As temperature increases, the tempered distribution becomes more and more representative of the prior distribution rather than the target distribution. 

Each chain explores its tempered distribution using the AIMCMC algorithm with stretch proposals. At specified intervals, chains can swap temperatures according to an M-H acceptance criterion given by \citep{Vousden2016},
\begin{equation}\label{eq:a_swap}
    a_\text{swap} = \text{min}\left( 1, \left[\frac{\Like(\vecTh_{i})}{\Like(\vecTh_{j})}\right]^{\beta_j-\beta_i} \right),
\end{equation}
where $\beta_i=1/T_i$ is the inverse temperature of chain $i$. The form of \mbox{Equation \ref{eq:a_swap}} indicates that temperature swaps usually occur between chains at adjacent rungs in the temperature ladder. We set the max temperature to infinity per the \texttt{ptemcee} documentation. Setting the max temperature to infinity means the highest temperature chain is exclusively exploring the prior distribution. We also use the adaptive tempering option. For more details on the adaptive method, please see the original paper. 

The parallel tempered run for each binary was performed under the same settings. With the highest temperature, $T_{10}$, set to $\infty$, the remaining temperatures are geometrically spaced set by \texttt{ptemcee} defaults using log-spacing between 1 and $(2.04807)^8$ (2.04807 is chosen based on the dimensionality of the parameter space). We set the scale factor ($a$) in \mbox{Equation \ref{eq:g_of_y}} to 1.15. We found the default setting of 2 was showing an undesirably small acceptance fraction for proposed jumps. We run a burn in of $10^4$ steps for each run. In \citet{Marsat2020LISAPE}, it is shown that we can generally expect 8 sky location posterior modes in the LISA reference frame (as opposed to the SSB reference frame). The longitudinal modes are located at values separated by $\pi/2$ from the true longitude value. Each of these four values has an associated latitudinal value that is either above or below the LISA orbital plane at $\pm\beta$. We use this information to ensure faster convergence in our sampler by placing 4 walkers in each temperature group at each of the 8 sky modes. It is clear from our analysis that our burn in phase allows the walkers to spread out accordingly away from modes that are not likely. We also verified that if we start all the walkers on the correct mode, the walkers spread out and locate the other modes if there is posterior weight there thanks to the tempering technique.

\subsection{MCMC Autocorrelation Time}\label{sec:autocorrelation}

The posterior samples generated in an MCMC analysis are not independent. Therefore, we estimate the autocorrelation within a chain, referred to as the autocorrelation time, $\tau_f$, following \citet{Sokal1997} and \citet{autocorrelation_2}. With multiple chains sampled based on methods described above, we work to estimate the mean, $\hat{\mu}$, and variance, $\hat{\sigma}^2$, of the posterior distribution. The sampling variance, $\sigma^2$, on these estimators is given by,
\begin{equation}
    \sigma^2 = \frac{\tau_f}{N}\hat{\sigma}^2.
\end{equation}
Therefore, the effective sample size (ESS) is N/$\tau_f$, representing the number of samples necessary to reduce the variance in the estimator to an acceptable value. 

We estimate the autocorrelation time  with the estimator for the normalized autocorrelation function, $\hat{\rho}_f(\tau)$. This is based on the chain generated for each walker, $\{f_n\}_{n=1}^N$, and is given by,
\begin{equation}
    \hat{\rho}_f(\tau) = \frac{\hat{c}_f(\tau)}{\hat{c}_f(0)},
\end{equation}
with
\begin{equation}
    \hat{c}_f(\tau) = \frac{1}{N-\tau}\sum_{n=1}^{N-\tau}(f_n-\mu_f)(f_{n+\tau} - \mu_f).
\end{equation}
The mean of the chain, $\mu_f$, is given by,
\begin{equation}
    \frac{1}{N}\sum_{n=1}^Nf_n.
\end{equation}
The integrated autocorrelation estimator is given by,
\begin{equation}\label{eq:autocorrelation_time}
    \hat{\tau}_f(M)=1+2\sum_{\tau=1}^M \hat{\rho}_f(\tau),
\end{equation}
where $M\ll N$. \citet{Sokal1997} suggests using the smallest value of $M$ that satisfies $M\geq 5\hat{\tau}_f(M)$. With the multiple chains generated with \texttt{emcee}, the variance in the estimator is reduced. Therefore, according to \citet{autocorrelation_2}, this procedure is satisfactory for chains longer than $50\hat{\tau}_f$. We thin the chains by this value before plotting posteriors to achieve a set of effectively independent samples. 

We run our sampler without checking the ESS. We ensure the ESS is greater than $10^4$ at the culmination of sampling. We report the autocorrelation time determined with \mbox{Equation \ref{eq:autocorrelation_time}}, the ESS, and the acceptance fraction for each run in \mbox{Table \ref{tb:mcmc_run_info}} in the Appendix. 

\section{Results and Discussion}\label{sec:results_discussion}

\begin{center}
\begin{table*}[tbh]
\caption{This table shows the injection parameters of each binary tested. The first row shows the parameter while the second row shows the units for each parameter. The final column displays the overall SNR for each source. It must be noted this table is meant to be a summary. The actual injection parameters are known to a much higher precision. As a reminder, the actual coalescence time is $t_\text{coal}=t_\text{ref}+t_0$, where $t_0$ is 1 year.
\newline $^1$ This binary was injected without a noise realization.
\newline $^2$ $a_1$ and $a_2$ are fixed during sampling.}\label{tb:testinfo}
\begin{tabular}{ C{1.2cm} C{1.2cm} C{1.2cm} C{1.2cm} C{1.2cm} C{1.2cm} C{1.2cm} C{1.2cm} C{1.2cm} C{1.2cm} C{1.2cm} C{1.2cm}  C{1.2cm}}
 \hline\hline
Binary & $M_T$ & $q$ & $a_1$ & $a_2$ & $D_L$ & $\phi_\text{ref}$ & $\iota$ & $\lambda$ & $\beta$ & $\psi$ & $t_\text{ref}$ & SNR \\ 
 - & $M_\odot$ & ($q\leq1$) & - & - & Gpc & - & - & - & - & - & sec & - \\
 1 & $2\times10^6$ & 1/3 & 0.0 & 0.0 & 36.59 & 2.14 & 1.05 & -0.024 & 0.62 & 2.03 & 50.25 & 588\\
  2$^1$ & $2\times10^6$ & 1/3 & 0.0 & 0.0 & 36.59& 2.14 & 1.05 & -0.024 & 0.62 & 2.03 & 50.25 & 588 \\
   3$^2$ & $2\times10^6$ & 1/3 & 0.0 & 0.0 & 36.59 & 2.14 & 1.05 & -0.024 & 0.62 & 2.03 & 50.25 & 588 \\
    4 & $2\times10^6$ & 1/3 & 0.85 & 0.88 & 36.59 & 2.14 & 1.05 & -0.024 & 0.62 & 2.03 & 50.25 & 878\\
     5 & $2\times10^6$ & 1/3 & -0.83 & -0.89 & 36.59 & 2.14 & 1.05 & -0.024 & 0.62 & 2.03 & 50.25 & 473\\
      6 & $4\times10^7$ & 1/5 & 0.0 & 0.0 & 15.93 & 3.10 & 1.34 & 4.21 & -0.73 & 0.14 & 23.93 & 310 \\
       7$^1$ & $3\times10^5$ & 7/10 & 0.0 & 0.0 & 70.58 & 1.22 & 1.32 & 2.24 & -0.22 & 2.73 & 94.33 & 40 \\
\hline
\end{tabular}
\end{table*}
\end{center}

\begin{center}
\begin{table*}[tbh]
\caption{The SNR contribution from each mode is shown below in addition to the full SNR of the source. The mode SNRs are computed in isolation to avoid confusion from mode mixing. This is to highlight in a general sense the contribution to the total SNR from each mode. Since the various modes mix with each other, the total SNR is not equal to the quadrature sum of all of the modes.}\label{tb:snr_contribution}
\begin{tabular}{ C{1.2cm} C{1.2cm} C{1.2cm} C{1.2cm} C{1.2cm} C{1.2cm} C{1.2cm} C{1.2cm}}
 \hline\hline
Binary & All & (2,2) & (3,3) & (4,4) & (2,1) & (3,2) & (4,3) \\ 
1-3 & 588  & 552 & 190 & 97 & 36 & 21 & 5\\
4 & 878 & 820 & 279 & 146 & 49 & 31 & 8\\
5 & 474  & 443 & 150 & 74 & 50 & 17 & 4\\
6 & 311 & 230 & 217  & 197 & 25 & 57 & 38 \\
7 & 40 & 40 & 3 & 2 & $<$1 & $<$1 & $<$1\\
\hline
\end{tabular}
\end{table*}
\end{center}

Our first aim of this paper is to build upon \citet{Marsat2020LISAPE} by adding aligned spins to our waveform and injecting noise within our fabricated data streams. To achieve this we run four different sets of binary parameters. We begin with the main source used in \citet{Marsat2020LISAPE} with $\{M_T=2\times10^6M_\odot,q=1/3,z=4\}$ (injection parameters for all binaries tested are shown in \mbox{Table \ref{tb:testinfo}}). Binary 1 is examined with Schwarzschild MBHs and injected noise. However, even with the $a_1=a_2=0$ injection, we allow the sampler to examine other spins maintaing a $D=11$ dimensional parameter space. 

Binary 2 is the same as binary 1, but with a pure waveform injection with no noise realization (also referred to as the zero-noise representation). This setup allows us to determine what the posterior would look like averaged over many runs with noise. 

For Binary 3, which is a similar injection to Binary 1, we fix the spins in the sampler so that we are sampling in a $D=9$ dimensional space, which allows for a closer comparison with \citet{Marsat2020LISAPE} in terms of the sampling methods (we do use a different waveform model). In addition to Schwarzchild injections, we want to examine how including spinning MBHs changes the posterior distributions while holding the other parameters constant. 

Binary 4 represents the same set of MBHs as Binaries 1-3 with $a_1\approx0.85$ and $a_2\approx0.88$. A set of antialigned spins is tested in Binary 5 with $a_1\approx-0.83$ and $a_2\approx-0.89$. The non-zero spins were chosen from a uniform distribution between +(-)0.75 and +(-)0.9 for the aligned (anti-aligned) binary.

We also test a larger and smaller binary in terms of the total mass. The larger binary has $\{M_T=4\times10^7M_\odot,q=1/5,z=2\}$ and the smaller binary has $\{M_T=3\times10^5M_\odot,q=7/10,z=7\}$. Both large and small binaries were injected with Schwarzschild MBHs, but the spins were allowed to vary in sampling. The large binary was chosen to examine MBH binaries that merge at lower frequencies than the center of the LISA band. It was also injected with noise.

The small binary was chosen to represent large seed MBHs at a higher redshift. During repeated attempts to analyze the small binary injected in noise with a variety of trial sampler settings, we were unable to attain a converged posterior distribution because the low temperature chains were unable to explore the likelihood surface efficiently and expand throughout the parameter space to the level expected. This is likely due to the combination of a lower SNR, as well as the inability to gain extra information from the higher order modes. When we lowered the temperatures in an effort to suppress the effect of the posterior tails on the ability for the walkers to maneuver the likelihood surface, we found the chains exhibited $\tau_f=61$ with an ESS of 15744, which are usually positive indicators. We believe the posteriors would converge to the proper distribution if we were able to run our sampler for a longer time. Even with our accelerated likelihood computations, this became difficult. Therefore, we present results for the small binary in the zero-noise representation. In future work we will further examine this problem and work towards analyzing noise injections at lower masses.

All extrinsic quantities for the large and small binaries were sampled from uniform distributions from the entirety of the prior domain. See \mbox{Table \ref{tb:testinfo}} for a full breakdown of the parameters used. \mbox{Table \ref{tb:snr_contribution}} shows the SNR of each harmonic mode for each binary tested. Please note the modes do not directly add in quadrature due to mode mixing. Therefore, the total SNR shown in the ``all'' column will not be the quadtrature sum of the singular mode values. This is purely to indicate the type of effect each mode has in the characterization of each binary. It is clear from the values seen in \mbox{Table \ref{tb:snr_contribution}} that more higher modes will be needed to fully describe MBH binary systems since the likelihood is related to the SNR$^2$ \citep{Marsat2020LISAPE}. 

\begin{table*}
\caption{The table below shows the recovered parameters in comparison to the injected parameters for Binaries 1, 2, and 3. The one-dimensional 1$\sigma$ errors for each parameter is shown in addition to the recovered mean. The sky position and orientation parameters are not shown because they are inherently multi-modal due to, at minimum, reflection across the LISA orbital plane (see \mbox{Figure \ref{fig:sky_map_example}}). Binary 1 represents similar paramters to those shown in \citet{Marsat2020LISAPE} with Schwarzschild MBHs injected; however, the sampler is free to examine spins other than zero. This binary also included a noise realization generated according to \mbox{Equation \ref{eq:noise_gen}}. Binary 2 is the same injection and sampler settings as Binary 1. This binary, however, does not include a noise realization allowing the likelihood to be calculated in a zero-noise representation. Binary 3 has the same injection as Binary 1 with a noise realization. However, the spin parameters are fixed during sampling, therefore, covering a parameter space of dimensionality $D=9$ rather than $D=11$. This is why ``N/A'' is shown for the recovery of these spin parameters.}\label{tb:recover1}
\begin{center}
\begin{tabular}{| C{1.8cm} | C{1.8cm} | C{3.0cm} | C{3.0cm} | C{3.0cm} |}
 \cline{3-5}
 \multicolumn{2}{c|}{} & Binary 1 & Binary 2 & Binary 3 \\
\hline
Parameter & Injection & Recovered & Recovered & Recovered \\
\hline
$\ln{M_T}$ & 14.50866 & 14.50913$\substack{+0.00075 \\ -0.00075}$ & 14.50857$\substack{+0.00080 \\ -0.00081}$ & 14.50787$\substack{+0.00061 \\ -0.00054}$  \\
\hline
$q$ & 0.33333 & 0.33296$\substack{+0.00085 \\ -0.00085}$ & 0.33330$\substack{+0.00095 \\ -0.00080}$ & 0.33439$\substack{+0.00072 \\ -0.00092}$  \\
\hline
$a_1$ & 0.0 & -0.0047$\substack{+0.0049 \\ -0.0046}$ & -0.0012$\substack{+0.0048 \\ -0.0047}$ & N/A \\
\hline
$a_2$ & 0.0 & 0.028$\substack{+0.018 \\ -0.019}$ & 0.003$\substack{+0.018 \\ -0.018}$ & N/A \\
\hline
$\ln{D_L}$ & 3.581 & 3.603687$\substack{+0.020 \\ -0.020}$ & 3.604$\substack{+0.021 \\ -0.021}$ & 3.617$\substack{+0.018 \\ -0.018}$\\
\hline
$\phi_\text{ref}$ & 2.140 & 2.36$\substack{+0.14 \\ -0.15}$ & 2.16$\substack{+0.14 \\ -0.14}$ & 2.116$\substack{+0.022 \\ -0.018}$ \\
\hline
$t_\text{ref}$ & 3.9169 & 3.69$\substack{+0.17 \\ -0.20}$ & 3.89$\substack{+0.13 \\ -0.15}$ & 3.9238$\substack{+0.0074 \\ -0.0081}$ \\

\hline
\end{tabular}
\end{center}
\end{table*}

\begin{table*}
\caption{Similar to \mbox{Table \ref{tb:recover1}}, the parameter recovery for Binaries 4 and 5 are shown. Binary 4 has the same injection parameters and sampler settings as Binary 1 with the exception of the injected spin parameters. This binary is injected in a spin configuration with high aligned spins for each MBH. Binary 5 is similar to Binary 4, but it is injected in a high anti-aligned spin configuration.}\label{tb:recover2}
\begin{center}
\begin{tabular}{| C{1.8cm} | C{1.8cm} | C{3.0cm} | C{1.8cm} | C{3.0cm} |}
 \cline{2-5}
 \multicolumn{1}{c|}{} & \multicolumn{2}{c|}{Binary 4} & \multicolumn{2}{c|}{Binary 5} \\
\hline
Parameter & Injection & Recovered & Injection & Recovered \\
\hline
$\ln{M_T}$ & 14.508658 & 14.50865$\substack{+0.00037 \\ -0.00042}$ & 14.5087 & 14.5090$\substack{+0.0013 \\ -0.0015}$ \\
\hline
$q$ & 0.33333 & 0.33335$\substack{+0.00052 \\ -0.00047}$ & 0.3333 & 0.3323$\substack{+0.0016 \\ -0.0014}$\\
\hline
$a_1$ & 0.85697 & 0.85678$\substack{+0.00059 \\ -0.00059}$ & -0.8298 & -0.8291 $\substack{+0.0080 \\ -0.0079}$ \\
\hline
$a_2$ & 0.8830 & 0.8846$\substack{+0.0033 \\ -0.0033}$ & -0.887 & -0.895 $\substack{+0.027 \\ -0.030}$ \\
\hline
$\ln{D_L}$ & 3.600 & 3.615$\substack{+0.015 \\ -0.015}$ & 3.600 & 3.591$\substack{+0.027 \\ -0.026}$ \\
\hline
$\phi_\text{ref}$  & 2.140 & 2.150$\substack{+0.029 \\ -0.029}$ & 2.14 & 2.11$\substack{+0.16 \\ -0.17}$ \\
\hline
$t_\text{ref}$ & 3.917 & 3.907$\substack{+0.022 \\ -0.023}$ & 3.92 & 3.95 $\substack{+0.18 \\ -0.20}$  \\

\hline
\end{tabular}
\end{center}
\end{table*}

\begin{table*}
\caption{Recovered parameters for Binaries 6 and 7 are shown below similar to \mbox{Table \ref{tb:recover1}}. Binary 6 represents a completely different binary from Binaries 1-5 in terms of injection parameters. Its total mass is chosen to represent a larger binary about an order of magnitude larger than Binaries 1-5. It is, however, also injected with Schwarzschild MBHs in a noise realization. Binary 7 represents a binary with total mass about an order of magnitude smaller than Binaries 1-5. It is also injected with Schwarzschild MBHs. However, this binary, like Binary 2, is analyzed in a zero-noise representation. This is due to issues achieving convergence with our sampler on this binary when it was injected with a noise realization. Achieving convergence in the presence of noise for this binary is a topic for future work. The spin parameters and $\phi_\text{ref}$ for Binary 7 show a multi-modal structure; therefore, their associated mean values and errors do not resemble the actual recovery that is observed in the posterior distributions. For this reason, these quantities are labeled with ``Multi.'' Please see the full posterior in \mbox{Figure \ref{fig:small_mass_posterior}} for more information.}\label{tb:recover3}
\begin{center}
\begin{tabular}{| C{1.8cm} | C{1.8cm} | C{3.0cm} | C{1.8cm} | p{3.0cm} |}
 \cline{2-5}
 \multicolumn{1}{c|}{} & \multicolumn{2}{c|}{Binary 6} & \multicolumn{2}{c|}{Binary 7} \\
\hline
Parameter & Injection & Recovered & Injection & Recovered \\
\hline
$\ln{M_T}$  & 17.5044 & 17.5033$\substack{+0.0017 \\ -0.0021}$ & 12.612 & 12.609$\substack{+0.012 \\ -0.009}$ \\
\hline
$q$  & 0.2000 & 0.1990$\substack{+0.0009 \\ -0.0010}$ & 0.700 & 0.716$\substack{+0.079 \\ -0.072}$\\
\hline
$a_1$ & 0.0 & -0.0044$\substack{+0.0042 \\ -0.0053}$ & 0.0 & Multi \\
\hline
$a_2$ & 0.0 & -0.0038$\substack{+0.0039 \\ -0.0051}$ & 0.0 & Multi \\
\hline
$\ln{D_L}$ & 2.77 & 2.81$\substack{+0.19 \\ -0.13}$ & 4.27 & 4.58$\substack{+0.35 \\ -0.24}$ \\
\hline
$\phi_\text{ref}$ & 3.098 & 3.103$\substack{+0.013\\ -0.036}$ & 2.4 & Multi \\
\hline
$t_\text{ref}$ & 3.2 & 2.6$\substack{+1.4 \\ -1.6}$ & 6.849 & 6.857$\substack{+0.010 \\ -0.014}$\\

\hline
\end{tabular}
\end{center}
\end{table*}

\begin{figure*}[tbh]
\begin{center}
\includegraphics[scale=0.45]{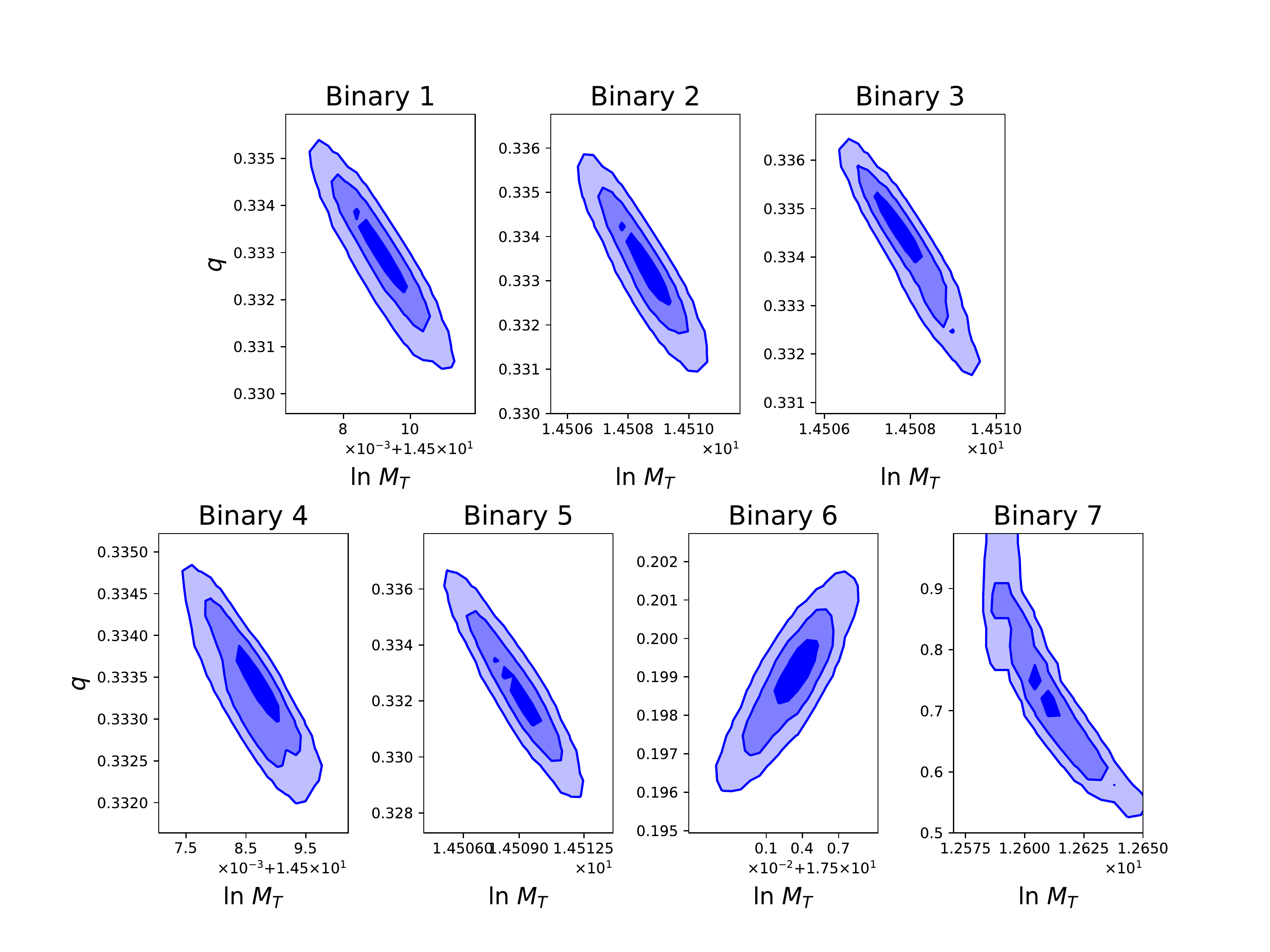}
ls \caption{The two-dimensional posterior distributions of $\ln{M_T}$ versus $q$ for all binaries tested are shown here. The binary is labeled in the title of each plot. All binaries, with the exception of Binary 6, display an anti-correlation between these parameters. Binary 6 shows a positive correlation. As explained in \mbox{Section \ref{sec:results_discussion}}, this is due to the slope of the sensitivity curve at frequencies around the merger of the binary where the signal is strongest.}\label{fig:mt_vs_mr}
\end{center}
\end{figure*}

In the following analysis, we will present subsections of the overall corner plots representing the parameter posterior distributions; we focus on specific interesting and/or important posterior distributions. First, we will discuss intrinsic parameters, followed by a discussion on sky localization constraints and other extrinsic quantities. The full corner plot for each binary is shown in \mbox{Section \ref{sec:corner_plots}} of the Appendix. Recovered parameter values for $\ln{M_T}$, $q$, $a_1$, $a_2$, $\ln{D_L}$, $\phi_\text{ref}$, and $\ln{t_\text{ref}}$, including their means and 1$\sigma$ errors, are shown in \mbox{Tables \ref{tb:recover1}-\ref{tb:recover3}}. Binaries 1, 2, and 3 are shown in \mbox{Table \ref{tb:recover1}}. \mbox{Table \ref{tb:recover2}} contains parameters for Binaries 4 and 5. Binaries 6 and 7 are shown in \mbox{Table \ref{tb:recover3}}. These tables do not include the sky location  and orientation because these parameters are inherently multi-modal rendering their one dimensional means and 1$\sigma$ errors not representative of the true recovery values. 

\subsection{Intrinsic Parameters}

\begin{figure*}[tbh]
\begin{center}
\includegraphics[scale=0.45]{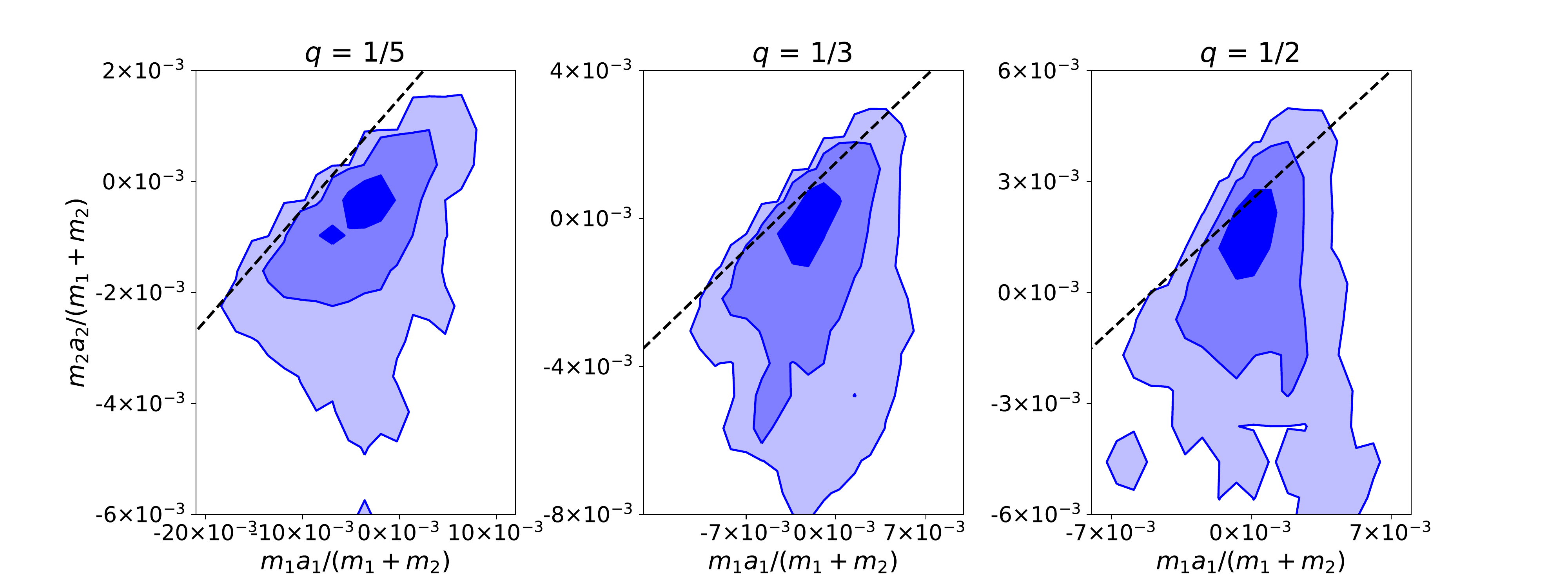}
\caption{This plot shows the two-dimensional posterior distributions for the mass-weighted spin parameters for the large binary (Binary 6). The posteriors show a unique non-ellipsoidal shape that is not observed for the other binaries. The dashed line in each plot has a slope equal to the mass ratio to show the steep likelihood dropoff towards the upper left is approximately parallel to this line. The large binary exhibits almost its entire signal in the merger and ringdown portion of coalescence. Therefore, it is more sensitive to quantities like the reference frequency that sets the location of the merger-ringdown. The reference frequency, $f_\text{ref}$, is set by the symmetric combination of the mass-weighted spins. Since this binary has similar values for $a_1$ and $a_2$, the injection has a larger mass-weighted contribution from its primary. For this reason, we see the likelihood is very sensitive to changes in the spins when the mass-weighted spin contribution from the secondary is larger than the primary's contribution.}\label{fig:large_spin_comp}
\end{center}
\end{figure*}

%\subsection{Total Mass and Mass Ratio}\label{sec:mt_vs_mr}
The first marginalized posterior we address is the two-dimensional posterior for $\ln{M_T}$ and $q$. \mbox{Figure \ref{fig:mt_vs_mr}} shows this parameter space for each binary. There are two aspects to point out. First, Binaries 1-5 with the same injection mass and mass ratio, as well as the small binary (Binary 7), show an anti-correlated behavior, while the posterior for the larger binary (Binary 6), shows a positively correlated behavior. The difference involves the position of the merger and ringdown over the sensitivity curve, or more specifically, where the majority of the signal is accumulated. For the binaries with $M_T=2\times10^6M_\odot$, the peak of the merger and ringdown occur at frequencies around or above $10^{-2}$Hz. In this region, the sensitivity curve has a positive slope (see \mbox{Figure \ref{fig:signal_examples}}). At small deviations from the highest likelihood point, the signal, as it changes with varying parameters, must follow the shape of the sensitivity in order to maintain a similar likelihood value, thereby creating this observed correlation. The effect of increasing the total mass by a small perturbation will move the signal slightly to higher strains and lower frequencies. Therefore, in order for this signal with a slightly larger mass to follow the sensitivity curve, we will need to decrease the mass ratio, which causes the signal to decrease in strain. For this reason, we see an anti-correlation between these two parameters. Once again, in the limit of small deviations, the actual signal shape will not change significantly. 

This description also applies to the small binary because it evolves at higher frequencies than Binaries 1-5, indicating the sensitivity curve is also sloping upwards where a majority of its signal is accumulated. However, since the mass ratio is not strongly constrained due to the lack of information from higher modes, the morphology of the signal does change slightly over the range of mass ratios observed in the posterior distribution. This effect adds curvature to the posterior ellipse. 

The opposite is true for the large binary. At the frequencies at which the large binary merges and rings down ($\sim10^{-3}$Hz), the sensitivity curve has a negative slope. As the total mass is increased, the signal will move to higher strains and lower frequencies. However, due to the slope of the sensitivity curve, an increase in the mass ratio is also needed to ensure the signal actually follows this slope. Hence, we see a positive correlation between the two quantities. In addition to the correlation observed, the small binary exhibits a curved ellipsoidal shape and a mass ratio that is not strongly constrained. This is most likely due to the low or negligible signal contributed from the higher modes (see \mbox{Table \ref{tb:snr_contribution}}). 

When comparing Binaries 1-5 with one another, we see that most of the posterior distributions per parameter set are similar in character. There are a few interesting differences. First, most parameters, for all of these binaries, have similar errors that vary in comparison to one another in proportion with the SNR for each source (see \mbox{Table \ref{tb:snr_contribution}}). However, this is not true for the reference values, $\phi_\text{ref}$ and $t_\text{ref}$, for Binary 3 (fixed spins during sampling with $D=9$). Even though this binary has the exact same injection waveform as Binaries 1 and 2, the reference values are constrained to a higher degree, with error values similar to the higher SNR Binary 4 (spin up). The reference frequency at which these reference values are set is determined in part by the symmetric combination of the mass-weighted spins \citep{Husa2016, Khan2016}. Therefore, when we fix the spins during sampling, we are ensuring that two of four quantities (the others are total mass and mass ratio) that determine the reference frequency are fixed. This allows for better constraints on these reference values. Similarly, in \mbox{Figures \ref{fig:sylvain_1}-\ref{fig:sylvain_5}} we can see that Binary 3 (fixed spins during sampling) has unique posterior shapes when comparing the total mass and mass ratio to these reference values. It can be seen these posteriors show a higher degree of correlation or anti-correlation. Once again, with the spins fixed, the mass ratio and total mass are the two quantities that determine the reference frequency, therefore, causing the magnitude of correlation with these quantities to increase. 

When examining the large binary, it is clear its posterior distributions are much less ellipsoidal than those shown in Binaries 1-5. The base reason for this is the frequency range over which we observe this signal. We receive minimal information from the inspiral of this binary, receiving all information from the merger and ringdown. One peculiar intrinsic parameter posterior of note for the larger binary is the $a_1$ versus $a_2$ posterior as it is far from elliptical, and actually strongly asymmetric in relation to an anti-symmetric combination of mass-weighted spins. To further investigate this observation, we analyzed two more injections with the same parameters as Binary 6, but with differing mass ratios. \mbox{Figure \ref{fig:large_spin_comp}} shows three binaries similar to Binary 6 in the plane of the mass-weighted spin values. The $q=1/5$ binary is Binary 6. The other two are injected with all the same parameters, but differing mass ratios. All three binaries exhibit similar behavior in this plane. Towards the upper left of the plot, the likelihood surface drops off steeply. On the contrary, towards the lower right, the likelihood surface displays a shallow fall. The dashed lines represent a line with the slope equal to the mass ratio. Above (below) this line, the mass-weighted spin of the primary is smaller (greater) than the secondary. The purpose of this line is to illustrate that this steep drop off is generally parallel to a line with a slope equal to the mass ratio. For this large binary, where we receive minimal information from the inspiral, the swapping of the mass weighted spins causes the signal to change in comparison to the injection with varying rapidity that indicates the broken symmetry of the masses greatly affects the ability of the template to match the injection. At $q=1/5$ and $q=1/3$, the posteriors generally stay parallel in character to that line. As the mass ratio is increased to $q=1/2$, the mass weighting tends toward a more even symmetry, causing this effect to decrease; however, it still shows this non-elliptical behavior.  

The small binary (binary 7) has many unique posterior distributions due to its lower SNR and minimal information contributed from its higher harmonic modes (see \mbox{Figure \ref{fig:small_mass_posterior}}). For example, the spin posteriors show a multi-modal structure. $a_1$ shows a virtually equivalent mode at a slightly higher spin of $\sim$0.2. Similarly, $a_2$ shows a corresponding second mode at a lower spin of $\sim-0.3$. The ratio of these secondary spin modes is approximately equal to the mass ratio due to the effect of the mass-weighted symmetric spin during waveform construction. Even with this multi-modal structure, the $a_1$ versus $a_2$ posterior shows an ellipsoidal shape surrounding the two local maxima. 

\subsection{Sky Location and other Extrinsic Parameters}

\begin{figure}[tbh]
\begin{center}
\includegraphics[scale=0.43]{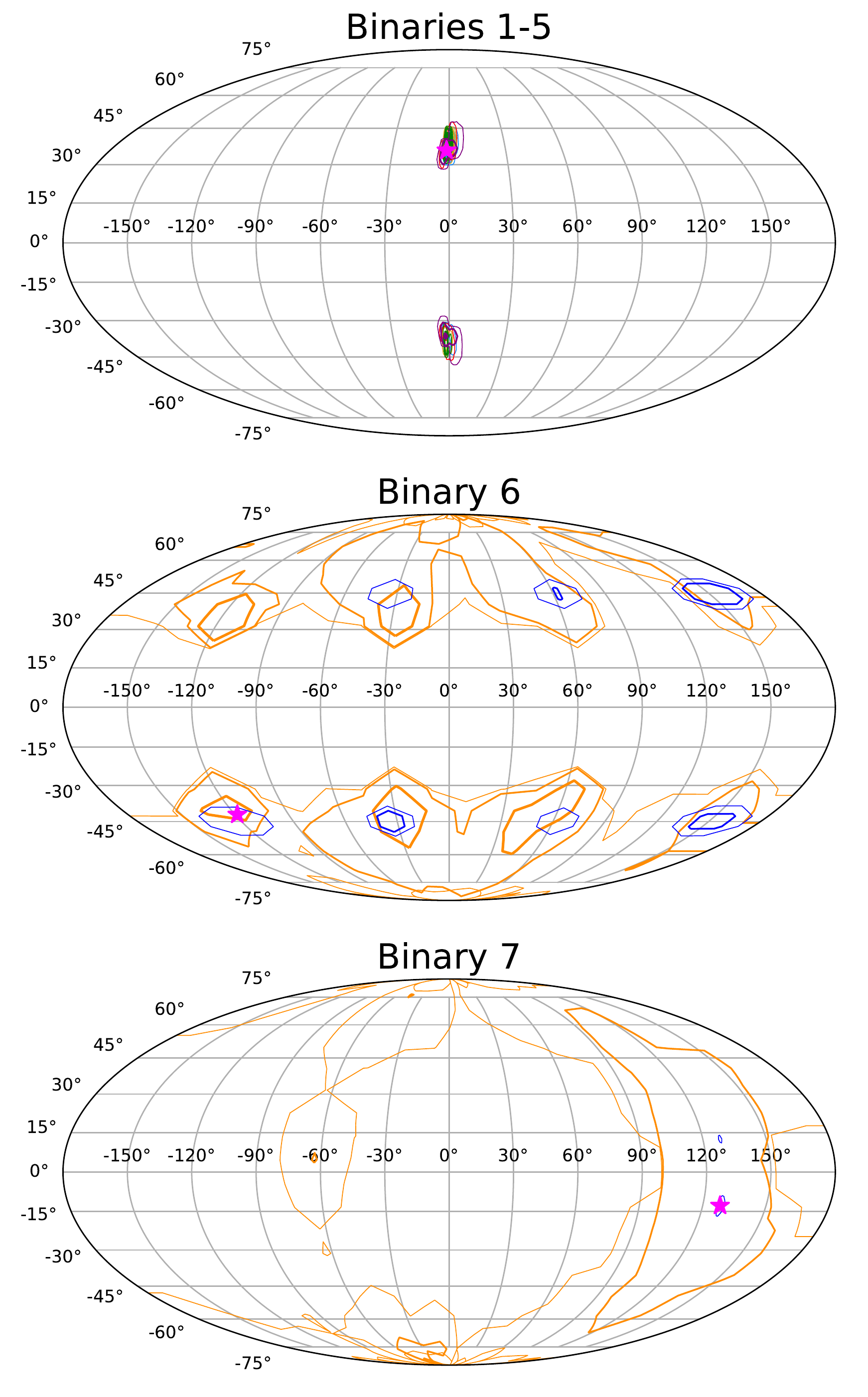}
\end{center}
\caption{The sky location from all binaries tested is shown above. Binaries 1-5 are shown in the top plot. Binaries 6 and 7 are shown in orange in the center and bottom plots, respectively. The 3$\sigma$, 2$\sigma$, and 1$\sigma$ contours are shown with increasing line thickness and the true injection point is marked by a magenta star. The large binary (Binary 6) has effectively no localization. Similarly, Binary 7 also has a large localization area. The inclination for both Binaries 6 and 7 indicate they are close to an edge-on configuration. To illustrate the difference compared to a more face-on configuration, we add sky locations in blue for both large and small binaries with the exact same parameters, but with an injection inclination of $\pi/8$. Binaries 1-5 are all concentrated in the center of the map, precisely around the correct longitude value for those sources. The degeneracy above and below the plane also exists. It cannot be seen easily, but there is more posterior weight around the correct ecliptic latitude for all Binaries 1-5 compared to the position opposite the LISA orbital plane.}\label{fig:sky_map_example}
\end{figure}

In terms of the ecliptic longitude and latitude, we observe what is generally expected. \mbox{Figure \ref{fig:sky_map_example}} shows the 1-3$\sigma$ contours for the sky localization of each binary. For Binaries 1-5, with strong SNRs, signal at higher frequencies, and measurements of higher modes, we are able to precisely identify the ecliptic longitude, while strongly recovering the ecliptic latitude. Also, as expected, we are unable to differentiate between between above and below the LISA orbital plane, even though more posterior weight is located at the correct latitudinal mode. The relative level to which we constrained the sky location for Binaries 1-5 is determined by their SNR, modulated slightly by the randomness of the noise. 

The sky localization achieved for the large binary is effectively minimal. This lack of localization in this specific case is due to Binary 6 existing in a near-edge-on configuration. In an edge-on configuration, the information from higher modes in terms of constraining the sky location is minimized. With the same binary injection parameters but an inclination in a more face-on configuration ($\iota=\pi/8$), the sky localization is reduced to the eight degenerate sky modes expected for binaries at lower frequencies (see \citet{Marsat2020LISAPE}). This configuration is shown for reference in \mbox{Figure \ref{fig:sky_map_example}}. In this specific case, seven of the eight modes are highlighted. The frequency spectrum for Binary 6 does not extend to high enough frequencies to constrain the longitude to one of the four longitudinal modes. However, if this binary was run in the ``spin up'' configuration performed for Binary 4, the frequency-domain spectrum would extend high enough allowing the localization to narrow to one longitudinal mode. We also observe uncertainty as to whether the source is truly above or below the plane.

Similar to the large binary, the sky location constraint for the small binary is weak. This is also due to the edge-on configuration. When analyzed in a more face-on configuration ($\iota=\pi/8$), the sky localization is much smaller, locating the correct longitudinal mode as is expected for a high-frequency binary. The sampler also located the correct latitude, but confusion as to whether the source is above or below the plane does exist. This face-on configuration is also shown in \mbox{Figure \ref{fig:sky_map_example}} for reference.

The other extrinsic orientation parameters follow a predictable structure related to the ecliptic longitude and latitude. The inclination is reflected at $-\cos{\iota}$ and $\psi$ is reflected at $\pi-\psi$. Additionally, since more than one longitudinal mode is observed for binary 6, $\psi$ is observed at deviations of $\pi/2$ from both the true and reflected values. Binaries 1-5 show Gaussian distributions for $\ln{D_L}$, $\ln{\phi_\text{ref}}$ and $\ln{t_\text{ref}}$. However, it can be seen that $t_\text{ref}$ correlates similarly to $\phi_\text{ref}$ as they both depend on the intrinsic parameters in terms of setting the reference frequency at which these values are set.

The large binary is the only injection where the reference time is not constrained well (see \mbox{Figure \ref{fig:large_mass_posterior}}). Since the merger and ringdown are short relative to an inspiral evolution, the movement of the LISA constellation, as well as its rotation, is small during the observation of this signal. We are effectively observing a snapshot of the temporal aspect of the response function; therefore, moving the binary forward or backward in time will not change the response of the spacecraft by much, especially since we are operating on the order of seconds for the reference time parameter. The behavior of $\phi_\text{ref}$ is reflective of the unique spin behavior detailed previously: $\phi_\text{ref}$ directly correlates with the abnormal distribution for $a_2$. Similar to the spin distributions, this is due to the setting of the reference frequency at which this phase is assigned. The luminosity distance for the large binary is constrained to within $\sim5$ Gpc and shows the mean value and main peak near the injection value. However, the distribution of the distance measurements is not Gaussian, tending towards higher distances. Without much inspiral signal, we lose information that helps to constrain the distance to the binary.

We are able to constrain $t_\text{ref}$ for the small binary quite well due to its long evolution in band (see \mbox{Figure \ref{fig:small_mass_posterior}}). $\phi_\text{ref}$ seems to show a ``split'' posterior. However, it is really a symptom of wrapping the phase at 2$\pi$. If we were to unwrap those posteriors, the areas that seem to be disjointed and separated would fit as one multi-modal posterior similar to those shown for the intrinsic parameters. The distance posterior for the small binary is unique. Similar to the large binary, this posterior tends towards larger distances than the injection. A difference with the small binary is that the mean value and main peak is not located at the true injection value. Since this binary is analyzed in a zero-noise representation, the random nature of noise did not cause this deviation. Without any noise, the maximum likelihood value would be found at the true distance value. However, since we do not see a main peak around this value, it means the parameter space volume centered around the true value does not contain a high percentage of posterior weight; our observations show there is more posterior weight around $\sim100$ Gpc.

\section{Conclusion}\label{sec:conclusion}

We present the first noise-infused analysis of MBH binaries with LISA employing full waveforms spanning inspiral, merger, and ringdown that include higher order harmonic modes. Additionally, we expand on previous work by adding the inclusion of aligned spins. This study was performed with \phenomhm \citep{London2018} and the fast frequency domain response from \citet{Marsat2018}. Both of these processes, as well as the likelihood computation, were accelerated with GPUs, which is a first for the LISA response function. We showed the extreme improvement in speed these GPU devices achieve. In instances where full data streams are analyzed with noise-infused injections, GPUs can achieve speeds $\sim500$ or more times faster than a single CPU.

With these accelerated likelihood computations, we used Bayesian inference techniques to extract full marginalized posterior distributions across the $D=11$ dimensional parameter space. In particular, we employed a parallel-tempered implementation of the stretch proposal MCMC sampler implemented in \citet{Vousden2016}. 

We obtained full posterior distributions for 7 test binaries, which can be seen in \mbox{Figures \ref{fig:sylvain_1} - \ref{fig:small_mass_posterior}} in \mbox{Section \ref{sec:corner_plots}} of the Appendix. The parameters of the test binaries can be seen in \mbox{Table \ref{tb:testinfo}}. For the intrinsic parameters, $\phi_\text{ref}$, $D_L$, and $t_\text{ref}$ we show the recovered parameter values and their 1$\sigma$ errors in \mbox{Tables \ref{tb:recover1}-\ref{tb:recover3}}. The errors generally follow proportionally to the SNR. However, when we fix the spins for Binary 3, we can further constrain the reference values because the spins help determine the frequency at which the reference values are set.

We highlighted the correlation between $\ln{M_T}$ and $q$ as it varied for the different binaries. They exhibited a positive (negative) correlation if the slope of the sensitivity curve near the merger, where most of the signal is accumulated, is negative (positive). 

The large binary (Binary 6) exhibits unique posteriors in $a_1$ versus $a_2$. We show this posterior exhibits a non-ellipsoidal structure that follows a slope similar to the mass ratio (see \mbox{Figure \ref{fig:large_spin_comp}}). Since the large binary signal is observed during only the merger and ringdown, this binary is more susceptible to likelihood changes when the secondary mass-weighted spin begins to dominate over the primary. 

The small binary (Binary 7) shows a multi-modal distribution most likely due to its lower SNR and lack of strong observation of higher order modes.

We also discuss our ability to localize each binary (see \mbox{Figure \ref{fig:sky_map_example}}). The inclination of the binary holds strong influence over its localization. Face-on configurations provided much better sky localization for all binaries compared to edge-on configurations.

In future work, we plan to expand on the findings shown here by growing the binary parameter space that we have looked at, focusing on astrophysically motivated catalogs. Additionally, we will examine how the sky localization for a binary changes over time with the new LISA sensitivity curve, as well as with newer, advanced waveforms like \phenomhm, such as the coming waveform family for \texttt{PhenomX} \citep{Pratten2020PhenomX, GarciaQuiros2020PhenomXHM, Pratten2020IMRPhenomXPHM}.

With this work, we see the aligned spins add a new dimension to the sampling of binaries resulting in interesting configurations in spin posteriors, as well as varying the constraints on the other parameters. We see that injected noise, as expected, has varied our ability to recover parameters. The analysis of these noise injections was made possible by the implementation of GPU-accelerated likelihood calculations, a tool that will help to further the reach of these parameter estimation studies in the run up to the launch of the LISA mission.

\begin{acknowledgments}
M.L.K. acknowledges support from the National Science Foundation under grant DGE-0948017 and the Chateaubriand Fellowship from the Office for Science \& Technology of the Embassy of France in the United States. M.L.K. would like to thank folks at the Labortatoire Astroparticule \& Cosmologie in Paris, including Edward K. Porter, Marc Arene, Calum Murray, and the LISA group at APC for being gracious hosts. A.J.K.C. acknowledges support from the Jet Propulsion Laboratory Research and Technology Development program. This research was supported in part through the computational resources and staff contributions provided for the Quest/Grail high performance computing facility at Northwestern University. Astropy, a community-developed core Python package for Astronomy, was used in this research \citep{Astropy}. Parts of this research were carried out at the Jet Propulsion Laboratory, California Institute of Technology, under a contract with the National Aeronautics and Space Administration (80NM0018D0004). This paper also employed use of Scipy \citep{scipy}, Numpy \citep{Numpy}, and Matplotlib \citep{Matplotlib}.
\end{acknowledgments}

%\appendix

%\section{Appendixes}


\appendix

\section{Posterior Distributions}\label{sec:corner_plots}

\begin{figure*}
\begin{center}
\includegraphics[scale=0.37]{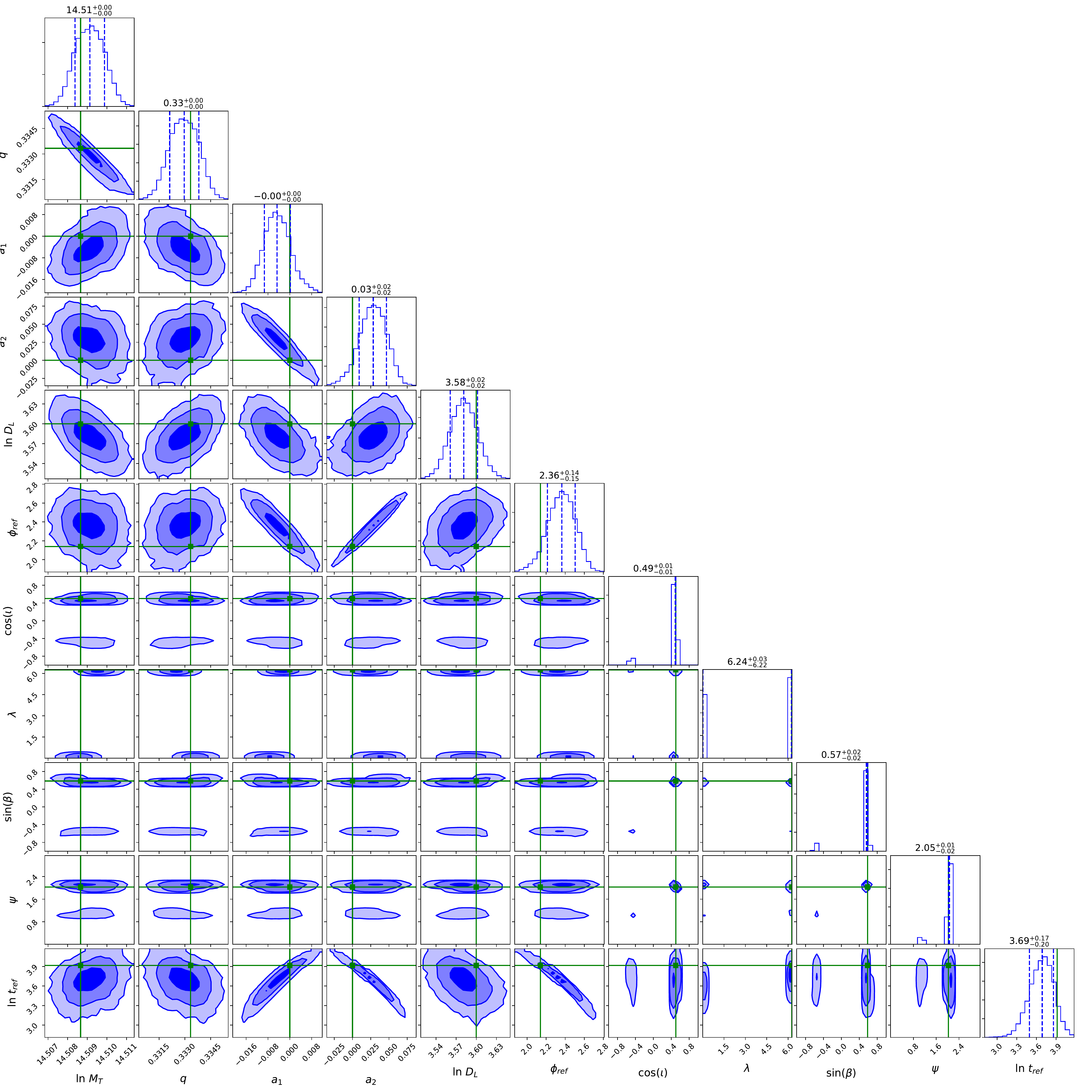}
\caption{The posterior distributions for all parameters for Binary 1 is shown above. Binary 1 was injected with Schwarzschild MBHs with a sampled noise realization. The sampler in this case was allowed to sample nonzero spins. Refer to the beginning of \mbox{Section \ref{sec:corner_plots}} of the Appendix for information on the corner plots.}\label{fig:sylvain_1}
\end{center}
\end{figure*}

\begin{figure*}
\begin{center}
\includegraphics[scale=0.37]{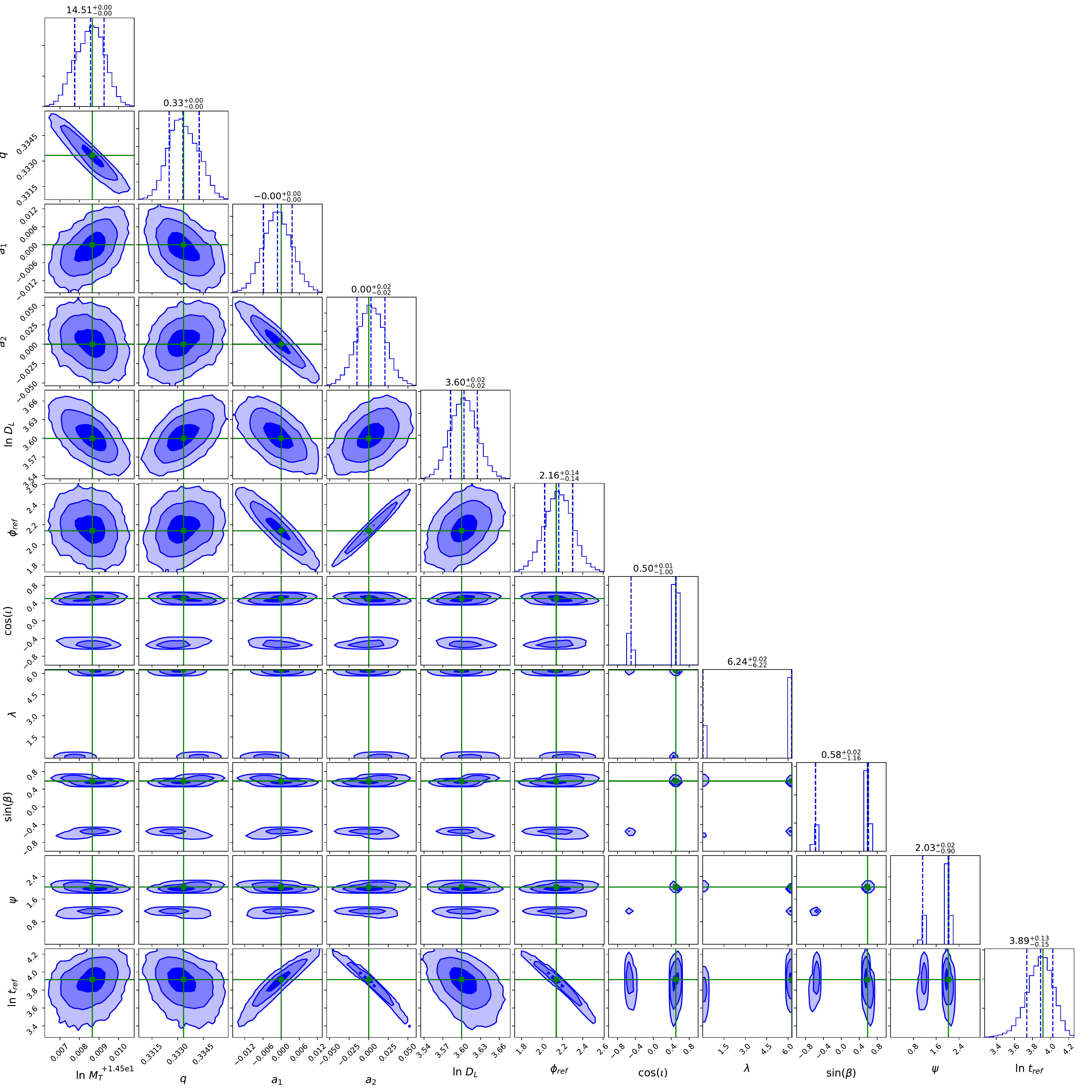}
\caption{Binary 2's posterior distributions are shown in this figure. Binary 2 was injected with Schwarzschild MBHs and analyzed in a zero-noise representation. Spins were allowed to very during sampling. Please see \mbox{Section \ref{sec:corner_plots}} of the Appendix for more information about the construction of the corner plots.}\label{fig:sylvain_2}
\end{center}
\end{figure*}

\begin{figure*}
\begin{center}
\includegraphics[scale=0.37]{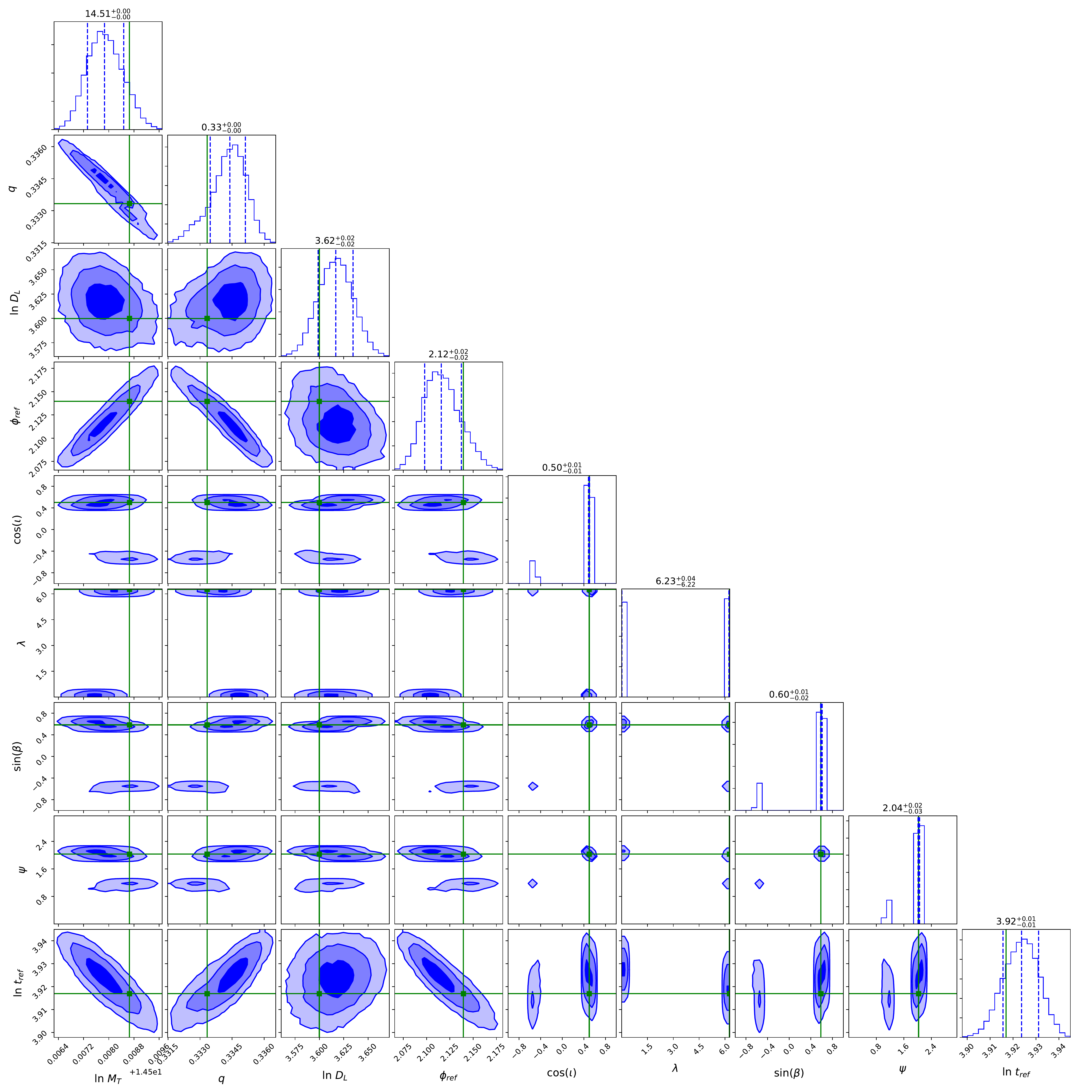}
\caption{The posterior distributions for Binary 3 are shown. Binary 3 was injected with Schwarzschild MBHs in a sampled noise realization. However, the sampler for this binary held the spins fixed, therefore, analyzing 9 parameters rather than 11. For this reason, spin distributions are not shown. \mbox{Section \ref{sec:corner_plots}} of the Appendix discusses the constructions used in the corner plots.}\label{fig:sylvain_3}
\end{center}
\end{figure*}

\begin{figure*}
\begin{center}
\includegraphics[scale=0.37]{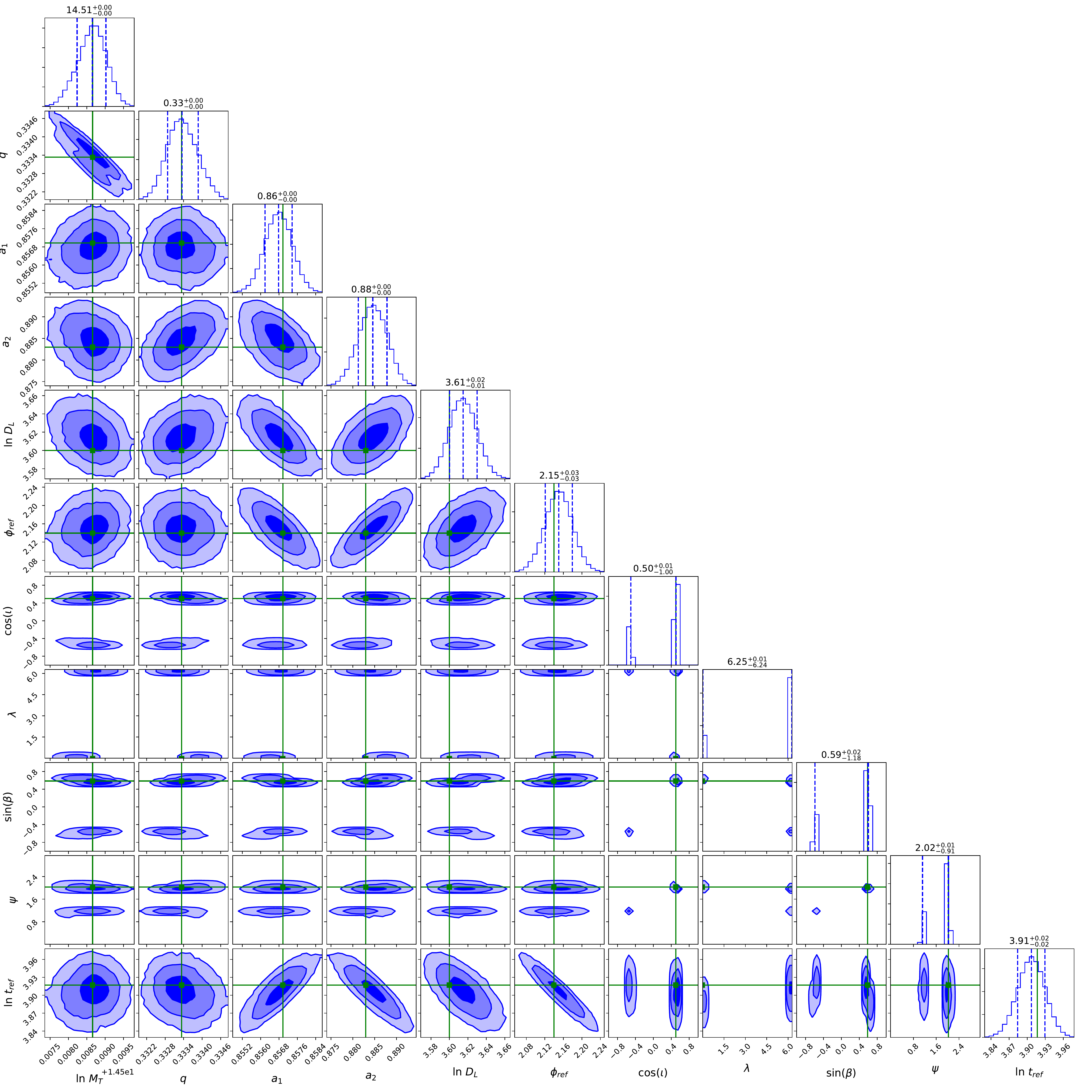}
\caption{The binary injected with strong aligned spins, Binary 4, is shown here. This binary was injected with a noise realization and the sampler was allowed to vary the spins during sampling. Please see \mbox{Section \ref{sec:corner_plots}} of the Appendix for information on the constructions involved in this plot.}\label{fig:sylvain_4}
\end{center}
\end{figure*}

\begin{figure*}
\begin{center}
\includegraphics[scale=0.37]{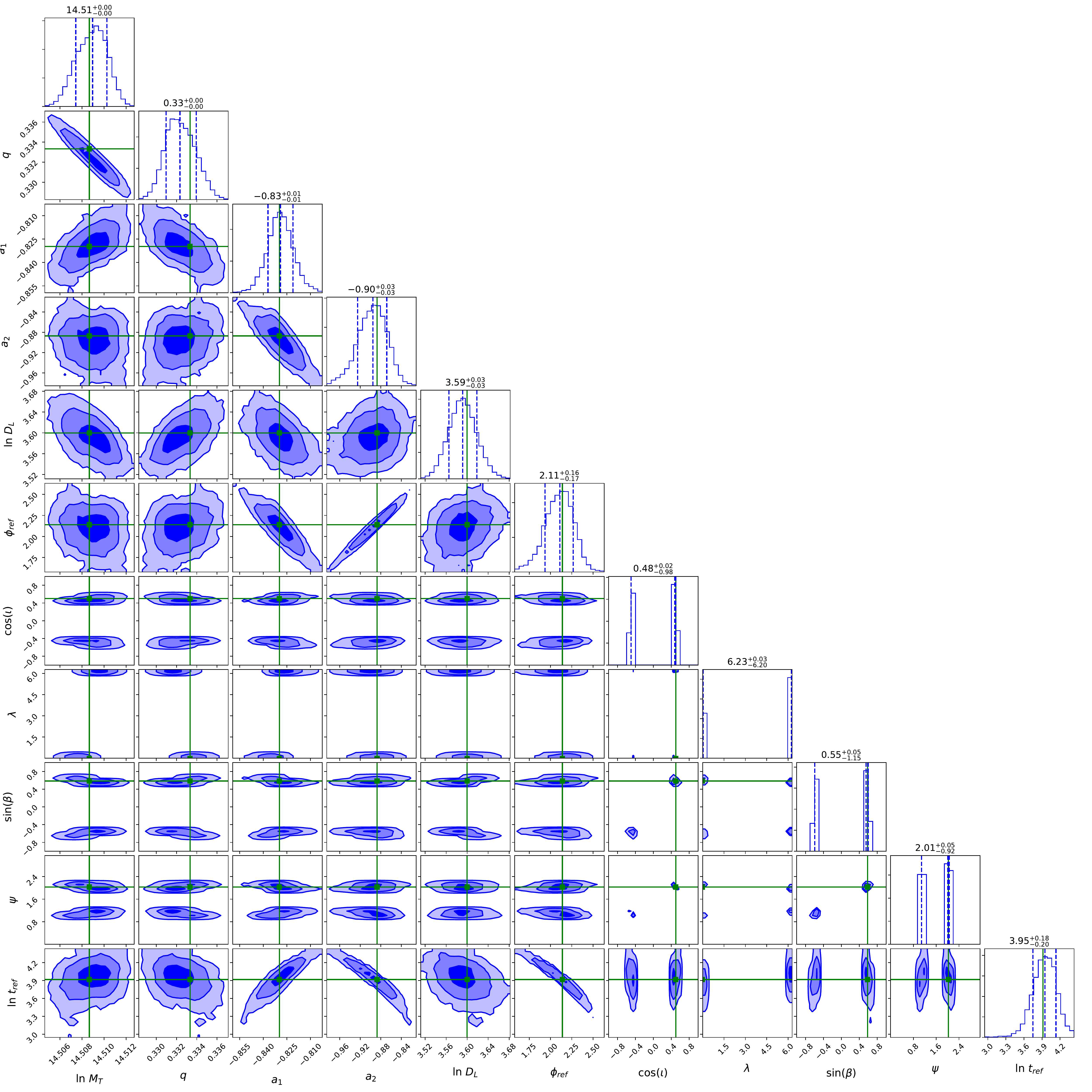}
\caption{The binary injected with strong anti-aligned spins, Binary 5, is shown in this plot. The sampler here was allowed to vary spins. The data stream contained a sampled noise realization. The constructions involved in this corner plot are explained in \mbox{Section \ref{sec:corner_plots}} of the Appendix.}\label{fig:sylvain_5}
\end{center}
\end{figure*}

\begin{figure*}
\begin{center}
\includegraphics[scale=0.37]{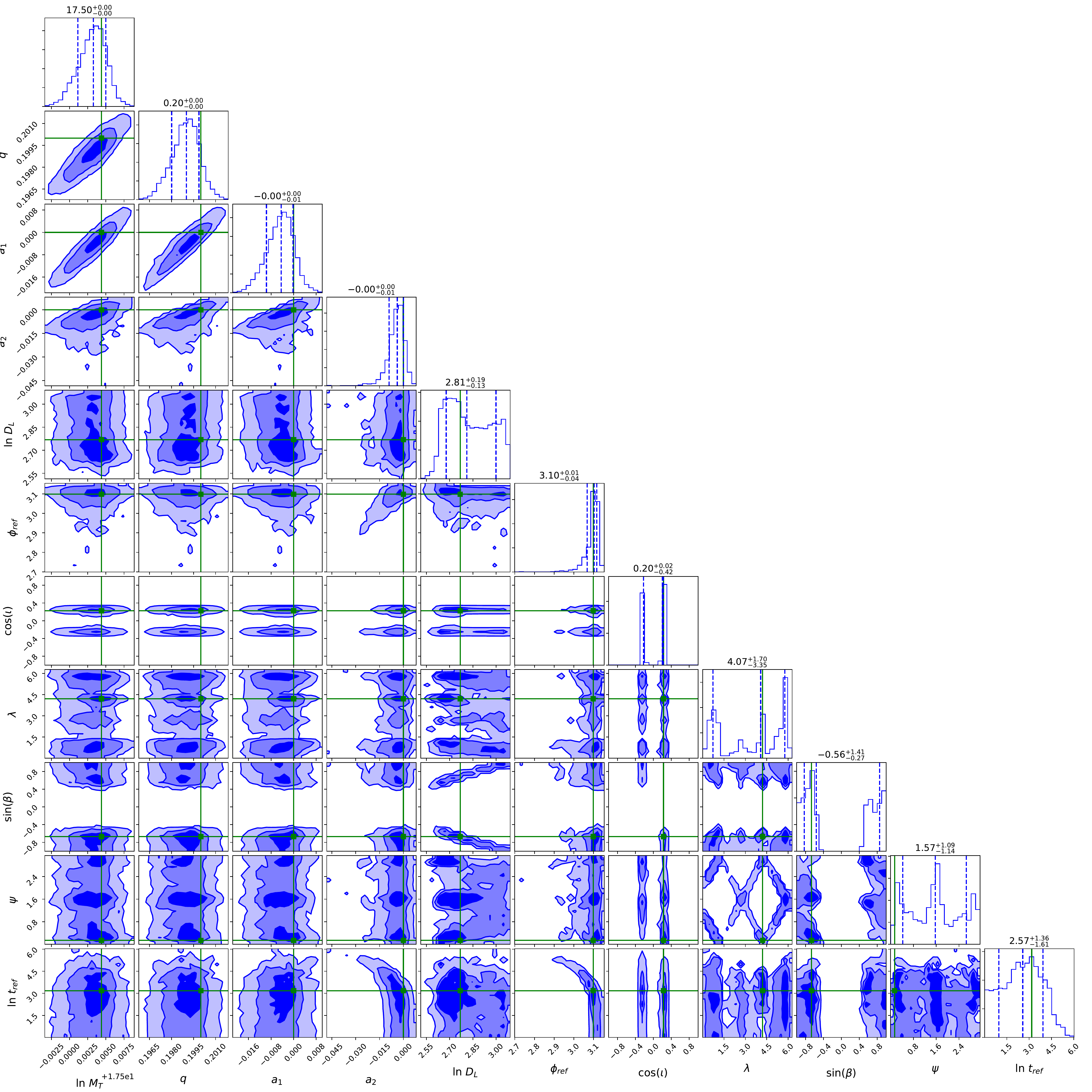}
\caption{The large mass binary, Binary 6, is shown above. The posterior distributions represent a signal with a noise realization and Schwarzschild MBHs. During sampling, the MCMC algorithm varied the spins, therefore, analyzing the full 11-dimensional parameter space. Please see \mbox{Section \ref{sec:corner_plots}} of the Appendix for information on the corner plot constructions.}\label{fig:large_mass_posterior}
\end{center}
\end{figure*}

\begin{figure*}
\begin{center}
\includegraphics[scale=0.37]{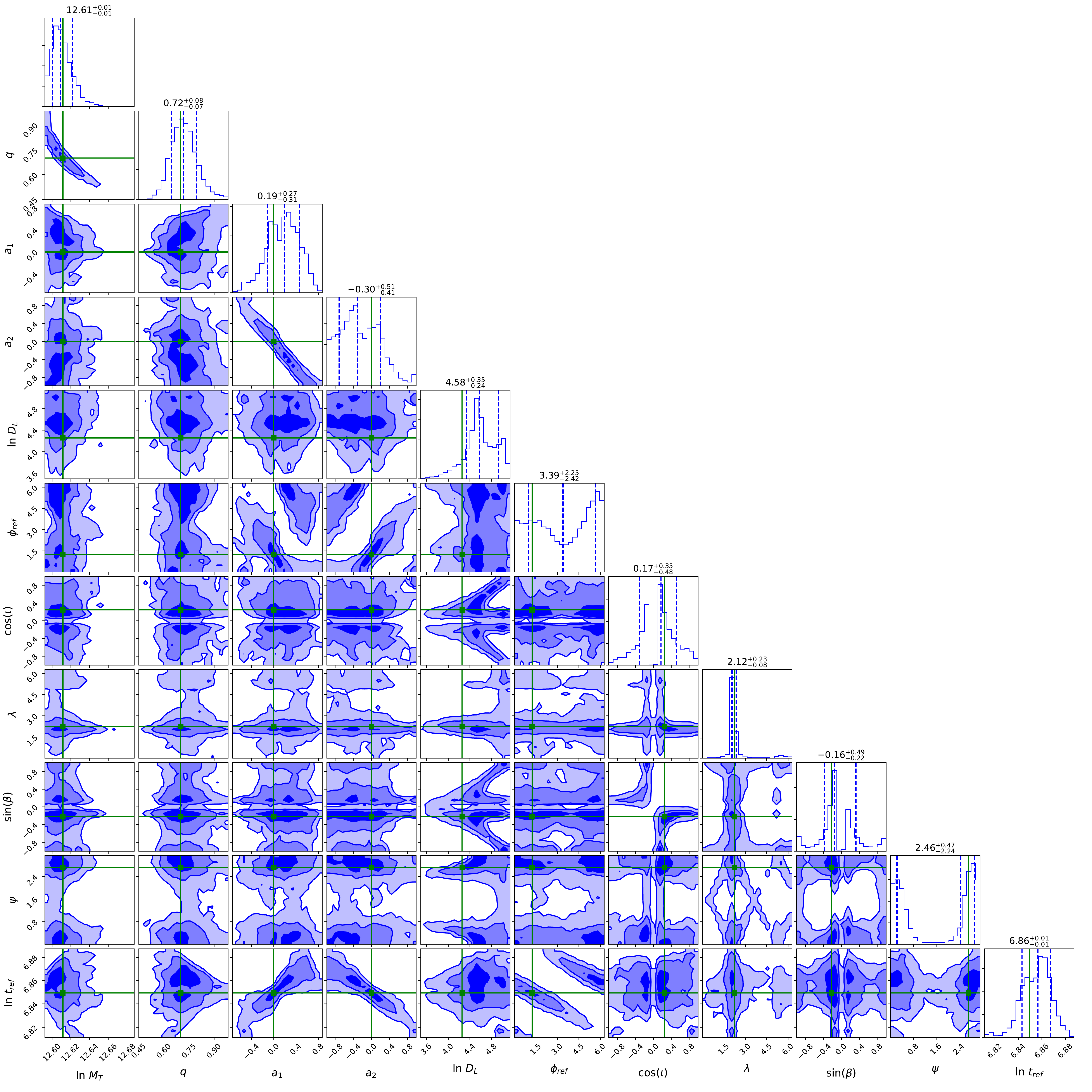}
\caption{The posterior distributions for the small mass binary, Binary 7, are shown in this plot. The small mass binary MCMC run was performed in a zero-noise representation due to issues with convergence of the sampler when adding noise. This is an issue that will be addressed in future work. This binary was injected with Schwarzschild MBHs and the spins were allowed to vary during sampling. \mbox{Section \ref{sec:corner_plots}} of the Appendix describes the constructions used in the corner plots.}\label{fig:small_mass_posterior}
\end{center}
\end{figure*}

Here we present the full corner plots for all binaries tested (see \mbox{Table \ref{tb:testinfo}}) to accompany subsections of the corner plots shown in \mbox{Section \ref{sec:results_discussion}}. Binaries 1-7 are shown in \mbox{Figures \ref{fig:sylvain_1} - \ref{fig:small_mass_posterior}}, respectively. The exact injection parameters for each binary can be seen in \mbox{Table \ref{tb:testinfo}}, and are represented by a green point in the full posteriors. All sky position and orientation parameters are displayed over their entire domain (or prior range) due to their inherent multi-modal structure. The other parameters are shown zooming in on the distributions that account for 99.9\% of marginalized posterior points. The inclination, $\iota$, and ecliptic latitude, $\beta$, are plotted as $\cos{\iota}$ and $\sin{\beta}$ because these are the parameters sampled during the MCMC runs. Similarly, the total mass, $M_T$; luminosity distance, $D_L$; and reference time, $t_\text{ref}$, are all sampled and plotted as the natural log of these values. \mbox{Table \ref{tb:mcmc_run_info}} shows the autocorrelation time ($\tau_f$) and effective sample size (ESS) according to \mbox{Section \ref{sec:autocorrelation}} for each MCMC run. It also gives the average acceptance fraction (AF) for each run.  

\begin{center}
\begin{table}[h]
\caption{The sampler output info for each run is shown below. $\tau_f$ is the autocorrelation time as discussed in \mbox{Section \ref{sec:autocorrelation}}. ESS is the effective sample size. This is the number of posterior density points used to create the plots shown below. The last column shows the average acceptance fraction (AF) for each run.}\label{tb:mcmc_run_info}
\begin{tabular}{ c c c c}
 \hline\hline
Binary & $\tau_f$ & ESS & AF \\
1 & 97 & 79200 & 0.20 \\
2 & 93 & 103232 & 0.19 \\
3 & 63 & 50816 & 0.25 \\
4 & 90 & 67584 & 0.24 \\
5 & 80 & 52000 & 0.19 \\
6 & 45 & 64000 & 0.18 \\
7 & 8 & 600000 & 0.21 \\
\hline
\end{tabular}
\end{table}
\end{center}

\begin{thebibliography}{120}%
\makeatletter
\providecommand \@ifxundefined [1]{%
 \@ifx{#1\undefined}
}%
\providecommand \@ifnum [1]{%
 \ifnum #1\expandafter \@firstoftwo
 \else \expandafter \@secondoftwo
 \fi
}%
\providecommand \@ifx [1]{%
 \ifx #1\expandafter \@firstoftwo
 \else \expandafter \@secondoftwo
 \fi
}%
\providecommand \natexlab [1]{#1}%
\providecommand \enquote  [1]{``#1''}%
\providecommand \bibnamefont  [1]{#1}%
\providecommand \bibfnamefont [1]{#1}%
\providecommand \citenamefont [1]{#1}%
\providecommand \href@noop [0]{\@secondoftwo}%
\providecommand \href [0]{\begingroup \@sanitize@url \@href}%
\providecommand \@href[1]{\@@startlink{#1}\@@href}%
\providecommand \@@href[1]{\endgroup#1\@@endlink}%
\providecommand \@sanitize@url [0]{\catcode `\\12\catcode `\$12\catcode
  `\&12\catcode `\#12\catcode `\^12\catcode `\_12\catcode `\%12\relax}%
\providecommand \@@startlink[1]{}%
\providecommand \@@endlink[0]{}%
\providecommand \url  [0]{\begingroup\@sanitize@url \@url }%
\providecommand \@url [1]{\endgroup\@href {#1}{\urlprefix }}%
\providecommand \urlprefix  [0]{URL }%
\providecommand \Eprint [0]{\href }%
\providecommand \doibase [0]{https://doi.org/}%
\providecommand \selectlanguage [0]{\@gobble}%
\providecommand \bibinfo  [0]{\@secondoftwo}%
\providecommand \bibfield  [0]{\@secondoftwo}%
\providecommand \translation [1]{[#1]}%
\providecommand \BibitemOpen [0]{}%
\providecommand \bibitemStop [0]{}%
\providecommand \bibitemNoStop [0]{.\EOS\space}%
\providecommand \EOS [0]{\spacefactor3000\relax}%
\providecommand \BibitemShut  [1]{\csname bibitem#1\endcsname}%
\let\auto@bib@innerbib\@empty
%</preamble>
\bibitem [{\citenamefont {{Amaro-Seoane et al}}(2017)}]{LISAMissionProposal_2}%
  \BibitemOpen
  \bibfield  {author} {\bibinfo {author} {\bibfnamefont {P.}~\bibnamefont
  {{Amaro-Seoane et al}}},\ }\bibfield  {title} {\bibinfo {title} {{Laser
  Interferometer Space Antenna}},\ }\href@noop {} {\bibfield  {journal}
  {\bibinfo  {journal} {ArXiv e-prints}\ } (\bibinfo {year} {2017})},\ \Eprint
  {https://arxiv.org/abs/1702.00786} {arXiv:1702.00786 [astro-ph.IM]}
  \BibitemShut {NoStop}%
\bibitem [{\citenamefont {{The LIGO Scientific Collaboration}}\ and\
  \citenamefont {{the Virgo Collaboration}}(2018)}]{O2_summary_2}%
  \BibitemOpen
  \bibfield  {author} {\bibinfo {author} {\bibnamefont {{The LIGO Scientific
  Collaboration}}}\ and\ \bibinfo {author} {\bibnamefont {{the Virgo
  Collaboration}}},\ }\bibfield  {title} {\bibinfo {title} {{GWTC-1: A
  Gravitational-Wave Transient Catalog of Compact Binary Mergers Observed by
  LIGO and Virgo during the First and Second Observing Runs}},\ }\href@noop {}
  {\bibfield  {journal} {\bibinfo  {journal} {arXiv e-prints}\ ,\ \bibinfo
  {eid} {arXiv:1811.12907}} (\bibinfo {year} {2018})},\ \Eprint
  {https://arxiv.org/abs/1811.12907} {arXiv:1811.12907 [astro-ph.HE]}
  \BibitemShut {NoStop}%
\bibitem [{\citenamefont {{LVK Collaboration}}(2018)}]{LVK2018LivingReview}%
  \BibitemOpen
  \bibfield  {author} {\bibinfo {author} {\bibnamefont {{LVK Collaboration}}},\
  }\bibfield  {title} {\bibinfo {title} {{Prospects for observing and
  localizing gravitational-wave transients with Advanced LIGO, Advanced Virgo
  and KAGRA}},\ }\href {https://doi.org/10.1007/s41114-018-0012-9} {\bibfield
  {journal} {\bibinfo  {journal} {Living Reviews in Relativity}\ }\textbf
  {\bibinfo {volume} {21}},\ \bibinfo {eid} {3} (\bibinfo {year} {2018})},\
  \Eprint {https://arxiv.org/abs/1304.0670} {arXiv:1304.0670 [gr-qc]}
  \BibitemShut {NoStop}%
\bibitem [{\citenamefont {Klein}\ \emph {et~al.}(2016)\citenamefont {Klein},
  \citenamefont {Barausse}, \citenamefont {Sesana}, \citenamefont {Petiteau},
  \citenamefont {Berti}, \citenamefont {Babak}, \citenamefont {Gair},
  \citenamefont {Aoudia}, \citenamefont {Hinder}, \citenamefont {Ohme},\ and\
  \citenamefont {Wardell}}]{Klein2016}%
  \BibitemOpen
  \bibfield  {author} {\bibinfo {author} {\bibfnamefont {A.}~\bibnamefont
  {Klein}}, \bibinfo {author} {\bibfnamefont {E.}~\bibnamefont {Barausse}},
  \bibinfo {author} {\bibfnamefont {A.}~\bibnamefont {Sesana}}, \bibinfo
  {author} {\bibfnamefont {A.}~\bibnamefont {Petiteau}}, \bibinfo {author}
  {\bibfnamefont {E.}~\bibnamefont {Berti}}, \bibinfo {author} {\bibfnamefont
  {S.}~\bibnamefont {Babak}}, \bibinfo {author} {\bibfnamefont
  {J.}~\bibnamefont {Gair}}, \bibinfo {author} {\bibfnamefont {S.}~\bibnamefont
  {Aoudia}}, \bibinfo {author} {\bibfnamefont {I.}~\bibnamefont {Hinder}},
  \bibinfo {author} {\bibfnamefont {F.}~\bibnamefont {Ohme}},\ and\ \bibinfo
  {author} {\bibfnamefont {B.}~\bibnamefont {Wardell}},\ }\bibfield  {title}
  {\bibinfo {title} {{Science with the space-based interferometer eLISA:
  Supermassive black hole binaries}},\ }\href
  {https://doi.org/10.1103/PhysRevD.93.024003} {\bibfield  {journal} {\bibinfo
  {journal} {\prd}\ }\textbf {\bibinfo {volume} {93}},\ \bibinfo {pages}
  {24003} (\bibinfo {year} {2016})},\ \Eprint
  {https://arxiv.org/abs/1511.05581} {arXiv:1511.05581 [gr-qc]} \BibitemShut
  {NoStop}%
\bibitem [{\citenamefont {Berti}\ \emph {et~al.}(2016)\citenamefont {Berti},
  \citenamefont {Sesana}, \citenamefont {Barausse}, \citenamefont {Cardoso},\
  and\ \citenamefont {Belczynski}}]{Berti2016}%
  \BibitemOpen
  \bibfield  {author} {\bibinfo {author} {\bibfnamefont {E.}~\bibnamefont
  {Berti}}, \bibinfo {author} {\bibfnamefont {A.}~\bibnamefont {Sesana}},
  \bibinfo {author} {\bibfnamefont {E.}~\bibnamefont {Barausse}}, \bibinfo
  {author} {\bibfnamefont {V.}~\bibnamefont {Cardoso}},\ and\ \bibinfo {author}
  {\bibfnamefont {K.}~\bibnamefont {Belczynski}},\ }\bibfield  {title}
  {\bibinfo {title} {{Spectroscopy of Kerr Black Holes with Earth- and
  Space-Based Interferometers}},\ }\href
  {https://doi.org/10.1103/PhysRevLett.117.101102} {\bibfield  {journal}
  {\bibinfo  {journal} {Phys. Rev. Lett.}\ }\textbf {\bibinfo {volume} {117}},\
  \bibinfo {pages} {101102} (\bibinfo {year} {2016})},\ \Eprint
  {https://arxiv.org/abs/1605.09286} {arXiv:1605.09286 [gr-qc]} \BibitemShut
  {NoStop}%
\bibitem [{\citenamefont {Salcido}\ \emph {et~al.}(2016)\citenamefont
  {Salcido}, \citenamefont {Bower}, \citenamefont {Theuns}, \citenamefont
  {McAlpine}, \citenamefont {Schaller}, \citenamefont {Crain}, \citenamefont
  {Schaye},\ and\ \citenamefont {Regan}}]{Salcido2016}%
  \BibitemOpen
  \bibfield  {author} {\bibinfo {author} {\bibfnamefont {J.}~\bibnamefont
  {Salcido}}, \bibinfo {author} {\bibfnamefont {R.}~\bibnamefont {Bower}},
  \bibinfo {author} {\bibfnamefont {T.}~\bibnamefont {Theuns}}, \bibinfo
  {author} {\bibfnamefont {S.}~\bibnamefont {McAlpine}}, \bibinfo {author}
  {\bibfnamefont {M.}~\bibnamefont {Schaller}}, \bibinfo {author}
  {\bibfnamefont {R.}~\bibnamefont {Crain}}, \bibinfo {author} {\bibfnamefont
  {J.}~\bibnamefont {Schaye}},\ and\ \bibinfo {author} {\bibfnamefont
  {J.}~\bibnamefont {Regan}},\ }\bibfield  {title} {\bibinfo {title} {{Music
  from the heavens - gravitational waves from supermassive black hole mergers
  in the EAGLE simulations}},\ }\href {https://doi.org/10.1093/mnras/stw2048}
  {\bibfield  {journal} {\bibinfo  {journal} {\mnras}\ }\textbf {\bibinfo
  {volume} {463}},\ \bibinfo {pages} {870} (\bibinfo {year} {2016})},\ \Eprint
  {https://arxiv.org/abs/1601.06156} {arXiv:1601.06156} \BibitemShut {NoStop}%
\bibitem [{\citenamefont {Katz}\ \emph {et~al.}(2019)\citenamefont {Katz},
  \citenamefont {Kelley}, \citenamefont {Dosopoulou}, \citenamefont {Berry},
  \citenamefont {Blecha},\ and\ \citenamefont {Larson}}]{Katz2019}%
  \BibitemOpen
  \bibfield  {author} {\bibinfo {author} {\bibfnamefont {M.~L.}\ \bibnamefont
  {Katz}}, \bibinfo {author} {\bibfnamefont {L.~Z.}\ \bibnamefont {Kelley}},
  \bibinfo {author} {\bibfnamefont {F.}~\bibnamefont {Dosopoulou}}, \bibinfo
  {author} {\bibfnamefont {S.}~\bibnamefont {Berry}}, \bibinfo {author}
  {\bibfnamefont {L.}~\bibnamefont {Blecha}},\ and\ \bibinfo {author}
  {\bibfnamefont {S.~L.}\ \bibnamefont {Larson}},\ }\bibfield  {title}
  {\bibinfo {title} {{Probing Massive Black Hole Binary Populations with
  LISA}},\ }\href@noop {} {\bibfield  {journal} {\bibinfo  {journal} {arXiv
  e-prints}\ ,\ \bibinfo {pages} {arXiv:1908.05779}} (\bibinfo {year}
  {2019})},\ \Eprint {https://arxiv.org/abs/1908.05779} {arXiv:1908.05779
  [astro-ph.HE]} \BibitemShut {NoStop}%
\bibitem [{\citenamefont {Bonetti}\ \emph {et~al.}(2019)\citenamefont
  {Bonetti}, \citenamefont {Sesana}, \citenamefont {Haardt}, \citenamefont
  {Barausse},\ and\ \citenamefont {Colpi}}]{Bonetti2019}%
  \BibitemOpen
  \bibfield  {author} {\bibinfo {author} {\bibfnamefont {M.}~\bibnamefont
  {Bonetti}}, \bibinfo {author} {\bibfnamefont {A.}~\bibnamefont {Sesana}},
  \bibinfo {author} {\bibfnamefont {F.}~\bibnamefont {Haardt}}, \bibinfo
  {author} {\bibfnamefont {E.}~\bibnamefont {Barausse}},\ and\ \bibinfo
  {author} {\bibfnamefont {M.}~\bibnamefont {Colpi}},\ }\bibfield  {title}
  {\bibinfo {title} {{Post-Newtonian evolution of massive black hole triplets
  in galactic nuclei - IV. Implications for LISA}},\ }\href
  {https://doi.org/10.1093/mnras/stz903} {\bibfield  {journal} {\bibinfo
  {journal} {\mnras}\ }\textbf {\bibinfo {volume} {486}},\ \bibinfo {pages}
  {4044} (\bibinfo {year} {2019})},\ \Eprint {https://arxiv.org/abs/1812.01011}
  {arXiv:1812.01011 [astro-ph.GA]} \BibitemShut {NoStop}%
\bibitem [{\citenamefont {{eLISA Consortium}}(2013)}]{GravitationalUniverse_2}%
  \BibitemOpen
  \bibfield  {author} {\bibinfo {author} {\bibnamefont {{eLISA Consortium}}},\
  }\bibfield  {title} {\bibinfo {title} {{The Gravitational Universe}},\
  }\href@noop {} {\bibfield  {journal} {\bibinfo  {journal} {arXiv e-prints}\
  ,\ \bibinfo {eid} {arXiv:1305.5720}} (\bibinfo {year} {2013})},\ \Eprint
  {https://arxiv.org/abs/1305.5720} {arXiv:1305.5720 [astro-ph.CO]}
  \BibitemShut {NoStop}%
\bibitem [{\citenamefont {Barausse}\ \emph {et~al.}(2015)\citenamefont
  {Barausse}, \citenamefont {Bellovary}, \citenamefont {Berti}, \citenamefont
  {{Holley-Bockelmann K. Farris}}, \citenamefont {Sathyaprakash},\ and\
  \citenamefont {Sesana}}]{Barausse2015}%
  \BibitemOpen
  \bibfield  {author} {\bibinfo {author} {\bibfnamefont {E.}~\bibnamefont
  {Barausse}}, \bibinfo {author} {\bibfnamefont {J.}~\bibnamefont {Bellovary}},
  \bibinfo {author} {\bibfnamefont {E.}~\bibnamefont {Berti}}, \bibinfo
  {author} {\bibfnamefont {B.}~\bibnamefont {{Holley-Bockelmann K. Farris}}},
  \bibinfo {author} {\bibfnamefont {B.}~\bibnamefont {Sathyaprakash}},\ and\
  \bibinfo {author} {\bibfnamefont {A.}~\bibnamefont {Sesana}},\ }\bibfield
  {title} {\bibinfo {title} {{Massive Black Hole Science with eLISA}},\
  }\href@noop {} {\bibfield  {journal} {\bibinfo  {journal} {J. Phys.: Conf.
  Ser.}\ }\textbf {\bibinfo {volume} {610}},\ \bibinfo {pages} {12001}
  (\bibinfo {year} {2015})}\BibitemShut {NoStop}%
\bibitem [{\citenamefont {Katz}\ and\ \citenamefont {Larson}(2019)}]{Katz2018}%
  \BibitemOpen
  \bibfield  {author} {\bibinfo {author} {\bibfnamefont {M.}~\bibnamefont
  {Katz}}\ and\ \bibinfo {author} {\bibfnamefont {S.}~\bibnamefont {Larson}},\
  }\bibfield  {title} {\bibinfo {title} {{Evaluating black hole detectability
  with LISA}},\ }\href {https://doi.org/10.1093/mnras/sty3321} {\bibfield
  {journal} {\bibinfo  {journal} {\mnras}\ }\textbf {\bibinfo {volume} {483}},\
  \bibinfo {pages} {3108} (\bibinfo {year} {2019})},\ \Eprint
  {https://arxiv.org/abs/1807.02511} {arXiv:1807.02511 [gr-qc]} \BibitemShut
  {NoStop}%
\bibitem [{\citenamefont {Loeb}\ and\ \citenamefont {Rasio}(1994)}]{Loeb1994}%
  \BibitemOpen
  \bibfield  {author} {\bibinfo {author} {\bibfnamefont {A.}~\bibnamefont
  {Loeb}}\ and\ \bibinfo {author} {\bibfnamefont {F.}~\bibnamefont {Rasio}},\
  }\bibfield  {title} {\bibinfo {title} {{Collapse of primordial gas clouds and
  the formation of quasar black holes}},\ }\href
  {https://doi.org/10.1086/174548} {\bibfield  {journal} {\bibinfo  {journal}
  {\apj}\ }\textbf {\bibinfo {volume} {432}},\ \bibinfo {pages} {52} (\bibinfo
  {year} {1994})}\BibitemShut {NoStop}%
\bibitem [{\citenamefont {Begelman}\ \emph {et~al.}(2006)\citenamefont
  {Begelman}, \citenamefont {Volonteri},\ and\ \citenamefont
  {Rees}}]{Begelman2006}%
  \BibitemOpen
  \bibfield  {author} {\bibinfo {author} {\bibfnamefont {M.}~\bibnamefont
  {Begelman}}, \bibinfo {author} {\bibfnamefont {M.}~\bibnamefont
  {Volonteri}},\ and\ \bibinfo {author} {\bibfnamefont {M.}~\bibnamefont
  {Rees}},\ }\bibfield  {title} {\bibinfo {title} {{Formation of supermassive
  black holes by direct collapse in pre-galactic haloes}},\ }\href
  {https://doi.org/10.1111/j.1365-2966.2006.10467.x} {\bibfield  {journal}
  {\bibinfo  {journal} {\mnras}\ }\textbf {\bibinfo {volume} {370}},\ \bibinfo
  {pages} {289} (\bibinfo {year} {2006})}\BibitemShut {NoStop}%
\bibitem [{\citenamefont {Latif}\ \emph {et~al.}(2013)\citenamefont {Latif},
  \citenamefont {Schleicher}, \citenamefont {Schmidt},\ and\ \citenamefont
  {Niemeyer}}]{Latif2013}%
  \BibitemOpen
  \bibfield  {author} {\bibinfo {author} {\bibfnamefont {M.}~\bibnamefont
  {Latif}}, \bibinfo {author} {\bibfnamefont {D.}~\bibnamefont {Schleicher}},
  \bibinfo {author} {\bibfnamefont {W.}~\bibnamefont {Schmidt}},\ and\ \bibinfo
  {author} {\bibfnamefont {J.}~\bibnamefont {Niemeyer}},\ }\bibfield  {title}
  {\bibinfo {title} {{The characteristic black hole mass resulting from direct
  collapse in the early Universe}},\ }\href
  {https://doi.org/10.1093/mnras/stt1786} {\bibfield  {journal} {\bibinfo
  {journal} {\mnras}\ }\textbf {\bibinfo {volume} {436}},\ \bibinfo {pages}
  {2989} (\bibinfo {year} {2013})},\ \Eprint {https://arxiv.org/abs/1309.1097}
  {arXiv:1309.1097} \BibitemShut {NoStop}%
\bibitem [{\citenamefont {Habouzit}\ \emph {et~al.}(2016)\citenamefont
  {Habouzit}, \citenamefont {Volonteri}, \citenamefont {Latif}, \citenamefont
  {Dubois},\ and\ \citenamefont {Peirani}}]{Habouzit2016}%
  \BibitemOpen
  \bibfield  {author} {\bibinfo {author} {\bibfnamefont {M.}~\bibnamefont
  {Habouzit}}, \bibinfo {author} {\bibfnamefont {M.}~\bibnamefont {Volonteri}},
  \bibinfo {author} {\bibfnamefont {M.}~\bibnamefont {Latif}}, \bibinfo
  {author} {\bibfnamefont {Y.}~\bibnamefont {Dubois}},\ and\ \bibinfo {author}
  {\bibfnamefont {S.}~\bibnamefont {Peirani}},\ }\bibfield  {title} {\bibinfo
  {title} {{On the number density of `direct collapse' black hole seeds}},\
  }\href {https://doi.org/10.1093/mnras/stw1924} {\bibfield  {journal}
  {\bibinfo  {journal} {\mnras}\ }\textbf {\bibinfo {volume} {463}},\ \bibinfo
  {pages} {529} (\bibinfo {year} {2016})},\ \Eprint
  {https://arxiv.org/abs/1601.00557} {arXiv:1601.00557} \BibitemShut {NoStop}%
\bibitem [{\citenamefont {Ardaneh}\ \emph {et~al.}(2018)\citenamefont
  {Ardaneh}, \citenamefont {Luo}, \citenamefont {Shlosman}, \citenamefont
  {Nagamine}, \citenamefont {Wise},\ and\ \citenamefont
  {Begelman}}]{Ardaneh2018}%
  \BibitemOpen
  \bibfield  {author} {\bibinfo {author} {\bibfnamefont {K.}~\bibnamefont
  {Ardaneh}}, \bibinfo {author} {\bibfnamefont {Y.}~\bibnamefont {Luo}},
  \bibinfo {author} {\bibfnamefont {I.}~\bibnamefont {Shlosman}}, \bibinfo
  {author} {\bibfnamefont {K.}~\bibnamefont {Nagamine}}, \bibinfo {author}
  {\bibfnamefont {J.}~\bibnamefont {Wise}},\ and\ \bibinfo {author}
  {\bibfnamefont {M.}~\bibnamefont {Begelman}},\ }\bibfield  {title} {\bibinfo
  {title} {{Direct Collapse to Supermassive Black Hole Seeds with Radiation
  Transfer: Cosmological Halos}},\ }\href@noop {} {\bibfield  {journal}
  {\bibinfo  {journal} {ArXiv e-prints}\ } (\bibinfo {year} {2018})},\ \Eprint
  {https://arxiv.org/abs/1803.03278} {arXiv:1803.03278} \BibitemShut {NoStop}%
\bibitem [{\citenamefont {Dunn}\ \emph {et~al.}(2018)\citenamefont {Dunn},
  \citenamefont {Bellovary}, \citenamefont {Holley-Bockelmann}, \citenamefont
  {Christensen},\ and\ \citenamefont {Quinn}}]{Dunn2018}%
  \BibitemOpen
  \bibfield  {author} {\bibinfo {author} {\bibfnamefont {G.}~\bibnamefont
  {Dunn}}, \bibinfo {author} {\bibfnamefont {J.}~\bibnamefont {Bellovary}},
  \bibinfo {author} {\bibfnamefont {K.}~\bibnamefont {Holley-Bockelmann}},
  \bibinfo {author} {\bibfnamefont {C.}~\bibnamefont {Christensen}},\ and\
  \bibinfo {author} {\bibfnamefont {T.}~\bibnamefont {Quinn}},\ }\bibfield
  {title} {\bibinfo {title} {{Sowing Black Hole Seeds: Direct Collapse Black
  Hole Formation with Realistic Lyman-Werner Radiation in Cosmological
  Simulations}},\ }\href {https://doi.org/10.3847/1538-4357/aac7c2} {\bibfield
  {journal} {\bibinfo  {journal} {\apj}\ }\textbf {\bibinfo {volume} {861}},\
  \bibinfo {pages} {39} (\bibinfo {year} {2018})},\ \Eprint
  {https://arxiv.org/abs/1803.01007} {arXiv:1803.01007} \BibitemShut {NoStop}%
\bibitem [{\citenamefont {Omukai}\ \emph {et~al.}(2008)\citenamefont {Omukai},
  \citenamefont {Schneider},\ and\ \citenamefont {Haiman}}]{Omukai2008}%
  \BibitemOpen
  \bibfield  {author} {\bibinfo {author} {\bibfnamefont {K.}~\bibnamefont
  {Omukai}}, \bibinfo {author} {\bibfnamefont {R.}~\bibnamefont {Schneider}},\
  and\ \bibinfo {author} {\bibfnamefont {Z.}~\bibnamefont {Haiman}},\
  }\bibfield  {title} {\bibinfo {title} {{Can Supermassive Black Holes Form in
  Metal-enriched High-Redshift Protogalaxies?}},\ }\href
  {https://doi.org/10.1086/591636} {\bibfield  {journal} {\bibinfo  {journal}
  {\apj}\ }\textbf {\bibinfo {volume} {686}},\ \bibinfo {pages} {801} (\bibinfo
  {year} {2008})},\ \Eprint {https://arxiv.org/abs/0804.3141} {arXiv:0804.3141}
  \BibitemShut {NoStop}%
\bibitem [{\citenamefont {Devecchi}\ and\ \citenamefont
  {Volonteri}(2009)}]{Devecchi2009}%
  \BibitemOpen
  \bibfield  {author} {\bibinfo {author} {\bibfnamefont {B.}~\bibnamefont
  {Devecchi}}\ and\ \bibinfo {author} {\bibfnamefont {M.}~\bibnamefont
  {Volonteri}},\ }\bibfield  {title} {\bibinfo {title} {{Formation of the First
  Nuclear Clusters and Massive Black Holes at High Redshift}},\ }\href
  {https://doi.org/10.1088/0004-637X/694/1/302} {\bibfield  {journal} {\bibinfo
   {journal} {\apj}\ }\textbf {\bibinfo {volume} {694}},\ \bibinfo {pages}
  {302} (\bibinfo {year} {2009})},\ \Eprint {https://arxiv.org/abs/0810.1057}
  {arXiv:0810.1057} \BibitemShut {NoStop}%
\bibitem [{\citenamefont {Davies}\ \emph {et~al.}(2011)\citenamefont {Davies},
  \citenamefont {Miller},\ and\ \citenamefont {Bellovary}}]{Davies2011}%
  \BibitemOpen
  \bibfield  {author} {\bibinfo {author} {\bibfnamefont {M.~B.}\ \bibnamefont
  {Davies}}, \bibinfo {author} {\bibfnamefont {M.~C.}\ \bibnamefont {Miller}},\
  and\ \bibinfo {author} {\bibfnamefont {J.~M.}\ \bibnamefont {Bellovary}},\
  }\bibfield  {title} {\bibinfo {title} {{Supermassive Black Hole Formation Via
  Gas Accretion in Nuclear Stellar Clusters}},\ }\href
  {https://doi.org/10.1088/2041-8205/740/2/L42} {\bibfield  {journal} {\bibinfo
   {journal} {\apjl}\ }\textbf {\bibinfo {volume} {740}},\ \bibinfo {pages}
  {L42} (\bibinfo {year} {2011})},\ \Eprint {https://arxiv.org/abs/1106.5943}
  {arXiv:1106.5943 [astro-ph.CO]} \BibitemShut {NoStop}%
\bibitem [{\citenamefont {Katz}\ \emph {et~al.}(2015)\citenamefont {Katz},
  \citenamefont {Sijacki},\ and\ \citenamefont {Haehnelt}}]{Katz2015}%
  \BibitemOpen
  \bibfield  {author} {\bibinfo {author} {\bibfnamefont {H.}~\bibnamefont
  {Katz}}, \bibinfo {author} {\bibfnamefont {D.}~\bibnamefont {Sijacki}},\ and\
  \bibinfo {author} {\bibfnamefont {M.}~\bibnamefont {Haehnelt}},\ }\bibfield
  {title} {\bibinfo {title} {{Seeding high-redshift QSOs by collisional runaway
  in primordial star clusters}},\ }\href
  {https://doi.org/10.1093/mnras/stv1048} {\bibfield  {journal} {\bibinfo
  {journal} {\mnras}\ }\textbf {\bibinfo {volume} {451}},\ \bibinfo {pages}
  {2352} (\bibinfo {year} {2015})},\ \Eprint {https://arxiv.org/abs/1502.03448}
  {arXiv:1502.03448} \BibitemShut {NoStop}%
\bibitem [{\citenamefont {Haiman}\ \emph {et~al.}(2000)\citenamefont {Haiman},
  \citenamefont {Abel},\ and\ \citenamefont {Rees}}]{Haiman2000}%
  \BibitemOpen
  \bibfield  {author} {\bibinfo {author} {\bibfnamefont {Z.}~\bibnamefont
  {Haiman}}, \bibinfo {author} {\bibfnamefont {T.}~\bibnamefont {Abel}},\ and\
  \bibinfo {author} {\bibfnamefont {M.}~\bibnamefont {Rees}},\ }\bibfield
  {title} {\bibinfo {title} {{The Radiative Feedback of the First Cosmological
  Objects}},\ }\href {https://doi.org/10.1086/308723} {\bibfield  {journal}
  {\bibinfo  {journal} {\apj}\ }\textbf {\bibinfo {volume} {534}},\ \bibinfo
  {pages} {11} (\bibinfo {year} {2000})}\BibitemShut {NoStop}%
\bibitem [{\citenamefont {Fryer}\ \emph {et~al.}(2001)\citenamefont {Fryer},
  \citenamefont {Woosley},\ and\ \citenamefont {Heger}}]{Fryer2001}%
  \BibitemOpen
  \bibfield  {author} {\bibinfo {author} {\bibfnamefont {C.}~\bibnamefont
  {Fryer}}, \bibinfo {author} {\bibfnamefont {S.}~\bibnamefont {Woosley}},\
  and\ \bibinfo {author} {\bibfnamefont {A.}~\bibnamefont {Heger}},\ }\bibfield
   {title} {\bibinfo {title} {{Pair-Instability Supernovae, Gravity Waves, and
  Gamma-Ray Transients}},\ }\href {https://doi.org/10.1086/319719} {\bibfield
  {journal} {\bibinfo  {journal} {\apj}\ }\textbf {\bibinfo {volume} {550}},\
  \bibinfo {pages} {372} (\bibinfo {year} {2001})}\BibitemShut {NoStop}%
\bibitem [{\citenamefont {Heger}\ \emph {et~al.}(2003)\citenamefont {Heger},
  \citenamefont {Fryer}, \citenamefont {Woosley}, \citenamefont {Langer},\ and\
  \citenamefont {Hartmann}}]{Heger2003}%
  \BibitemOpen
  \bibfield  {author} {\bibinfo {author} {\bibfnamefont {A.}~\bibnamefont
  {Heger}}, \bibinfo {author} {\bibfnamefont {C.}~\bibnamefont {Fryer}},
  \bibinfo {author} {\bibfnamefont {S.}~\bibnamefont {Woosley}}, \bibinfo
  {author} {\bibfnamefont {N.}~\bibnamefont {Langer}},\ and\ \bibinfo {author}
  {\bibfnamefont {D.}~\bibnamefont {Hartmann}},\ }\bibfield  {title} {\bibinfo
  {title} {{How Massive Single Stars End Their Life}},\ }\href
  {https://doi.org/10.1086/375341} {\bibfield  {journal} {\bibinfo  {journal}
  {\apj}\ }\textbf {\bibinfo {volume} {591}},\ \bibinfo {pages} {288} (\bibinfo
  {year} {2003})}\BibitemShut {NoStop}%
\bibitem [{\citenamefont {Volonteri}\ \emph {et~al.}(2003)\citenamefont
  {Volonteri}, \citenamefont {Madau},\ and\ \citenamefont
  {Haardt}}]{Volonteri2003}%
  \BibitemOpen
  \bibfield  {author} {\bibinfo {author} {\bibfnamefont {M.}~\bibnamefont
  {Volonteri}}, \bibinfo {author} {\bibfnamefont {P.}~\bibnamefont {Madau}},\
  and\ \bibinfo {author} {\bibfnamefont {F.}~\bibnamefont {Haardt}},\
  }\bibfield  {title} {\bibinfo {title} {{The Formation of Galaxy Stellar Cores
  by the Hierarchical Merging of Supermassive Black Holes}},\ }\href
  {https://doi.org/10.1086/376722} {\bibfield  {journal} {\bibinfo  {journal}
  {\apj}\ }\textbf {\bibinfo {volume} {593}},\ \bibinfo {pages} {661} (\bibinfo
  {year} {2003})}\BibitemShut {NoStop}%
\bibitem [{\citenamefont {Tanaka}\ and\ \citenamefont
  {Haiman}(2009)}]{Tanaka2009}%
  \BibitemOpen
  \bibfield  {author} {\bibinfo {author} {\bibfnamefont {T.}~\bibnamefont
  {Tanaka}}\ and\ \bibinfo {author} {\bibfnamefont {Z.}~\bibnamefont
  {Haiman}},\ }\bibfield  {title} {\bibinfo {title} {{The Assembly of
  Supermassive Black Holes at High Redshifts}},\ }\href
  {https://doi.org/10.1088/0004-637X/696/2/1798} {\bibfield  {journal}
  {\bibinfo  {journal} {\apj}\ }\textbf {\bibinfo {volume} {696}},\ \bibinfo
  {pages} {1798} (\bibinfo {year} {2009})},\ \Eprint
  {https://arxiv.org/abs/0807.4702} {arXiv:0807.4702} \BibitemShut {NoStop}%
\bibitem [{\citenamefont {Alvarez}\ \emph {et~al.}(2009)\citenamefont
  {Alvarez}, \citenamefont {Wise},\ and\ \citenamefont {Abel}}]{Alvarez2009}%
  \BibitemOpen
  \bibfield  {author} {\bibinfo {author} {\bibfnamefont {M.~A.}\ \bibnamefont
  {Alvarez}}, \bibinfo {author} {\bibfnamefont {J.~H.}\ \bibnamefont {Wise}},\
  and\ \bibinfo {author} {\bibfnamefont {T.}~\bibnamefont {Abel}},\ }\bibfield
  {title} {\bibinfo {title} {{{ACCRETION} {ONTO} {THE} {FIRST} {STELLAR}-{MASS}
  {BLACK} {HOLES}}},\ }\href {https://doi.org/10.1088/0004-637x/701/2/l133}
  {\bibfield  {journal} {\bibinfo  {journal} {The Astrophysical Journal}\
  }\textbf {\bibinfo {volume} {701}},\ \bibinfo {pages} {L133} (\bibinfo {year}
  {2009})}\BibitemShut {NoStop}%
\bibitem [{\citenamefont {Petiteau}\ \emph {et~al.}(2011)\citenamefont
  {Petiteau}, \citenamefont {Babak},\ and\ \citenamefont
  {Sesana}}]{Petiteau2011}%
  \BibitemOpen
  \bibfield  {author} {\bibinfo {author} {\bibfnamefont {A.}~\bibnamefont
  {Petiteau}}, \bibinfo {author} {\bibfnamefont {S.}~\bibnamefont {Babak}},\
  and\ \bibinfo {author} {\bibfnamefont {A.}~\bibnamefont {Sesana}},\
  }\bibfield  {title} {\bibinfo {title} {{Constraining the Dark Energy Equation
  of State Using LISA Observations of Spinning Massive Black Hole Binaries}},\
  }\href {https://doi.org/10.1088/0004-637X/732/2/82} {\bibfield  {journal}
  {\bibinfo  {journal} {\apj}\ }\textbf {\bibinfo {volume} {732}},\ \bibinfo
  {pages} {82} (\bibinfo {year} {2011})},\ \Eprint
  {https://arxiv.org/abs/1102.0769} {arXiv:1102.0769} \BibitemShut {NoStop}%
\bibitem [{\citenamefont {Gair}\ \emph {et~al.}(2013)\citenamefont {Gair},
  \citenamefont {Vallisneri}, \citenamefont {Larson},\ and\ \citenamefont
  {Baker}}]{Gair2013}%
  \BibitemOpen
  \bibfield  {author} {\bibinfo {author} {\bibfnamefont {J.}~\bibnamefont
  {Gair}}, \bibinfo {author} {\bibfnamefont {M.}~\bibnamefont {Vallisneri}},
  \bibinfo {author} {\bibfnamefont {S.}~\bibnamefont {Larson}},\ and\ \bibinfo
  {author} {\bibfnamefont {J.}~\bibnamefont {Baker}},\ }\bibfield  {title}
  {\bibinfo {title} {{Testing General Relativity with Low-Frequency,
  Space-Based Gravitational-Wave Detectors}},\ }\href
  {https://doi.org/10.12942/lrr-2013-7} {\bibfield  {journal} {\bibinfo
  {journal} {Living Reviews in Relativity}\ }\textbf {\bibinfo {volume} {16}},\
  \bibinfo {pages} {7} (\bibinfo {year} {2013})},\ \Eprint
  {https://arxiv.org/abs/1212.5575} {arXiv:1212.5575 [gr-qc]} \BibitemShut
  {NoStop}%
\bibitem [{\citenamefont {Schutz}(1986)}]{Schutz1986}%
  \BibitemOpen
  \bibfield  {author} {\bibinfo {author} {\bibfnamefont {B.}~\bibnamefont
  {Schutz}},\ }\bibfield  {title} {\bibinfo {title} {{Determining the Hubble
  constant from gravitational wave observations}},\ }\href
  {https://doi.org/10.1038/323310a0} {\bibfield  {journal} {\bibinfo  {journal}
  {\nat}\ }\textbf {\bibinfo {volume} {323}},\ \bibinfo {pages} {310} (\bibinfo
  {year} {1986})}\BibitemShut {NoStop}%
\bibitem [{\citenamefont {Holz}\ and\ \citenamefont {Hughes}(2005)}]{Holz2005}%
  \BibitemOpen
  \bibfield  {author} {\bibinfo {author} {\bibfnamefont {D.~E.}\ \bibnamefont
  {Holz}}\ and\ \bibinfo {author} {\bibfnamefont {S.~A.}\ \bibnamefont
  {Hughes}},\ }\bibfield  {title} {\bibinfo {title} {{Using Gravitational-Wave
  Standard Sirens}},\ }\href {https://doi.org/10.1086/431341} {\bibfield
  {journal} {\bibinfo  {journal} {\apj}\ }\textbf {\bibinfo {volume} {629}},\
  \bibinfo {pages} {15} (\bibinfo {year} {2005})},\ \Eprint
  {https://arxiv.org/abs/astro-ph/0504616} {arXiv:astro-ph/0504616 [astro-ph]}
  \BibitemShut {NoStop}%
\bibitem [{\citenamefont {Armitage}\ and\ \citenamefont
  {Natarajan}(2002)}]{Armitage2002}%
  \BibitemOpen
  \bibfield  {author} {\bibinfo {author} {\bibfnamefont {P.~J.}\ \bibnamefont
  {Armitage}}\ and\ \bibinfo {author} {\bibfnamefont {P.}~\bibnamefont
  {Natarajan}},\ }\bibfield  {title} {\bibinfo {title} {{Accretion during the
  Merger of Supermassive Black Holes}},\ }\href
  {https://doi.org/10.1086/339770} {\bibfield  {journal} {\bibinfo  {journal}
  {\apjl}\ }\textbf {\bibinfo {volume} {567}},\ \bibinfo {pages} {L9} (\bibinfo
  {year} {2002})},\ \Eprint {https://arxiv.org/abs/astro-ph/0201318}
  {arXiv:astro-ph/0201318 [astro-ph]} \BibitemShut {NoStop}%
\bibitem [{\citenamefont {Burke-Spolaor}(2013)}]{Burke-Spolaor2013}%
  \BibitemOpen
  \bibfield  {author} {\bibinfo {author} {\bibfnamefont {S.}~\bibnamefont
  {Burke-Spolaor}},\ }\bibfield  {title} {\bibinfo {title} {{Multi-messenger
  approaches to binary supermassive black holes in the
  {\textquoteleft}continuous-wave{\textquoteright} regime}},\ }\href
  {https://doi.org/10.1088/0264-9381/30/22/224013} {\bibfield  {journal}
  {\bibinfo  {journal} {Classical and Quantum Gravity}\ }\textbf {\bibinfo
  {volume} {30}},\ \bibinfo {pages} {224013} (\bibinfo {year} {2013})},\
  \Eprint {https://arxiv.org/abs/1308.4408} {arXiv:1308.4408 [astro-ph.CO]}
  \BibitemShut {NoStop}%
\bibitem [{\citenamefont {Bogdanovi{\'{c}}}(2015)}]{Bogdanovic2015}%
  \BibitemOpen
  \bibfield  {author} {\bibinfo {author} {\bibfnamefont {T.}~\bibnamefont
  {Bogdanovi{\'{c}}}},\ }\bibfield  {title} {\bibinfo {title} {{Supermassive
  Black Hole Binaries: The Search Continues}},\ }in\ \href
  {https://doi.org/10.1007/978-3-319-10488-1_9} {\emph {\bibinfo {booktitle}
  {Gravitational Wave Astrophysics}}},\ Vol.~\bibinfo {volume} {40}\ (\bibinfo
  {year} {2015})\ p.\ \bibinfo {pages} {103},\ \Eprint
  {https://arxiv.org/abs/1406.5193} {arXiv:1406.5193 [astro-ph.HE]}
  \BibitemShut {NoStop}%
\bibitem [{\citenamefont {{Dal Canton}}\ \emph {et~al.}(2019)\citenamefont
  {{Dal Canton}}, \citenamefont {Mangiagli}, \citenamefont {Noble},
  \citenamefont {Schnittman}, \citenamefont {Ptak}, \citenamefont {Klein},
  \citenamefont {Sesana},\ and\ \citenamefont {Camp}}]{DalCanton2019}%
  \BibitemOpen
  \bibfield  {author} {\bibinfo {author} {\bibfnamefont {T.}~\bibnamefont {{Dal
  Canton}}}, \bibinfo {author} {\bibfnamefont {A.}~\bibnamefont {Mangiagli}},
  \bibinfo {author} {\bibfnamefont {S.~C.}\ \bibnamefont {Noble}}, \bibinfo
  {author} {\bibfnamefont {J.}~\bibnamefont {Schnittman}}, \bibinfo {author}
  {\bibfnamefont {A.}~\bibnamefont {Ptak}}, \bibinfo {author} {\bibfnamefont
  {A.}~\bibnamefont {Klein}}, \bibinfo {author} {\bibfnamefont
  {A.}~\bibnamefont {Sesana}},\ and\ \bibinfo {author} {\bibfnamefont
  {J.}~\bibnamefont {Camp}},\ }\bibfield  {title} {\bibinfo {title}
  {{Detectability of Modulated X-Rays from LISA{\textquoteright}s Supermassive
  Black Hole Mergers}},\ }\href {https://doi.org/10.3847/1538-4357/ab505a}
  {\bibfield  {journal} {\bibinfo  {journal} {\apj}\ }\textbf {\bibinfo
  {volume} {886}},\ \bibinfo {pages} {146} (\bibinfo {year} {2019})},\ \Eprint
  {https://arxiv.org/abs/1902.01538} {arXiv:1902.01538 [astro-ph.HE]}
  \BibitemShut {NoStop}%
\bibitem [{\citenamefont {Cornish}\ and\ \citenamefont
  {Porter}(2006)}]{Cornish2006}%
  \BibitemOpen
  \bibfield  {author} {\bibinfo {author} {\bibfnamefont {N.~J.}\ \bibnamefont
  {Cornish}}\ and\ \bibinfo {author} {\bibfnamefont {E.~K.}\ \bibnamefont
  {Porter}},\ }\bibfield  {title} {\bibinfo {title} {{MCMC exploration of
  supermassive black hole binary inspirals}},\ }\href
  {https://doi.org/10.1088/0264-9381/23/19/S15} {\bibfield  {journal} {\bibinfo
   {journal} {Classical and Quantum Gravity}\ }\textbf {\bibinfo {volume}
  {23}},\ \bibinfo {pages} {S761} (\bibinfo {year} {2006})},\ \Eprint
  {https://arxiv.org/abs/gr-qc/0605085} {arXiv:gr-qc/0605085 [gr-qc]}
  \BibitemShut {NoStop}%
\bibitem [{\citenamefont {Cornish}\ and\ \citenamefont
  {Porter}(2007)}]{Cornish2007}%
  \BibitemOpen
  \bibfield  {author} {\bibinfo {author} {\bibfnamefont {N.~J.}\ \bibnamefont
  {Cornish}}\ and\ \bibinfo {author} {\bibfnamefont {E.~K.}\ \bibnamefont
  {Porter}},\ }\bibfield  {title} {\bibinfo {title} {{Catching supermassive
  black hole binaries without a net}},\ }\href
  {https://doi.org/10.1103/PhysRevD.75.021301} {\bibfield  {journal} {\bibinfo
  {journal} {\prd}\ }\textbf {\bibinfo {volume} {75}},\ \bibinfo {pages}
  {21301} (\bibinfo {year} {2007})},\ \Eprint
  {https://arxiv.org/abs/gr-qc/0605135} {arXiv:gr-qc/0605135 [gr-qc]}
  \BibitemShut {NoStop}%
\bibitem [{\citenamefont {Porter}\ and\ \citenamefont
  {Carr{\'{e}}}(2014)}]{Porter2014}%
  \BibitemOpen
  \bibfield  {author} {\bibinfo {author} {\bibfnamefont {E.~K.}\ \bibnamefont
  {Porter}}\ and\ \bibinfo {author} {\bibfnamefont {J.}~\bibnamefont
  {Carr{\'{e}}}},\ }\bibfield  {title} {\bibinfo {title} {{A Hamiltonian
  Monte-Carlo method for Bayesian inference of supermassive black hole
  binaries}},\ }\href {https://doi.org/10.1088/0264-9381/31/14/145004}
  {\bibfield  {journal} {\bibinfo  {journal} {Classical and Quantum Gravity}\
  }\textbf {\bibinfo {volume} {31}},\ \bibinfo {pages} {145004} (\bibinfo
  {year} {2014})},\ \Eprint {https://arxiv.org/abs/1311.7539} {arXiv:1311.7539
  [gr-qc]} \BibitemShut {NoStop}%
\bibitem [{\citenamefont {Porter}\ and\ \citenamefont
  {Cornish}(2015)}]{Porter2015}%
  \BibitemOpen
  \bibfield  {author} {\bibinfo {author} {\bibfnamefont {E.~K.}\ \bibnamefont
  {Porter}}\ and\ \bibinfo {author} {\bibfnamefont {N.~J.}\ \bibnamefont
  {Cornish}},\ }\bibfield  {title} {\bibinfo {title} {{Fisher versus Bayes: A
  comparison of parameter estimation techniques for massive black hole binaries
  to high redshifts with eLISA}},\ }\href
  {https://doi.org/10.1103/PhysRevD.91.104001} {\bibfield  {journal} {\bibinfo
  {journal} {\prd}\ }\textbf {\bibinfo {volume} {91}},\ \bibinfo {pages}
  {104001} (\bibinfo {year} {2015})},\ \Eprint
  {https://arxiv.org/abs/1502.05735} {arXiv:1502.05735 [gr-qc]} \BibitemShut
  {NoStop}%
\bibitem [{\citenamefont {Marsat}\ \emph {et~al.}(2020)\citenamefont {Marsat},
  \citenamefont {Baker},\ and\ \citenamefont {{Dal
  Canton}}}]{Marsat2020LISAPE}%
  \BibitemOpen
  \bibfield  {author} {\bibinfo {author} {\bibfnamefont {S.}~\bibnamefont
  {Marsat}}, \bibinfo {author} {\bibfnamefont {J.~G.}\ \bibnamefont {Baker}},\
  and\ \bibinfo {author} {\bibfnamefont {T.}~\bibnamefont {{Dal Canton}}},\
  }\bibfield  {title} {\bibinfo {title} {{Exploring the Bayesian parameter
  estimation of binary black holes with LISA}},\ }\href@noop {} {\bibfield
  {journal} {\bibinfo  {journal} {arXiv e-prints}\ ,\ \bibinfo {pages}
  {arXiv:2003.00357}} (\bibinfo {year} {2020})},\ \Eprint
  {https://arxiv.org/abs/2003.00357} {arXiv:2003.00357 [gr-qc]} \BibitemShut
  {NoStop}%
\bibitem [{\citenamefont {Christensen}\ and\ \citenamefont
  {Meyer}(1998)}]{ChristensenMeyer1998MCMCOrig}%
  \BibitemOpen
  \bibfield  {author} {\bibinfo {author} {\bibfnamefont {N.}~\bibnamefont
  {Christensen}}\ and\ \bibinfo {author} {\bibfnamefont {R.}~\bibnamefont
  {Meyer}},\ }\bibfield  {title} {\bibinfo {title} {{Markov chain Monte Carlo
  methods for Bayesian gravitational radiation data analysis}},\ }\href
  {https://doi.org/10.1103/PhysRevD.58.082001} {\bibfield  {journal} {\bibinfo
  {journal} {\prd}\ }\textbf {\bibinfo {volume} {58}},\ \bibinfo {pages}
  {82001} (\bibinfo {year} {1998})}\BibitemShut {NoStop}%
\bibitem [{\citenamefont {Porter}\ and\ \citenamefont
  {Cornish}(2008)}]{Porter2008}%
  \BibitemOpen
  \bibfield  {author} {\bibinfo {author} {\bibfnamefont {E.~K.}\ \bibnamefont
  {Porter}}\ and\ \bibinfo {author} {\bibfnamefont {N.~J.}\ \bibnamefont
  {Cornish}},\ }\bibfield  {title} {\bibinfo {title} {{Effect of higher
  harmonic corrections on the detection of massive black hole binaries with
  LISA}},\ }\href {https://doi.org/10.1103/PhysRevD.78.064005} {\bibfield
  {journal} {\bibinfo  {journal} {\prd}\ }\textbf {\bibinfo {volume} {78}},\
  \bibinfo {pages} {64005} (\bibinfo {year} {2008})},\ \Eprint
  {https://arxiv.org/abs/0804.0332} {arXiv:0804.0332 [gr-qc]} \BibitemShut
  {NoStop}%
\bibitem [{\citenamefont {Vallisneri}(2008)}]{Vallisneri2008}%
  \BibitemOpen
  \bibfield  {author} {\bibinfo {author} {\bibfnamefont {M.}~\bibnamefont
  {Vallisneri}},\ }\bibfield  {title} {\bibinfo {title} {{Use and abuse of the
  Fisher information matrix in the assessment of gravitational-wave
  parameter-estimation prospects}},\ }\href
  {https://doi.org/10.1103/PhysRevD.77.042001} {\bibfield  {journal} {\bibinfo
  {journal} {\prd}\ }\textbf {\bibinfo {volume} {77}},\ \bibinfo {pages}
  {42001} (\bibinfo {year} {2008})},\ \Eprint
  {https://arxiv.org/abs/gr-qc/0703086} {arXiv:gr-qc/0703086 [gr-qc]}
  \BibitemShut {NoStop}%
\bibitem [{\citenamefont {Cutler}\ and\ \citenamefont
  {Vallisneri}(2007)}]{Cutler2007ParameterExtracionErrors}%
  \BibitemOpen
  \bibfield  {author} {\bibinfo {author} {\bibfnamefont {C.}~\bibnamefont
  {Cutler}}\ and\ \bibinfo {author} {\bibfnamefont {M.}~\bibnamefont
  {Vallisneri}},\ }\bibfield  {title} {\bibinfo {title} {{LISA detections of
  massive black hole inspirals: Parameter extraction errors due to inaccurate
  template waveforms}},\ }\href {https://doi.org/10.1103/PhysRevD.76.104018}
  {\bibfield  {journal} {\bibinfo  {journal} {\prd}\ }\textbf {\bibinfo
  {volume} {76}},\ \bibinfo {pages} {104018} (\bibinfo {year} {2007})},\
  \Eprint {https://arxiv.org/abs/0707.2982} {arXiv:0707.2982 [gr-qc]}
  \BibitemShut {NoStop}%
\bibitem [{\citenamefont {Team}(2018)}]{SciRD1}%
  \BibitemOpen
  \bibfield  {author} {\bibinfo {author} {\bibfnamefont {L.~S.~S.}\
  \bibnamefont {Team}},\ }\href@noop {} {\bibinfo {title} {{LISA Science
  Requirements Document}}} (\bibinfo {year} {2018})\BibitemShut {NoStop}%
\bibitem [{\citenamefont {Cutler}\ and\ \citenamefont
  {Flanagan}(1994)}]{Cutler1994}%
  \BibitemOpen
  \bibfield  {author} {\bibinfo {author} {\bibfnamefont {C.}~\bibnamefont
  {Cutler}}\ and\ \bibinfo {author} {\bibfnamefont {{\'{E}}.}~\bibnamefont
  {Flanagan}},\ }\bibfield  {title} {\bibinfo {title} {{Gravitational waves
  from merging compact binaries: How accurately can one extract the binary's
  parameters from the inspiral waveform?}},\ }\href
  {https://doi.org/10.1103/PhysRevD.49.2658} {\bibfield  {journal} {\bibinfo
  {journal} {\prd}\ }\textbf {\bibinfo {volume} {49}},\ \bibinfo {pages} {2658}
  (\bibinfo {year} {1994})}\BibitemShut {NoStop}%
\bibitem [{\citenamefont {Vecchio}(2004)}]{Vecchio2004}%
  \BibitemOpen
  \bibfield  {author} {\bibinfo {author} {\bibfnamefont {A.}~\bibnamefont
  {Vecchio}},\ }\bibfield  {title} {\bibinfo {title} {{LISA observations of
  rapidly spinning massive black hole binary systems}},\ }\href
  {https://doi.org/10.1103/PhysRevD.70.042001} {\bibfield  {journal} {\bibinfo
  {journal} {\prd}\ }\textbf {\bibinfo {volume} {70}},\ \bibinfo {pages}
  {42001} (\bibinfo {year} {2004})},\ \Eprint
  {https://arxiv.org/abs/astro-ph/0304051} {arXiv:astro-ph/0304051 [astro-ph]}
  \BibitemShut {NoStop}%
\bibitem [{\citenamefont {Arun}(2006)}]{Arun2006}%
  \BibitemOpen
  \bibfield  {author} {\bibinfo {author} {\bibfnamefont {K.}~\bibnamefont
  {Arun}},\ }\bibfield  {title} {\bibinfo {title} {{Parameter estimation of
  coalescing supermassive black hole binaries with LISA}},\ }\href
  {https://doi.org/10.1103/PhysRevD.74.024025} {\bibfield  {journal} {\bibinfo
  {journal} {\prd}\ }\textbf {\bibinfo {volume} {74}},\ \bibinfo {pages}
  {24025} (\bibinfo {year} {2006})},\ \Eprint
  {https://arxiv.org/abs/gr-qc/0605021} {arXiv:gr-qc/0605021 [gr-qc]}
  \BibitemShut {NoStop}%
\bibitem [{\citenamefont {Berti}\ \emph {et~al.}(2005)\citenamefont {Berti},
  \citenamefont {Buonanno},\ and\ \citenamefont {Will}}]{Berti2005}%
  \BibitemOpen
  \bibfield  {author} {\bibinfo {author} {\bibfnamefont {E.}~\bibnamefont
  {Berti}}, \bibinfo {author} {\bibfnamefont {A.}~\bibnamefont {Buonanno}},\
  and\ \bibinfo {author} {\bibfnamefont {C.~M.}\ \bibnamefont {Will}},\
  }\bibfield  {title} {\bibinfo {title} {{Estimating spinning binary parameters
  and testing alternative theories of gravity with LISA}},\ }\href
  {https://doi.org/10.1103/PhysRevD.71.084025} {\bibfield  {journal} {\bibinfo
  {journal} {\prd}\ }\textbf {\bibinfo {volume} {71}},\ \bibinfo {pages}
  {84025} (\bibinfo {year} {2005})},\ \Eprint
  {https://arxiv.org/abs/gr-qc/0411129} {arXiv:gr-qc/0411129 [gr-qc]}
  \BibitemShut {NoStop}%
\bibitem [{\citenamefont {Lang}\ and\ \citenamefont {Hughes}(2006)}]{Lang2006}%
  \BibitemOpen
  \bibfield  {author} {\bibinfo {author} {\bibfnamefont {R.~N.}\ \bibnamefont
  {Lang}}\ and\ \bibinfo {author} {\bibfnamefont {S.~A.}\ \bibnamefont
  {Hughes}},\ }\bibfield  {title} {\bibinfo {title} {{Measuring coalescing
  massive binary black holes with gravitational waves: The impact of
  spin-induced precession}},\ }\href
  {https://doi.org/10.1103/PhysRevD.74.122001} {\bibfield  {journal} {\bibinfo
  {journal} {\prd}\ }\textbf {\bibinfo {volume} {74}},\ \bibinfo {pages}
  {122001} (\bibinfo {year} {2006})},\ \Eprint
  {https://arxiv.org/abs/gr-qc/0608062} {arXiv:gr-qc/0608062 [gr-qc]}
  \BibitemShut {NoStop}%
\bibitem [{\citenamefont {Babak}\ \emph {et~al.}(2010)\citenamefont {Babak},
  \citenamefont {Baker}, \citenamefont {Benacquista}, \citenamefont {Cornish},
  \citenamefont {Larson}, \citenamefont {Mandel}, \citenamefont {McWilliams},
  \citenamefont {Petiteau}, \citenamefont {Porter}, \citenamefont {Robinson},
  \citenamefont {Vallisneri}, \citenamefont {Vecchio}, \citenamefont {{Data
  Challenge Task Force}}, \citenamefont {Adams}, \citenamefont {Arnaud},
  \citenamefont {B{\l}aut}, \citenamefont {Bridges}, \citenamefont {Cohen},
  \citenamefont {Cutler}, \citenamefont {Feroz}, \citenamefont {Gair},
  \citenamefont {Graff}, \citenamefont {Hobson}, \citenamefont {{Shapiro Key}},
  \citenamefont {Kr{\'{o}}lak}, \citenamefont {Lasenby}, \citenamefont {Prix},
  \citenamefont {Shang}, \citenamefont {Trias}, \citenamefont {Veitch},
  \citenamefont {Whelan},\ and\ \citenamefont {Participants}}]{Babak2010}%
  \BibitemOpen
  \bibfield  {author} {\bibinfo {author} {\bibfnamefont {S.}~\bibnamefont
  {Babak}}, \bibinfo {author} {\bibfnamefont {J.~G.}\ \bibnamefont {Baker}},
  \bibinfo {author} {\bibfnamefont {M.~J.}\ \bibnamefont {Benacquista}},
  \bibinfo {author} {\bibfnamefont {N.~J.}\ \bibnamefont {Cornish}}, \bibinfo
  {author} {\bibfnamefont {S.~L.}\ \bibnamefont {Larson}}, \bibinfo {author}
  {\bibfnamefont {I.}~\bibnamefont {Mandel}}, \bibinfo {author} {\bibfnamefont
  {S.~T.}\ \bibnamefont {McWilliams}}, \bibinfo {author} {\bibfnamefont
  {A.}~\bibnamefont {Petiteau}}, \bibinfo {author} {\bibfnamefont {E.~K.}\
  \bibnamefont {Porter}}, \bibinfo {author} {\bibfnamefont {E.~L.}\
  \bibnamefont {Robinson}}, \bibinfo {author} {\bibfnamefont {M.}~\bibnamefont
  {Vallisneri}}, \bibinfo {author} {\bibfnamefont {A.}~\bibnamefont {Vecchio}},
  \bibinfo {author} {\bibfnamefont {t.~M.~L.}\ \bibnamefont {{Data Challenge
  Task Force}}}, \bibinfo {author} {\bibfnamefont {M.}~\bibnamefont {Adams}},
  \bibinfo {author} {\bibfnamefont {K.~A.}\ \bibnamefont {Arnaud}}, \bibinfo
  {author} {\bibfnamefont {A.}~\bibnamefont {B{\l}aut}}, \bibinfo {author}
  {\bibfnamefont {M.}~\bibnamefont {Bridges}}, \bibinfo {author} {\bibfnamefont
  {M.}~\bibnamefont {Cohen}}, \bibinfo {author} {\bibfnamefont
  {C.}~\bibnamefont {Cutler}}, \bibinfo {author} {\bibfnamefont
  {F.}~\bibnamefont {Feroz}}, \bibinfo {author} {\bibfnamefont {J.~R.}\
  \bibnamefont {Gair}}, \bibinfo {author} {\bibfnamefont {P.}~\bibnamefont
  {Graff}}, \bibinfo {author} {\bibfnamefont {M.}~\bibnamefont {Hobson}},
  \bibinfo {author} {\bibfnamefont {J.}~\bibnamefont {{Shapiro Key}}}, \bibinfo
  {author} {\bibfnamefont {A.}~\bibnamefont {Kr{\'{o}}lak}}, \bibinfo {author}
  {\bibfnamefont {A.}~\bibnamefont {Lasenby}}, \bibinfo {author} {\bibfnamefont
  {R.}~\bibnamefont {Prix}}, \bibinfo {author} {\bibfnamefont {Y.}~\bibnamefont
  {Shang}}, \bibinfo {author} {\bibfnamefont {M.}~\bibnamefont {Trias}},
  \bibinfo {author} {\bibfnamefont {J.}~\bibnamefont {Veitch}}, \bibinfo
  {author} {\bibfnamefont {J.~T.}\ \bibnamefont {Whelan}},\ and\ \bibinfo
  {author} {\bibfnamefont {t.~C.~.}\ \bibnamefont {Participants}},\ }\bibfield
  {title} {\bibinfo {title} {{The Mock LISA Data Challenges: from challenge 3
  to challenge 4}},\ }\href {https://doi.org/10.1088/0264-9381/27/8/084009}
  {\bibfield  {journal} {\bibinfo  {journal} {Classical and Quantum Gravity}\
  }\textbf {\bibinfo {volume} {27}},\ \bibinfo {pages} {84009} (\bibinfo {year}
  {2010})},\ \Eprint {https://arxiv.org/abs/0912.0548} {arXiv:0912.0548
  [gr-qc]} \BibitemShut {NoStop}%
\bibitem [{\citenamefont {Brown}\ \emph {et~al.}(2007)\citenamefont {Brown},
  \citenamefont {Crowder}, \citenamefont {Cutler}, \citenamefont {Mand~el},\
  and\ \citenamefont {Vallisneri}}]{Brown2007}%
  \BibitemOpen
  \bibfield  {author} {\bibinfo {author} {\bibfnamefont {D.~A.}\ \bibnamefont
  {Brown}}, \bibinfo {author} {\bibfnamefont {J.}~\bibnamefont {Crowder}},
  \bibinfo {author} {\bibfnamefont {C.}~\bibnamefont {Cutler}}, \bibinfo
  {author} {\bibfnamefont {I.}~\bibnamefont {Mand~el}},\ and\ \bibinfo {author}
  {\bibfnamefont {M.}~\bibnamefont {Vallisneri}},\ }\bibfield  {title}
  {\bibinfo {title} {{A three-stage search for supermassive black-hole binaries
  in LISA data}},\ }\href {https://doi.org/10.1088/0264-9381/24/19/S22}
  {\bibfield  {journal} {\bibinfo  {journal} {Classical and Quantum Gravity}\
  }\textbf {\bibinfo {volume} {24}},\ \bibinfo {pages} {S595} (\bibinfo {year}
  {2007})},\ \Eprint {https://arxiv.org/abs/0704.2447} {arXiv:0704.2447
  [gr-qc]} \BibitemShut {NoStop}%
\bibitem [{\citenamefont {Crowder}\ \emph {et~al.}(2006)\citenamefont
  {Crowder}, \citenamefont {Cornish},\ and\ \citenamefont
  {Reddinger}}]{Crowder2006}%
  \BibitemOpen
  \bibfield  {author} {\bibinfo {author} {\bibfnamefont {J.}~\bibnamefont
  {Crowder}}, \bibinfo {author} {\bibfnamefont {N.~J.}\ \bibnamefont
  {Cornish}},\ and\ \bibinfo {author} {\bibfnamefont {J.~L.}\ \bibnamefont
  {Reddinger}},\ }\bibfield  {title} {\bibinfo {title} {{LISA data analysis
  using genetic algorithms}},\ }\href
  {https://doi.org/10.1103/PhysRevD.73.063011} {\bibfield  {journal} {\bibinfo
  {journal} {\prd}\ }\textbf {\bibinfo {volume} {73}},\ \bibinfo {pages}
  {63011} (\bibinfo {year} {2006})},\ \Eprint
  {https://arxiv.org/abs/gr-qc/0601036} {arXiv:gr-qc/0601036 [gr-qc]}
  \BibitemShut {NoStop}%
\bibitem [{\citenamefont {Wickham}\ \emph {et~al.}(2006)\citenamefont
  {Wickham}, \citenamefont {Stroeer},\ and\ \citenamefont
  {Vecchio}}]{Wickham2006}%
  \BibitemOpen
  \bibfield  {author} {\bibinfo {author} {\bibfnamefont {E.}~\bibnamefont
  {Wickham}}, \bibinfo {author} {\bibfnamefont {A.}~\bibnamefont {Stroeer}},\
  and\ \bibinfo {author} {\bibfnamefont {A.}~\bibnamefont {Vecchio}},\
  }\bibfield  {title} {\bibinfo {title} {{A Markov chain Monte Carlo approach
  to the study of massive black hole binary systems with LISA}},\ }\href
  {https://doi.org/10.1088/0264-9381/23/19/S20} {\bibfield  {journal} {\bibinfo
   {journal} {Classical and Quantum Gravity}\ }\textbf {\bibinfo {volume}
  {23}},\ \bibinfo {pages} {S819} (\bibinfo {year} {2006})},\ \Eprint
  {https://arxiv.org/abs/gr-qc/0605071} {arXiv:gr-qc/0605071 [gr-qc]}
  \BibitemShut {NoStop}%
\bibitem [{\citenamefont {R{\"{o}}ver}\ \emph {et~al.}(2007)\citenamefont
  {R{\"{o}}ver}, \citenamefont {Stroeer}, \citenamefont {Bloomer},
  \citenamefont {Christensen}, \citenamefont {Clark}, \citenamefont {Hendry},
  \citenamefont {Messenger}, \citenamefont {Meyer}, \citenamefont {Pitkin},
  \citenamefont {Toher}, \citenamefont {Umst{\"{a}}tter}, \citenamefont
  {Vecchio}, \citenamefont {Veitch},\ and\ \citenamefont {Woan}}]{Rover2007}%
  \BibitemOpen
  \bibfield  {author} {\bibinfo {author} {\bibfnamefont {C.}~\bibnamefont
  {R{\"{o}}ver}}, \bibinfo {author} {\bibfnamefont {A.}~\bibnamefont
  {Stroeer}}, \bibinfo {author} {\bibfnamefont {E.}~\bibnamefont {Bloomer}},
  \bibinfo {author} {\bibfnamefont {N.}~\bibnamefont {Christensen}}, \bibinfo
  {author} {\bibfnamefont {J.}~\bibnamefont {Clark}}, \bibinfo {author}
  {\bibfnamefont {M.}~\bibnamefont {Hendry}}, \bibinfo {author} {\bibfnamefont
  {C.}~\bibnamefont {Messenger}}, \bibinfo {author} {\bibfnamefont
  {R.}~\bibnamefont {Meyer}}, \bibinfo {author} {\bibfnamefont
  {M.}~\bibnamefont {Pitkin}}, \bibinfo {author} {\bibfnamefont
  {J.}~\bibnamefont {Toher}}, \bibinfo {author} {\bibfnamefont
  {R.}~\bibnamefont {Umst{\"{a}}tter}}, \bibinfo {author} {\bibfnamefont
  {A.}~\bibnamefont {Vecchio}}, \bibinfo {author} {\bibfnamefont
  {J.}~\bibnamefont {Veitch}},\ and\ \bibinfo {author} {\bibfnamefont
  {G.}~\bibnamefont {Woan}},\ }\bibfield  {title} {\bibinfo {title} {{Inference
  on inspiral signals using LISA MLDC data}},\ }\href
  {https://doi.org/10.1088/0264-9381/24/19/S15} {\bibfield  {journal} {\bibinfo
   {journal} {Classical and Quantum Gravity}\ }\textbf {\bibinfo {volume}
  {24}},\ \bibinfo {pages} {S521} (\bibinfo {year} {2007})},\ \Eprint
  {https://arxiv.org/abs/0707.3969} {arXiv:0707.3969 [gr-qc]} \BibitemShut
  {NoStop}%
\bibitem [{\citenamefont {Feroz}\ \emph {et~al.}(2009)\citenamefont {Feroz},
  \citenamefont {Gair}, \citenamefont {Hobson},\ and\ \citenamefont
  {Porter}}]{Feroz2009}%
  \BibitemOpen
  \bibfield  {author} {\bibinfo {author} {\bibfnamefont {F.}~\bibnamefont
  {Feroz}}, \bibinfo {author} {\bibfnamefont {J.~R.}\ \bibnamefont {Gair}},
  \bibinfo {author} {\bibfnamefont {M.~P.}\ \bibnamefont {Hobson}},\ and\
  \bibinfo {author} {\bibfnamefont {E.~K.}\ \bibnamefont {Porter}},\ }\bibfield
   {title} {\bibinfo {title} {{Use of the MULTINEST algorithm for gravitational
  wave data analysis}},\ }\href
  {https://doi.org/10.1088/0264-9381/26/21/215003} {\bibfield  {journal}
  {\bibinfo  {journal} {Classical and Quantum Gravity}\ }\textbf {\bibinfo
  {volume} {26}},\ \bibinfo {pages} {215003} (\bibinfo {year} {2009})},\
  \Eprint {https://arxiv.org/abs/0904.1544} {arXiv:0904.1544 [gr-qc]}
  \BibitemShut {NoStop}%
\bibitem [{\citenamefont {Gair}\ and\ \citenamefont {Porter}(2009)}]{Gair2009}%
  \BibitemOpen
  \bibfield  {author} {\bibinfo {author} {\bibfnamefont {J.~R.}\ \bibnamefont
  {Gair}}\ and\ \bibinfo {author} {\bibfnamefont {E.~K.}\ \bibnamefont
  {Porter}},\ }\bibfield  {title} {\bibinfo {title} {{Cosmic swarms: a search
  for supermassive black holes in the LISA data stream with a hybrid
  evolutionary algorithm}},\ }\href
  {https://doi.org/10.1088/0264-9381/26/22/225004} {\bibfield  {journal}
  {\bibinfo  {journal} {Classical and Quantum Gravity}\ }\textbf {\bibinfo
  {volume} {26}},\ \bibinfo {pages} {225004} (\bibinfo {year} {2009})},\
  \Eprint {https://arxiv.org/abs/0903.3733} {arXiv:0903.3733 [gr-qc]}
  \BibitemShut {NoStop}%
\bibitem [{\citenamefont {Petiteau}\ \emph {et~al.}(2009)\citenamefont
  {Petiteau}, \citenamefont {Shang},\ and\ \citenamefont
  {Babak}}]{Petiteau2009}%
  \BibitemOpen
  \bibfield  {author} {\bibinfo {author} {\bibfnamefont {A.}~\bibnamefont
  {Petiteau}}, \bibinfo {author} {\bibfnamefont {Y.}~\bibnamefont {Shang}},\
  and\ \bibinfo {author} {\bibfnamefont {S.}~\bibnamefont {Babak}},\ }\bibfield
   {title} {\bibinfo {title} {{The search for black hole binaries using a
  genetic algorithm}},\ }\href {https://doi.org/10.1088/0264-9381/26/20/204011}
  {\bibfield  {journal} {\bibinfo  {journal} {Classical and Quantum Gravity}\
  }\textbf {\bibinfo {volume} {26}},\ \bibinfo {pages} {204011} (\bibinfo
  {year} {2009})},\ \Eprint {https://arxiv.org/abs/0905.1785} {arXiv:0905.1785
  [gr-qc]} \BibitemShut {NoStop}%
\bibitem [{\citenamefont {Arun}\ \emph {et~al.}(2007)\citenamefont {Arun},
  \citenamefont {Iyer}, \citenamefont {Sathyaprakash}, \citenamefont {Sinha},\
  and\ \citenamefont {van~den Broeck}}]{Arun2007}%
  \BibitemOpen
  \bibfield  {author} {\bibinfo {author} {\bibfnamefont {K.}~\bibnamefont
  {Arun}}, \bibinfo {author} {\bibfnamefont {B.~R.}\ \bibnamefont {Iyer}},
  \bibinfo {author} {\bibfnamefont {B.}~\bibnamefont {Sathyaprakash}}, \bibinfo
  {author} {\bibfnamefont {S.}~\bibnamefont {Sinha}},\ and\ \bibinfo {author}
  {\bibfnamefont {C.}~\bibnamefont {van~den Broeck}},\ }\bibfield  {title}
  {\bibinfo {title} {{Higher signal harmonics, LISA's angular resolution, and
  dark energy}},\ }\href {https://doi.org/10.1103/PhysRevD.76.104016}
  {\bibfield  {journal} {\bibinfo  {journal} {\prd}\ }\textbf {\bibinfo
  {volume} {76}},\ \bibinfo {pages} {104016} (\bibinfo {year} {2007})},\
  \Eprint {https://arxiv.org/abs/0707.3920} {arXiv:0707.3920 [astro-ph]}
  \BibitemShut {NoStop}%
\bibitem [{\citenamefont {Trias}\ and\ \citenamefont
  {Sintes}(2008)}]{Trias2008}%
  \BibitemOpen
  \bibfield  {author} {\bibinfo {author} {\bibfnamefont {M.}~\bibnamefont
  {Trias}}\ and\ \bibinfo {author} {\bibfnamefont {A.~M.}\ \bibnamefont
  {Sintes}},\ }\bibfield  {title} {\bibinfo {title} {{LISA observations of
  supermassive black holes: Parameter estimation using full post-Newtonian
  inspiral waveforms}},\ }\href {https://doi.org/10.1103/PhysRevD.77.024030}
  {\bibfield  {journal} {\bibinfo  {journal} {\prd}\ }\textbf {\bibinfo
  {volume} {77}},\ \bibinfo {pages} {24030} (\bibinfo {year} {2008})},\ \Eprint
  {https://arxiv.org/abs/0707.4434} {arXiv:0707.4434 [gr-qc]} \BibitemShut
  {NoStop}%
\bibitem [{\citenamefont {McWilliams}\ \emph
  {et~al.}(2010{\natexlab{a}})\citenamefont {McWilliams}, \citenamefont
  {Thorpe}, \citenamefont {Baker},\ and\ \citenamefont
  {Kelly}}]{McWilliams2010}%
  \BibitemOpen
  \bibfield  {author} {\bibinfo {author} {\bibfnamefont {S.~T.}\ \bibnamefont
  {McWilliams}}, \bibinfo {author} {\bibfnamefont {J.~I.}\ \bibnamefont
  {Thorpe}}, \bibinfo {author} {\bibfnamefont {J.~G.}\ \bibnamefont {Baker}},\
  and\ \bibinfo {author} {\bibfnamefont {B.~J.}\ \bibnamefont {Kelly}},\
  }\bibfield  {title} {\bibinfo {title} {{Impact of mergers on LISA parameter
  estimation for nonspinning black hole binaries}},\ }\href
  {https://doi.org/10.1103/PhysRevD.81.064014} {\bibfield  {journal} {\bibinfo
  {journal} {\prd}\ }\textbf {\bibinfo {volume} {81}},\ \bibinfo {pages}
  {64014} (\bibinfo {year} {2010}{\natexlab{a}})},\ \Eprint
  {https://arxiv.org/abs/0911.1078} {arXiv:0911.1078 [gr-qc]} \BibitemShut
  {NoStop}%
\bibitem [{\citenamefont {Thorpe}\ \emph {et~al.}(2009)\citenamefont {Thorpe},
  \citenamefont {McWilliams}, \citenamefont {Kelly}, \citenamefont {Fahey},
  \citenamefont {Arnaud},\ and\ \citenamefont {Baker}}]{Thorpe2009}%
  \BibitemOpen
  \bibfield  {author} {\bibinfo {author} {\bibfnamefont {J.}~\bibnamefont
  {Thorpe}}, \bibinfo {author} {\bibfnamefont {S.}~\bibnamefont {McWilliams}},
  \bibinfo {author} {\bibfnamefont {B.}~\bibnamefont {Kelly}}, \bibinfo
  {author} {\bibfnamefont {R.}~\bibnamefont {Fahey}}, \bibinfo {author}
  {\bibfnamefont {K.}~\bibnamefont {Arnaud}},\ and\ \bibinfo {author}
  {\bibfnamefont {J.}~\bibnamefont {Baker}},\ }\bibfield  {title} {\bibinfo
  {title} {{LISA parameter estimation using numerical merger waveforms}},\
  }\href {https://doi.org/10.1088/0264-9381/26/9/094026} {\bibfield  {journal}
  {\bibinfo  {journal} {Classical and Quantum Gravity}\ }\textbf {\bibinfo
  {volume} {26}},\ \bibinfo {pages} {94026} (\bibinfo {year} {2009})},\ \Eprint
  {https://arxiv.org/abs/0811.0833} {arXiv:0811.0833 [astro-ph]} \BibitemShut
  {NoStop}%
\bibitem [{\citenamefont {McWilliams}\ \emph
  {et~al.}(2010{\natexlab{b}})\citenamefont {McWilliams}, \citenamefont
  {Kelly},\ and\ \citenamefont {Baker}}]{McWilliams2010b}%
  \BibitemOpen
  \bibfield  {author} {\bibinfo {author} {\bibfnamefont {S.~T.}\ \bibnamefont
  {McWilliams}}, \bibinfo {author} {\bibfnamefont {B.~J.}\ \bibnamefont
  {Kelly}},\ and\ \bibinfo {author} {\bibfnamefont {J.~G.}\ \bibnamefont
  {Baker}},\ }\bibfield  {title} {\bibinfo {title} {{Observing mergers of
  nonspinning black-hole binaries}},\ }\href
  {https://doi.org/10.1103/PhysRevD.82.024014} {\bibfield  {journal} {\bibinfo
  {journal} {\prd}\ }\textbf {\bibinfo {volume} {82}},\ \bibinfo {pages}
  {24014} (\bibinfo {year} {2010}{\natexlab{b}})},\ \Eprint
  {https://arxiv.org/abs/1004.0961} {arXiv:1004.0961 [gr-qc]} \BibitemShut
  {NoStop}%
\bibitem [{\citenamefont {McWilliams}\ \emph {et~al.}(2011)\citenamefont
  {McWilliams}, \citenamefont {Lang}, \citenamefont {Baker},\ and\
  \citenamefont {Thorpe}}]{McWilliams2011}%
  \BibitemOpen
  \bibfield  {author} {\bibinfo {author} {\bibfnamefont {S.}~\bibnamefont
  {McWilliams}}, \bibinfo {author} {\bibfnamefont {R.}~\bibnamefont {Lang}},
  \bibinfo {author} {\bibfnamefont {J.}~\bibnamefont {Baker}},\ and\ \bibinfo
  {author} {\bibfnamefont {J.}~\bibnamefont {Thorpe}},\ }\bibfield  {title}
  {\bibinfo {title} {{Sky localization of complete inspiral-merger-ringdown
  signals for nonspinning massive black hole binaries}},\ }\href
  {https://doi.org/10.1103/PhysRevD.84.064003} {\bibfield  {journal} {\bibinfo
  {journal} {\prd}\ }\textbf {\bibinfo {volume} {84}},\ \bibinfo {pages}
  {64003} (\bibinfo {year} {2011})},\ \Eprint {https://arxiv.org/abs/1104.5650}
  {arXiv:1104.5650 [gr-qc]} \BibitemShut {NoStop}%
\bibitem [{\citenamefont {Babak}\ \emph {et~al.}(2008)\citenamefont {Babak},
  \citenamefont {Hannam}, \citenamefont {Husa},\ and\ \citenamefont
  {Schutz}}]{Babak2008}%
  \BibitemOpen
  \bibfield  {author} {\bibinfo {author} {\bibfnamefont {S.}~\bibnamefont
  {Babak}}, \bibinfo {author} {\bibfnamefont {M.}~\bibnamefont {Hannam}},
  \bibinfo {author} {\bibfnamefont {S.}~\bibnamefont {Husa}},\ and\ \bibinfo
  {author} {\bibfnamefont {B.}~\bibnamefont {Schutz}},\ }\bibfield  {title}
  {\bibinfo {title} {{Resolving Super Massive Black Holes with LISA}},\
  }\href@noop {} {\bibfield  {journal} {\bibinfo  {journal} {arXiv e-prints}\
  ,\ \bibinfo {pages} {arXiv:0806.1591}} (\bibinfo {year} {2008})},\ \Eprint
  {https://arxiv.org/abs/0806.1591} {arXiv:0806.1591 [gr-qc]} \BibitemShut
  {NoStop}%
\bibitem [{\citenamefont {Babak}\ \emph {et~al.}(2014)\citenamefont {Babak},
  \citenamefont {Gair},\ and\ \citenamefont {Cole}}]{Babak2014}%
  \BibitemOpen
  \bibfield  {author} {\bibinfo {author} {\bibfnamefont {S.}~\bibnamefont
  {Babak}}, \bibinfo {author} {\bibfnamefont {J.}~\bibnamefont {Gair}},\ and\
  \bibinfo {author} {\bibfnamefont {R.}~\bibnamefont {Cole}},\ }\bibfield
  {title} {\bibinfo {title} {{Extreme mass ratio inspirals: perspectives for
  their detection}},\ }\href@noop {} {\bibfield  {journal} {\bibinfo  {journal}
  {ArXiv e-prints}\ } (\bibinfo {year} {2014})},\ \Eprint
  {https://arxiv.org/abs/1411.5253} {arXiv:1411.5253 [gr-qc]} \BibitemShut
  {NoStop}%
\bibitem [{\citenamefont {Baibhav}\ \emph {et~al.}(2020)\citenamefont
  {Baibhav}, \citenamefont {Berti},\ and\ \citenamefont
  {Cardoso}}]{Baibhav2020}%
  \BibitemOpen
  \bibfield  {author} {\bibinfo {author} {\bibfnamefont {V.}~\bibnamefont
  {Baibhav}}, \bibinfo {author} {\bibfnamefont {E.}~\bibnamefont {Berti}},\
  and\ \bibinfo {author} {\bibfnamefont {V.}~\bibnamefont {Cardoso}},\
  }\bibfield  {title} {\bibinfo {title} {{LISA parameter estimation and source
  localization with higher harmonics of the ringdown}},\ }\href@noop {}
  {\bibfield  {journal} {\bibinfo  {journal} {arXiv e-prints}\ ,\ \bibinfo
  {pages} {arXiv:2001.10011}} (\bibinfo {year} {2020})},\ \Eprint
  {https://arxiv.org/abs/2001.10011} {arXiv:2001.10011 [gr-qc]} \BibitemShut
  {NoStop}%
\bibitem [{\citenamefont {Marsat}\ and\ \citenamefont
  {Baker}(2018)}]{Marsat2018}%
  \BibitemOpen
  \bibfield  {author} {\bibinfo {author} {\bibfnamefont {S.}~\bibnamefont
  {Marsat}}\ and\ \bibinfo {author} {\bibfnamefont {J.~G.}\ \bibnamefont
  {Baker}},\ }\bibfield  {title} {\bibinfo {title} {{Fourier-domain modulations
  and delays of gravitational-wave signals}},\ }\href@noop {} {\bibfield
  {journal} {\bibinfo  {journal} {arXiv e-prints}\ ,\ \bibinfo {pages}
  {arXiv:1806.10734}} (\bibinfo {year} {2018})},\ \Eprint
  {https://arxiv.org/abs/1806.10734} {arXiv:1806.10734 [gr-qc]} \BibitemShut
  {NoStop}%
\bibitem [{\citenamefont {Larson}\ \emph {et~al.}(2000)\citenamefont {Larson},
  \citenamefont {Hiscock},\ and\ \citenamefont {Hellings}}]{Larson2000}%
  \BibitemOpen
  \bibfield  {author} {\bibinfo {author} {\bibfnamefont {S.}~\bibnamefont
  {Larson}}, \bibinfo {author} {\bibfnamefont {W.}~\bibnamefont {Hiscock}},\
  and\ \bibinfo {author} {\bibfnamefont {R.}~\bibnamefont {Hellings}},\
  }\bibfield  {title} {\bibinfo {title} {{Sensitivity curves for spaceborne
  gravitational wave interferometers}},\ }\href
  {https://doi.org/10.1103/PhysRevD.62.062001} {\bibfield  {journal} {\bibinfo
  {journal} {\prd}\ }\textbf {\bibinfo {volume} {62}},\ \bibinfo {pages}
  {62001} (\bibinfo {year} {2000})}\BibitemShut {NoStop}%
\bibitem [{\citenamefont {Husa}\ \emph {et~al.}(2016)\citenamefont {Husa},
  \citenamefont {Khan}, \citenamefont {Hannam}, \citenamefont {P{\"{u}}rrer},
  \citenamefont {Ohme}, \citenamefont {Forteza},\ and\ \citenamefont
  {Boh{\'{e}}}}]{Husa2016}%
  \BibitemOpen
  \bibfield  {author} {\bibinfo {author} {\bibfnamefont {S.}~\bibnamefont
  {Husa}}, \bibinfo {author} {\bibfnamefont {S.}~\bibnamefont {Khan}}, \bibinfo
  {author} {\bibfnamefont {M.}~\bibnamefont {Hannam}}, \bibinfo {author}
  {\bibfnamefont {M.}~\bibnamefont {P{\"{u}}rrer}}, \bibinfo {author}
  {\bibfnamefont {F.}~\bibnamefont {Ohme}}, \bibinfo {author} {\bibfnamefont
  {X.}~\bibnamefont {Forteza}},\ and\ \bibinfo {author} {\bibfnamefont
  {A.}~\bibnamefont {Boh{\'{e}}}},\ }\bibfield  {title} {\bibinfo {title}
  {{Frequency-domain gravitational waves from nonprecessing black-hole
  binaries. I. New numerical waveforms and anatomy of the signal}},\ }\href
  {https://doi.org/10.1103/PhysRevD.93.044006} {\bibfield  {journal} {\bibinfo
  {journal} {\prd}\ }\textbf {\bibinfo {volume} {93}},\ \bibinfo {pages}
  {44006} (\bibinfo {year} {2016})},\ \Eprint
  {https://arxiv.org/abs/1508.07250} {arXiv:1508.07250 [gr-qc]} \BibitemShut
  {NoStop}%
\bibitem [{\citenamefont {Khan}\ \emph {et~al.}(2016)\citenamefont {Khan},
  \citenamefont {Husa}, \citenamefont {Hannam}, \citenamefont {Ohme},
  \citenamefont {P{\"{u}}rrer}, \citenamefont {Forteza},\ and\ \citenamefont
  {Boh{\'{e}}}}]{Khan2016}%
  \BibitemOpen
  \bibfield  {author} {\bibinfo {author} {\bibfnamefont {S.}~\bibnamefont
  {Khan}}, \bibinfo {author} {\bibfnamefont {S.}~\bibnamefont {Husa}}, \bibinfo
  {author} {\bibfnamefont {M.}~\bibnamefont {Hannam}}, \bibinfo {author}
  {\bibfnamefont {F.}~\bibnamefont {Ohme}}, \bibinfo {author} {\bibfnamefont
  {M.}~\bibnamefont {P{\"{u}}rrer}}, \bibinfo {author} {\bibfnamefont
  {X.}~\bibnamefont {Forteza}},\ and\ \bibinfo {author} {\bibfnamefont
  {A.}~\bibnamefont {Boh{\'{e}}}},\ }\bibfield  {title} {\bibinfo {title}
  {{Frequency-domain gravitational waves from nonprecessing black-hole
  binaries. II. A phenomenological model for the advanced detector era}},\
  }\href {https://doi.org/10.1103/PhysRevD.93.044007} {\bibfield  {journal}
  {\bibinfo  {journal} {\prd}\ }\textbf {\bibinfo {volume} {93}},\ \bibinfo
  {pages} {44007} (\bibinfo {year} {2016})},\ \Eprint
  {https://arxiv.org/abs/1508.07253} {arXiv:1508.07253 [gr-qc]} \BibitemShut
  {NoStop}%
\bibitem [{\citenamefont {Cornish}\ and\ \citenamefont
  {Robson}(2018)}]{Cornish2018}%
  \BibitemOpen
  \bibfield  {author} {\bibinfo {author} {\bibfnamefont {N.}~\bibnamefont
  {Cornish}}\ and\ \bibinfo {author} {\bibfnamefont {T.}~\bibnamefont
  {Robson}},\ }\bibfield  {title} {\bibinfo {title} {{The construction and use
  of LISA sensitivity curves}},\ }\href@noop {} {\bibfield  {journal} {\bibinfo
   {journal} {ArXiv e-prints}\ } (\bibinfo {year} {2018})},\ \Eprint
  {https://arxiv.org/abs/1803.01944} {arXiv:1803.01944 [astro-ph.HE]}
  \BibitemShut {NoStop}%
\bibitem [{\citenamefont {Amaro-Seoane}\ \emph {et~al.}(2017)\citenamefont
  {Amaro-Seoane}, \citenamefont {Audley}, \citenamefont {Babak}, \citenamefont
  {Baker}, \citenamefont {Barausse}, \citenamefont {Bender}, \citenamefont
  {Berti}, \citenamefont {Binetruy}, \citenamefont {Born}, \citenamefont
  {Bortoluzzi}, \citenamefont {Camp}, \citenamefont {Caprini}, \citenamefont
  {Cardoso}, \citenamefont {Colpi}, \citenamefont {Conklin}, \citenamefont
  {Cornish}, \citenamefont {Cutler}, \citenamefont {Danzmann}, \citenamefont
  {Dolesi}, \citenamefont {Ferraioli}, \citenamefont {Ferroni}, \citenamefont
  {Fitzsimons}, \citenamefont {Gair}, \citenamefont {{Gesa Bote}},
  \citenamefont {Giardini}, \citenamefont {Gibert}, \citenamefont {Grimani},
  \citenamefont {Halloin}, \citenamefont {Heinzel}, \citenamefont {Hertog},
  \citenamefont {Hewitson}, \citenamefont {Holley-Bockelmann}, \citenamefont
  {Hollington}, \citenamefont {Hueller}, \citenamefont {Inchauspe},
  \citenamefont {Jetzer}, \citenamefont {Karnesis}, \citenamefont {Killow},
  \citenamefont {Klein}, \citenamefont {Klipstein}, \citenamefont {Korsakova},
  \citenamefont {Larson}, \citenamefont {Livas}, \citenamefont {Lloro},
  \citenamefont {Man}, \citenamefont {Mance}, \citenamefont {Martino},
  \citenamefont {Mateos}, \citenamefont {McKenzie}, \citenamefont {McWilliams},
  \citenamefont {Miller}, \citenamefont {Mueller}, \citenamefont {Nardini},
  \citenamefont {Nelemans}, \citenamefont {Nofrarias}, \citenamefont
  {Petiteau}, \citenamefont {Pivato}, \citenamefont {Plagnol}, \citenamefont
  {Porter}, \citenamefont {Reiche}, \citenamefont {Robertson}, \citenamefont
  {Robertson}, \citenamefont {Rossi}, \citenamefont {Russano}, \citenamefont
  {Schutz}, \citenamefont {Sesana}, \citenamefont {Shoemaker}, \citenamefont
  {Slutsky}, \citenamefont {Sopuerta}, \citenamefont {Sumner}, \citenamefont
  {Tamanini}, \citenamefont {Thorpe}, \citenamefont {Troebs}, \citenamefont
  {Vallisneri}, \citenamefont {Vecchio}, \citenamefont {Vetrugno},
  \citenamefont {Vitale}, \citenamefont {Volonteri}, \citenamefont {Wanner},
  \citenamefont {Ward}, \citenamefont {Wass}, \citenamefont {Weber},
  \citenamefont {Ziemer},\ and\ \citenamefont {Zweifel}}]{LISAMissionProposal}%
  \BibitemOpen
  \bibfield  {author} {\bibinfo {author} {\bibfnamefont {P.}~\bibnamefont
  {Amaro-Seoane}}, \bibinfo {author} {\bibfnamefont {H.}~\bibnamefont
  {Audley}}, \bibinfo {author} {\bibfnamefont {S.}~\bibnamefont {Babak}},
  \bibinfo {author} {\bibfnamefont {J.}~\bibnamefont {Baker}}, \bibinfo
  {author} {\bibfnamefont {E.}~\bibnamefont {Barausse}}, \bibinfo {author}
  {\bibfnamefont {P.}~\bibnamefont {Bender}}, \bibinfo {author} {\bibfnamefont
  {E.}~\bibnamefont {Berti}}, \bibinfo {author} {\bibfnamefont
  {P.}~\bibnamefont {Binetruy}}, \bibinfo {author} {\bibfnamefont
  {M.}~\bibnamefont {Born}}, \bibinfo {author} {\bibfnamefont {D.}~\bibnamefont
  {Bortoluzzi}}, \bibinfo {author} {\bibfnamefont {J.}~\bibnamefont {Camp}},
  \bibinfo {author} {\bibfnamefont {C.}~\bibnamefont {Caprini}}, \bibinfo
  {author} {\bibfnamefont {V.}~\bibnamefont {Cardoso}}, \bibinfo {author}
  {\bibfnamefont {M.}~\bibnamefont {Colpi}}, \bibinfo {author} {\bibfnamefont
  {J.}~\bibnamefont {Conklin}}, \bibinfo {author} {\bibfnamefont
  {N.}~\bibnamefont {Cornish}}, \bibinfo {author} {\bibfnamefont
  {C.}~\bibnamefont {Cutler}}, \bibinfo {author} {\bibfnamefont
  {K.}~\bibnamefont {Danzmann}}, \bibinfo {author} {\bibfnamefont
  {R.}~\bibnamefont {Dolesi}}, \bibinfo {author} {\bibfnamefont
  {L.}~\bibnamefont {Ferraioli}}, \bibinfo {author} {\bibfnamefont
  {V.}~\bibnamefont {Ferroni}}, \bibinfo {author} {\bibfnamefont
  {E.}~\bibnamefont {Fitzsimons}}, \bibinfo {author} {\bibfnamefont
  {J.}~\bibnamefont {Gair}}, \bibinfo {author} {\bibfnamefont {L.}~\bibnamefont
  {{Gesa Bote}}}, \bibinfo {author} {\bibfnamefont {D.}~\bibnamefont
  {Giardini}}, \bibinfo {author} {\bibfnamefont {F.}~\bibnamefont {Gibert}},
  \bibinfo {author} {\bibfnamefont {C.}~\bibnamefont {Grimani}}, \bibinfo
  {author} {\bibfnamefont {H.}~\bibnamefont {Halloin}}, \bibinfo {author}
  {\bibfnamefont {G.}~\bibnamefont {Heinzel}}, \bibinfo {author} {\bibfnamefont
  {T.}~\bibnamefont {Hertog}}, \bibinfo {author} {\bibfnamefont
  {M.}~\bibnamefont {Hewitson}}, \bibinfo {author} {\bibfnamefont
  {K.}~\bibnamefont {Holley-Bockelmann}}, \bibinfo {author} {\bibfnamefont
  {D.}~\bibnamefont {Hollington}}, \bibinfo {author} {\bibfnamefont
  {M.}~\bibnamefont {Hueller}}, \bibinfo {author} {\bibfnamefont
  {H.}~\bibnamefont {Inchauspe}}, \bibinfo {author} {\bibfnamefont
  {P.}~\bibnamefont {Jetzer}}, \bibinfo {author} {\bibfnamefont
  {N.}~\bibnamefont {Karnesis}}, \bibinfo {author} {\bibfnamefont
  {C.}~\bibnamefont {Killow}}, \bibinfo {author} {\bibfnamefont
  {A.}~\bibnamefont {Klein}}, \bibinfo {author} {\bibfnamefont
  {B.}~\bibnamefont {Klipstein}}, \bibinfo {author} {\bibfnamefont
  {N.}~\bibnamefont {Korsakova}}, \bibinfo {author} {\bibfnamefont
  {S.}~\bibnamefont {Larson}}, \bibinfo {author} {\bibfnamefont
  {J.}~\bibnamefont {Livas}}, \bibinfo {author} {\bibfnamefont
  {I.}~\bibnamefont {Lloro}}, \bibinfo {author} {\bibfnamefont
  {N.}~\bibnamefont {Man}}, \bibinfo {author} {\bibfnamefont {D.}~\bibnamefont
  {Mance}}, \bibinfo {author} {\bibfnamefont {J.}~\bibnamefont {Martino}},
  \bibinfo {author} {\bibfnamefont {I.}~\bibnamefont {Mateos}}, \bibinfo
  {author} {\bibfnamefont {K.}~\bibnamefont {McKenzie}}, \bibinfo {author}
  {\bibfnamefont {S.}~\bibnamefont {McWilliams}}, \bibinfo {author}
  {\bibfnamefont {C.}~\bibnamefont {Miller}}, \bibinfo {author} {\bibfnamefont
  {G.}~\bibnamefont {Mueller}}, \bibinfo {author} {\bibfnamefont
  {G.}~\bibnamefont {Nardini}}, \bibinfo {author} {\bibfnamefont
  {G.}~\bibnamefont {Nelemans}}, \bibinfo {author} {\bibfnamefont
  {M.}~\bibnamefont {Nofrarias}}, \bibinfo {author} {\bibfnamefont
  {A.}~\bibnamefont {Petiteau}}, \bibinfo {author} {\bibfnamefont
  {P.}~\bibnamefont {Pivato}}, \bibinfo {author} {\bibfnamefont
  {E.}~\bibnamefont {Plagnol}}, \bibinfo {author} {\bibfnamefont
  {E.}~\bibnamefont {Porter}}, \bibinfo {author} {\bibfnamefont
  {J.}~\bibnamefont {Reiche}}, \bibinfo {author} {\bibfnamefont
  {D.}~\bibnamefont {Robertson}}, \bibinfo {author} {\bibfnamefont
  {N.}~\bibnamefont {Robertson}}, \bibinfo {author} {\bibfnamefont
  {E.}~\bibnamefont {Rossi}}, \bibinfo {author} {\bibfnamefont
  {G.}~\bibnamefont {Russano}}, \bibinfo {author} {\bibfnamefont
  {B.}~\bibnamefont {Schutz}}, \bibinfo {author} {\bibfnamefont
  {A.}~\bibnamefont {Sesana}}, \bibinfo {author} {\bibfnamefont
  {D.}~\bibnamefont {Shoemaker}}, \bibinfo {author} {\bibfnamefont
  {J.}~\bibnamefont {Slutsky}}, \bibinfo {author} {\bibfnamefont
  {C.}~\bibnamefont {Sopuerta}}, \bibinfo {author} {\bibfnamefont
  {T.}~\bibnamefont {Sumner}}, \bibinfo {author} {\bibfnamefont
  {N.}~\bibnamefont {Tamanini}}, \bibinfo {author} {\bibfnamefont
  {I.}~\bibnamefont {Thorpe}}, \bibinfo {author} {\bibfnamefont
  {M.}~\bibnamefont {Troebs}}, \bibinfo {author} {\bibfnamefont
  {M.}~\bibnamefont {Vallisneri}}, \bibinfo {author} {\bibfnamefont
  {A.}~\bibnamefont {Vecchio}}, \bibinfo {author} {\bibfnamefont
  {D.}~\bibnamefont {Vetrugno}}, \bibinfo {author} {\bibfnamefont
  {S.}~\bibnamefont {Vitale}}, \bibinfo {author} {\bibfnamefont
  {M.}~\bibnamefont {Volonteri}}, \bibinfo {author} {\bibfnamefont
  {G.}~\bibnamefont {Wanner}}, \bibinfo {author} {\bibfnamefont
  {H.}~\bibnamefont {Ward}}, \bibinfo {author} {\bibfnamefont {P.}~\bibnamefont
  {Wass}}, \bibinfo {author} {\bibfnamefont {W.}~\bibnamefont {Weber}},
  \bibinfo {author} {\bibfnamefont {J.}~\bibnamefont {Ziemer}},\ and\ \bibinfo
  {author} {\bibfnamefont {P.}~\bibnamefont {Zweifel}},\ }\bibfield  {title}
  {\bibinfo {title} {{Laser Interferometer Space Antenna}},\ }\href@noop {}
  {\bibfield  {journal} {\bibinfo  {journal} {ArXiv e-prints}\ } (\bibinfo
  {year} {2017})},\ \Eprint {https://arxiv.org/abs/1702.00786}
  {arXiv:1702.00786 [astro-ph.IM]} \BibitemShut {NoStop}%
\bibitem [{\citenamefont {Finn}\ and\ \citenamefont {Thorne}(2000)}]{Finn2000}%
  \BibitemOpen
  \bibfield  {author} {\bibinfo {author} {\bibfnamefont {L.}~\bibnamefont
  {Finn}}\ and\ \bibinfo {author} {\bibfnamefont {K.}~\bibnamefont {Thorne}},\
  }\bibfield  {title} {\bibinfo {title} {{Gravitational waves from a compact
  star in a circular, inspiral orbit, in the equatorial plane of a massive,
  spinning black hole, as observed by LISA}},\ }\href
  {https://doi.org/10.1103/PhysRevD.62.124021} {\bibfield  {journal} {\bibinfo
  {journal} {\prd}\ }\textbf {\bibinfo {volume} {62}},\ \bibinfo {pages}
  {124021} (\bibinfo {year} {2000})}\BibitemShut {NoStop}%
\bibitem [{\citenamefont {Moore}\ \emph {et~al.}(2015)\citenamefont {Moore},
  \citenamefont {Cole},\ and\ \citenamefont {Berry}}]{Moore2015}%
  \BibitemOpen
  \bibfield  {author} {\bibinfo {author} {\bibfnamefont {C.~J.}\ \bibnamefont
  {Moore}}, \bibinfo {author} {\bibfnamefont {R.~H.}\ \bibnamefont {Cole}},\
  and\ \bibinfo {author} {\bibfnamefont {C.~P.~L.}\ \bibnamefont {Berry}},\
  }\bibfield  {title} {\bibinfo {title} {{Gravitational-wave sensitivity
  curves}},\ }\href {http://stacks.iop.org/0264-9381/32/i=1/a=015014}
  {\bibfield  {journal} {\bibinfo  {journal} {Classical and Quantum Gravity}\
  }\textbf {\bibinfo {volume} {32}},\ \bibinfo {pages} {15014} (\bibinfo {year}
  {2015})}\BibitemShut {NoStop}%
\bibitem [{\citenamefont {London}\ \emph {et~al.}(2018)\citenamefont {London},
  \citenamefont {Khan}, \citenamefont {Fauchon-Jones}, \citenamefont
  {Garc\'ia}, \citenamefont {Hannam}, \citenamefont {Husa}, \citenamefont
  {Jim{\'{e}}nez-Forteza}, \citenamefont {Kalaghatgi}, \citenamefont {Ohme},\
  and\ \citenamefont {Pannarale}}]{London2018}%
  \BibitemOpen
  \bibfield  {author} {\bibinfo {author} {\bibfnamefont {L.}~\bibnamefont
  {London}}, \bibinfo {author} {\bibfnamefont {S.}~\bibnamefont {Khan}},
  \bibinfo {author} {\bibfnamefont {E.}~\bibnamefont {Fauchon-Jones}}, \bibinfo
  {author} {\bibfnamefont {C.}~\bibnamefont {Garc\'ia}}, \bibinfo {author}
  {\bibfnamefont {M.}~\bibnamefont {Hannam}}, \bibinfo {author} {\bibfnamefont
  {S.}~\bibnamefont {Husa}}, \bibinfo {author} {\bibfnamefont {X.}~\bibnamefont
  {Jim{\'{e}}nez-Forteza}}, \bibinfo {author} {\bibfnamefont {C.}~\bibnamefont
  {Kalaghatgi}}, \bibinfo {author} {\bibfnamefont {F.}~\bibnamefont {Ohme}},\
  and\ \bibinfo {author} {\bibfnamefont {F.}~\bibnamefont {Pannarale}},\
  }\bibfield  {title} {\bibinfo {title} {{First Higher-Multipole Model of
  Gravitational Waves from Spinning and Coalescing Black-Hole Binaries}},\
  }\href {https://doi.org/10.1103/PhysRevLett.120.161102} {\bibfield  {journal}
  {\bibinfo  {journal} {Physical Review Letters}\ }\textbf {\bibinfo {volume}
  {120}},\ \bibinfo {pages} {161102} (\bibinfo {year} {2018})},\ \Eprint
  {https://arxiv.org/abs/1708.00404} {arXiv:1708.00404 [gr-qc]} \BibitemShut
  {NoStop}%
\bibitem [{\citenamefont {Babak}\ \emph {et~al.}(2017)\citenamefont {Babak},
  \citenamefont {Gair}, \citenamefont {Sesana}, \citenamefont {Barausse},
  \citenamefont {Sopuerta}, \citenamefont {Berry}, \citenamefont {Berti},
  \citenamefont {Amaro-Seoane}, \citenamefont {Petiteau},\ and\ \citenamefont
  {Klein}}]{Babak2017}%
  \BibitemOpen
  \bibfield  {author} {\bibinfo {author} {\bibfnamefont {S.}~\bibnamefont
  {Babak}}, \bibinfo {author} {\bibfnamefont {J.}~\bibnamefont {Gair}},
  \bibinfo {author} {\bibfnamefont {A.}~\bibnamefont {Sesana}}, \bibinfo
  {author} {\bibfnamefont {E.}~\bibnamefont {Barausse}}, \bibinfo {author}
  {\bibfnamefont {C.}~\bibnamefont {Sopuerta}}, \bibinfo {author}
  {\bibfnamefont {C.}~\bibnamefont {Berry}}, \bibinfo {author} {\bibfnamefont
  {E.}~\bibnamefont {Berti}}, \bibinfo {author} {\bibfnamefont
  {P.}~\bibnamefont {Amaro-Seoane}}, \bibinfo {author} {\bibfnamefont
  {A.}~\bibnamefont {Petiteau}},\ and\ \bibinfo {author} {\bibfnamefont
  {A.}~\bibnamefont {Klein}},\ }\bibfield  {title} {\bibinfo {title} {{Science
  with the space-based interferometer LISA. V. Extreme mass-ratio inspirals}},\
  }\href@noop {} {\bibfield  {journal} {\bibinfo  {journal} {Phys. Rev. D}\
  }\textbf {\bibinfo {volume} {95}},\ \bibinfo {pages} {3012} (\bibinfo {year}
  {2017})}\BibitemShut {NoStop}%
\bibitem [{\citenamefont {Edwards}\ \emph {et~al.}(2015)\citenamefont
  {Edwards}, \citenamefont {Meyer},\ and\ \citenamefont
  {Christensen}}]{EdwardsBayesPSDEstimate}%
  \BibitemOpen
  \bibfield  {author} {\bibinfo {author} {\bibfnamefont {M.~C.}\ \bibnamefont
  {Edwards}}, \bibinfo {author} {\bibfnamefont {R.}~\bibnamefont {Meyer}},\
  and\ \bibinfo {author} {\bibfnamefont {N.}~\bibnamefont {Christensen}},\
  }\bibfield  {title} {\bibinfo {title} {Bayesian semiparametric power spectral
  density estimation with applications in gravitational wave data analysis},\
  }\href {https://doi.org/10.1103/PhysRevD.92.064011} {\bibfield  {journal}
  {\bibinfo  {journal} {Phys. Rev. D}\ }\textbf {\bibinfo {volume} {92}},\
  \bibinfo {pages} {064011} (\bibinfo {year} {2015})}\BibitemShut {NoStop}%
\bibitem [{\citenamefont {Littenberg}\ and\ \citenamefont
  {Cornish}(2015)}]{Littenberg2015BayesSpectralEsitmationNoise}%
  \BibitemOpen
  \bibfield  {author} {\bibinfo {author} {\bibfnamefont {T.~B.}\ \bibnamefont
  {Littenberg}}\ and\ \bibinfo {author} {\bibfnamefont {N.~J.}\ \bibnamefont
  {Cornish}},\ }\bibfield  {title} {\bibinfo {title} {Bayesian inference for
  spectral estimation of gravitational wave detector noise},\ }\href
  {https://doi.org/10.1103/PhysRevD.91.084034} {\bibfield  {journal} {\bibinfo
  {journal} {Phys. Rev. D}\ }\textbf {\bibinfo {volume} {91}},\ \bibinfo
  {pages} {084034} (\bibinfo {year} {2015})}\BibitemShut {NoStop}%
\bibitem [{\citenamefont {{Biscoveanu}}\ \emph {et~al.}(2020)\citenamefont
  {{Biscoveanu}}, \citenamefont {{Haster}}, \citenamefont {{Vitale}},\ and\
  \citenamefont {{Davies}}}]{Biscoveanu2020PSDEstimation}%
  \BibitemOpen
  \bibfield  {author} {\bibinfo {author} {\bibfnamefont {S.}~\bibnamefont
  {{Biscoveanu}}}, \bibinfo {author} {\bibfnamefont {C.-J.}\ \bibnamefont
  {{Haster}}}, \bibinfo {author} {\bibfnamefont {S.}~\bibnamefont {{Vitale}}},\
  and\ \bibinfo {author} {\bibfnamefont {J.}~\bibnamefont {{Davies}}},\
  }\bibfield  {title} {\bibinfo {title} {{Quantifying the Effect of Power
  Spectral Density Uncertainty on Gravitational-Wave Parameter Estimation for
  Compact Binary Sources}},\ }\href@noop {} {\bibfield  {journal} {\bibinfo
  {journal} {arXiv e-prints}\ ,\ \bibinfo {eid} {arXiv:2004.05149}} (\bibinfo
  {year} {2020})},\ \Eprint {https://arxiv.org/abs/2004.05149}
  {arXiv:2004.05149 [astro-ph.HE]} \BibitemShut {NoStop}%
\bibitem [{\citenamefont {Goggans}\ and\ \citenamefont
  {Chi}(2004)}]{Goggans2004ThermoInt}%
  \BibitemOpen
  \bibfield  {author} {\bibinfo {author} {\bibfnamefont {P.~M.}\ \bibnamefont
  {Goggans}}\ and\ \bibinfo {author} {\bibfnamefont {Y.}~\bibnamefont {Chi}},\
  }\bibfield  {title} {\bibinfo {title} {Using thermodynamic integration to
  calculate the posterior probability in bayesian model selection problems},\
  }\href {https://doi.org/10.1063/1.1751356} {\bibfield  {journal} {\bibinfo
  {journal} {AIP Conference Proceedings}\ }\textbf {\bibinfo {volume} {707}},\
  \bibinfo {pages} {59} (\bibinfo {year} {2004})},\ \Eprint
  {https://arxiv.org/abs/https://aip.scitation.org/doi/pdf/10.1063/1.1751356}
  {https://aip.scitation.org/doi/pdf/10.1063/1.1751356} \BibitemShut {NoStop}%
\bibitem [{\citenamefont {Lartillot}\ and\ \citenamefont
  {Philippe}(2006)}]{Lartillot2006ThermoInt}%
  \BibitemOpen
  \bibfield  {author} {\bibinfo {author} {\bibfnamefont {N.}~\bibnamefont
  {Lartillot}}\ and\ \bibinfo {author} {\bibfnamefont {H.}~\bibnamefont
  {Philippe}},\ }\bibfield  {title} {\bibinfo {title} {{Computing Bayes Factors
  Using Thermodynamic Integration}},\ }\href
  {https://doi.org/10.1080/10635150500433722} {\bibfield  {journal} {\bibinfo
  {journal} {Systematic Biology}\ }\textbf {\bibinfo {volume} {55}},\ \bibinfo
  {pages} {195} (\bibinfo {year} {2006})},\ \Eprint
  {https://arxiv.org/abs/https://academic.oup.com/sysbio/article-pdf/55/2/195/26557316/10635150500433722.pdf}
  {https://academic.oup.com/sysbio/article-pdf/55/2/195/26557316/10635150500433722.pdf}
  \BibitemShut {NoStop}%
\bibitem [{\citenamefont {{Maturana-Russel}}\ \emph {et~al.}(2019)\citenamefont
  {{Maturana-Russel}}, \citenamefont {{Meyer}}, \citenamefont {{Veitch}},\ and\
  \citenamefont {{Christensen}}}]{Maturana-Russel2019SteppingStone}%
  \BibitemOpen
  \bibfield  {author} {\bibinfo {author} {\bibfnamefont {P.}~\bibnamefont
  {{Maturana-Russel}}}, \bibinfo {author} {\bibfnamefont {R.}~\bibnamefont
  {{Meyer}}}, \bibinfo {author} {\bibfnamefont {J.}~\bibnamefont {{Veitch}}},\
  and\ \bibinfo {author} {\bibfnamefont {N.}~\bibnamefont {{Christensen}}},\
  }\bibfield  {title} {\bibinfo {title} {{Stepping-stone sampling algorithm for
  calculating the evidence of gravitational wave models}},\ }\href
  {https://doi.org/10.1103/PhysRevD.99.084006} {\bibfield  {journal} {\bibinfo
  {journal} {\prd}\ }\textbf {\bibinfo {volume} {99}},\ \bibinfo {eid} {084006}
  (\bibinfo {year} {2019})},\ \Eprint {https://arxiv.org/abs/1810.04488}
  {arXiv:1810.04488 [physics.data-an]} \BibitemShut {NoStop}%
\bibitem [{\citenamefont {Kalaghatgi}\ \emph {et~al.}(2019)\citenamefont
  {Kalaghatgi}, \citenamefont {Hannam},\ and\ \citenamefont
  {Raymond}}]{Kalaghatgi2019}%
  \BibitemOpen
  \bibfield  {author} {\bibinfo {author} {\bibfnamefont {C.}~\bibnamefont
  {Kalaghatgi}}, \bibinfo {author} {\bibfnamefont {M.}~\bibnamefont {Hannam}},\
  and\ \bibinfo {author} {\bibfnamefont {V.}~\bibnamefont {Raymond}},\
  }\bibfield  {title} {\bibinfo {title} {{Parameter Estimation with a spinning
  multi-mode waveform model: IMRPhenomHM}},\ }\href@noop {} {\bibfield
  {journal} {\bibinfo  {journal} {arXiv e-prints}\ ,\ \bibinfo {pages}
  {arXiv:1909.10010}} (\bibinfo {year} {2019})},\ \Eprint
  {https://arxiv.org/abs/1909.10010} {arXiv:1909.10010 [gr-qc]} \BibitemShut
  {NoStop}%
\bibitem [{\citenamefont {Cornish}\ and\ \citenamefont
  {Rubbo}(2003)}]{Cornish2003}%
  \BibitemOpen
  \bibfield  {author} {\bibinfo {author} {\bibfnamefont {N.~J.}\ \bibnamefont
  {Cornish}}\ and\ \bibinfo {author} {\bibfnamefont {L.~J.}\ \bibnamefont
  {Rubbo}},\ }\bibfield  {title} {\bibinfo {title} {{Publisher's Note: LISA
  response function [Phys. Rev. D 67, 022001 (2003)]}},\ }\href
  {https://doi.org/10.1103/PhysRevD.67.029905} {\bibfield  {journal} {\bibinfo
  {journal} {\prd}\ }\textbf {\bibinfo {volume} {67}},\ \bibinfo {pages}
  {29905} (\bibinfo {year} {2003})},\ \Eprint
  {https://arxiv.org/abs/gr-qc/0209011} {arXiv:gr-qc/0209011 [gr-qc]}
  \BibitemShut {NoStop}%
\bibitem [{\citenamefont {Tinto}\ and\ \citenamefont
  {Armstrong}(1999)}]{Tinto1999}%
  \BibitemOpen
  \bibfield  {author} {\bibinfo {author} {\bibfnamefont {M.}~\bibnamefont
  {Tinto}}\ and\ \bibinfo {author} {\bibfnamefont {J.~W.}\ \bibnamefont
  {Armstrong}},\ }\bibfield  {title} {\bibinfo {title} {{Cancellation of laser
  noise in an unequal-arm interferometer detector of gravitational
  radiation}},\ }\href {https://doi.org/10.1103/PhysRevD.59.102003} {\bibfield
  {journal} {\bibinfo  {journal} {Phys. Rev. D}\ }\textbf {\bibinfo {volume}
  {59}},\ \bibinfo {pages} {102003} (\bibinfo {year} {1999})}\BibitemShut
  {NoStop}%
\bibitem [{\citenamefont {Armstrong}\ \emph {et~al.}(1999)\citenamefont
  {Armstrong}, \citenamefont {Estabrook},\ and\ \citenamefont
  {Tinto}}]{Armstrong1999}%
  \BibitemOpen
  \bibfield  {author} {\bibinfo {author} {\bibfnamefont {J.}~\bibnamefont
  {Armstrong}}, \bibinfo {author} {\bibfnamefont {F.}~\bibnamefont
  {Estabrook}},\ and\ \bibinfo {author} {\bibfnamefont {M.}~\bibnamefont
  {Tinto}},\ }\bibfield  {title} {\bibinfo {title} {{Time-Delay Interferometry
  for Space-based Gravitational Wave Searches}},\ }\href
  {https://doi.org/10.1086/308110} {\bibfield  {journal} {\bibinfo  {journal}
  {\apj}\ }\textbf {\bibinfo {volume} {527}},\ \bibinfo {pages} {814} (\bibinfo
  {year} {1999})}\BibitemShut {NoStop}%
\bibitem [{\citenamefont {Estabrook}\ \emph {et~al.}(2000)\citenamefont
  {Estabrook}, \citenamefont {Tinto},\ and\ \citenamefont
  {Armstrong}}]{Estabrook2000}%
  \BibitemOpen
  \bibfield  {author} {\bibinfo {author} {\bibfnamefont {F.~B.}\ \bibnamefont
  {Estabrook}}, \bibinfo {author} {\bibfnamefont {M.}~\bibnamefont {Tinto}},\
  and\ \bibinfo {author} {\bibfnamefont {J.~W.}\ \bibnamefont {Armstrong}},\
  }\bibfield  {title} {\bibinfo {title} {{Time-delay analysis of LISA
  gravitational wave data: Elimination of spacecraft motion effects}},\ }\href
  {https://doi.org/10.1103/PhysRevD.62.042002} {\bibfield  {journal} {\bibinfo
  {journal} {Phys. Rev. D}\ }\textbf {\bibinfo {volume} {62}},\ \bibinfo
  {pages} {42002} (\bibinfo {year} {2000})}\BibitemShut {NoStop}%
\bibitem [{\citenamefont {Dhurandhar}\ \emph {et~al.}(2002)\citenamefont
  {Dhurandhar}, \citenamefont {Nayak},\ and\ \citenamefont
  {Vinet}}]{Dhurandhar2002}%
  \BibitemOpen
  \bibfield  {author} {\bibinfo {author} {\bibfnamefont {S.~V.}\ \bibnamefont
  {Dhurandhar}}, \bibinfo {author} {\bibfnamefont {K.~R.}\ \bibnamefont
  {Nayak}},\ and\ \bibinfo {author} {\bibfnamefont {J.-Y.}\ \bibnamefont
  {Vinet}},\ }\bibfield  {title} {\bibinfo {title} {{Algebraic approach to
  time-delay data analysis for LISA}},\ }\href
  {https://doi.org/10.1103/PhysRevD.65.102002} {\bibfield  {journal} {\bibinfo
  {journal} {Phys. Rev. D}\ }\textbf {\bibinfo {volume} {65}},\ \bibinfo
  {pages} {102002} (\bibinfo {year} {2002})}\BibitemShut {NoStop}%
\bibitem [{\citenamefont {Tinto}\ and\ \citenamefont
  {Dhurandhar}(2005)}]{Tinto2005}%
  \BibitemOpen
  \bibfield  {author} {\bibinfo {author} {\bibfnamefont {M.}~\bibnamefont
  {Tinto}}\ and\ \bibinfo {author} {\bibfnamefont {S.~V.}\ \bibnamefont
  {Dhurandhar}},\ }\bibfield  {title} {\bibinfo {title} {{Time-Delay
  Interferometry}},\ }\href {https://doi.org/10.12942/lrr-2005-4} {\bibfield
  {journal} {\bibinfo  {journal} {Living Reviews in Relativity}\ }\textbf
  {\bibinfo {volume} {8}},\ \bibinfo {pages} {4} (\bibinfo {year}
  {2005})}\BibitemShut {NoStop}%
\bibitem [{\citenamefont {Vallisneri}(2005)}]{Vallisneri2005}%
  \BibitemOpen
  \bibfield  {author} {\bibinfo {author} {\bibfnamefont {M.}~\bibnamefont
  {Vallisneri}},\ }\bibfield  {title} {\bibinfo {title} {{Synthetic LISA:
  Simulating time delay interferometry in a model LISA}},\ }\href
  {https://doi.org/10.1103/PhysRevD.71.022001} {\bibfield  {journal} {\bibinfo
  {journal} {\prd}\ }\textbf {\bibinfo {volume} {71}},\ \bibinfo {pages}
  {22001} (\bibinfo {year} {2005})},\ \Eprint
  {https://arxiv.org/abs/gr-qc/0407102} {arXiv:gr-qc/0407102 [gr-qc]}
  \BibitemShut {NoStop}%
\bibitem [{\citenamefont {Tinto}\ and\ \citenamefont
  {Dhurandhar}(2014)}]{Tinto2014}%
  \BibitemOpen
  \bibfield  {author} {\bibinfo {author} {\bibfnamefont {M.}~\bibnamefont
  {Tinto}}\ and\ \bibinfo {author} {\bibfnamefont {S.~V.}\ \bibnamefont
  {Dhurandhar}},\ }\bibfield  {title} {\bibinfo {title} {{Time-Delay
  Interferometry}},\ }\href {https://doi.org/10.12942/lrr-2014-6} {\bibfield
  {journal} {\bibinfo  {journal} {Living Reviews in Relativity}\ }\textbf
  {\bibinfo {volume} {17}},\ \bibinfo {pages} {6} (\bibinfo {year}
  {2014})}\BibitemShut {NoStop}%
\bibitem [{\citenamefont {Muratore}\ \emph {et~al.}(2020)\citenamefont
  {Muratore}, \citenamefont {Vetrugno},\ and\ \citenamefont
  {Vitale}}]{Muratore2020}%
  \BibitemOpen
  \bibfield  {author} {\bibinfo {author} {\bibfnamefont {M.}~\bibnamefont
  {Muratore}}, \bibinfo {author} {\bibfnamefont {D.}~\bibnamefont {Vetrugno}},\
  and\ \bibinfo {author} {\bibfnamefont {S.}~\bibnamefont {Vitale}},\
  }\bibfield  {title} {\bibinfo {title} {{Revisitation of time delay
  interferometry combinations that suppress laser noise in LISA}},\ }\href@noop
  {} {\bibfield  {journal} {\bibinfo  {journal} {arXiv e-prints}\ ,\ \bibinfo
  {pages} {arXiv:2001.11221}} (\bibinfo {year} {2020})},\ \Eprint
  {https://arxiv.org/abs/2001.11221} {arXiv:2001.11221 [astro-ph.IM]}
  \BibitemShut {NoStop}%
\bibitem [{\citenamefont {Nickolls}\ \emph {et~al.}(2008)\citenamefont
  {Nickolls}, \citenamefont {Buck}, \citenamefont {Garland},\ and\
  \citenamefont {Skadron}}]{CUDA}%
  \BibitemOpen
  \bibfield  {author} {\bibinfo {author} {\bibfnamefont {J.}~\bibnamefont
  {Nickolls}}, \bibinfo {author} {\bibfnamefont {I.}~\bibnamefont {Buck}},
  \bibinfo {author} {\bibfnamefont {M.}~\bibnamefont {Garland}},\ and\ \bibinfo
  {author} {\bibfnamefont {K.}~\bibnamefont {Skadron}},\ }\bibfield  {title}
  {\bibinfo {title} {{Scalable Parallel Programming with CUDA}},\ }\href
  {https://doi.org/10.1145/1365490.1365500} {\bibfield  {journal} {\bibinfo
  {journal} {Queue}\ }\textbf {\bibinfo {volume} {6}},\ \bibinfo {pages} {40}
  (\bibinfo {year} {2008})}\BibitemShut {NoStop}%
\bibitem [{\citenamefont {Talbot}\ \emph {et~al.}(2019)\citenamefont {Talbot},
  \citenamefont {Smith}, \citenamefont {Thrane},\ and\ \citenamefont
  {Poole}}]{Talbot2019}%
  \BibitemOpen
  \bibfield  {author} {\bibinfo {author} {\bibfnamefont {C.}~\bibnamefont
  {Talbot}}, \bibinfo {author} {\bibfnamefont {R.}~\bibnamefont {Smith}},
  \bibinfo {author} {\bibfnamefont {E.}~\bibnamefont {Thrane}},\ and\ \bibinfo
  {author} {\bibfnamefont {G.~B.}\ \bibnamefont {Poole}},\ }\bibfield  {title}
  {\bibinfo {title} {{Parallelized Inference for Gravitational-Wave
  Astronomy}},\ }\href@noop {} {\bibfield  {journal} {\bibinfo  {journal}
  {arXiv e-prints}\ ,\ \bibinfo {pages} {arXiv:1904.02863}} (\bibinfo {year}
  {2019})},\ \Eprint {https://arxiv.org/abs/1904.02863} {arXiv:1904.02863
  [astro-ph.IM]} \BibitemShut {NoStop}%
\bibitem [{\citenamefont {Hannam}\ \emph {et~al.}(2014)\citenamefont {Hannam},
  \citenamefont {Schmidt}, \citenamefont {Boh{\'{e}}}, \citenamefont {Haegel},
  \citenamefont {Husa}, \citenamefont {Ohme}, \citenamefont {Pratten},\ and\
  \citenamefont {P{\"{u}}rrer}}]{Hannam2014}%
  \BibitemOpen
  \bibfield  {author} {\bibinfo {author} {\bibfnamefont {M.}~\bibnamefont
  {Hannam}}, \bibinfo {author} {\bibfnamefont {P.}~\bibnamefont {Schmidt}},
  \bibinfo {author} {\bibfnamefont {A.}~\bibnamefont {Boh{\'{e}}}}, \bibinfo
  {author} {\bibfnamefont {L.}~\bibnamefont {Haegel}}, \bibinfo {author}
  {\bibfnamefont {S.}~\bibnamefont {Husa}}, \bibinfo {author} {\bibfnamefont
  {F.}~\bibnamefont {Ohme}}, \bibinfo {author} {\bibfnamefont {G.}~\bibnamefont
  {Pratten}},\ and\ \bibinfo {author} {\bibfnamefont {M.}~\bibnamefont
  {P{\"{u}}rrer}},\ }\bibfield  {title} {\bibinfo {title} {{Simple Model of
  Complete Precessing Black-Hole-Binary Gravitational Waveforms}},\ }\href
  {https://doi.org/10.1103/PhysRevLett.113.151101} {\bibfield  {journal}
  {\bibinfo  {journal} {Phys. Rev. Lett.}\ }\textbf {\bibinfo {volume} {113}},\
  \bibinfo {pages} {151101} (\bibinfo {year} {2014})}\BibitemShut {NoStop}%
\bibitem [{\citenamefont {{LIGO Scientific Collaboration}}(2018)}]{lalsuite}%
  \BibitemOpen
  \bibfield  {author} {\bibinfo {author} {\bibnamefont {{LIGO Scientific
  Collaboration}}},\ }\href {https://doi.org/10.7935/GT1W-FZ16} {\bibinfo
  {title} {{{LIGO} {A}lgorithm {L}ibrary - {LALS}uite}}},\ \bibinfo
  {howpublished} {free software (GPL)} (\bibinfo {year} {2018})\BibitemShut
  {NoStop}%
\bibitem [{\citenamefont {Jones}\ \emph {et~al.}()\citenamefont {Jones},
  \citenamefont {Oliphant}, \citenamefont {Peterson},\ and\ \citenamefont
  {Others}}]{scipy}%
  \BibitemOpen
  \bibfield  {author} {\bibinfo {author} {\bibfnamefont {E.}~\bibnamefont
  {Jones}}, \bibinfo {author} {\bibfnamefont {T.}~\bibnamefont {Oliphant}},
  \bibinfo {author} {\bibfnamefont {P.}~\bibnamefont {Peterson}},\ and\
  \bibinfo {author} {\bibnamefont {Others}},\ }\href {http://www.scipy.org/}
  {\bibinfo {title} {{{SciPy}: Open source scientific tools for
  {Python}}}}\BibitemShut {NoStop}%
\bibitem [{cuS()}]{cuSparse_2}%
  \BibitemOpen
  \href@noop {} {\bibinfo {title} {cusparse}},\ \bibinfo {howpublished}
  {\url{https://docs.nvidia.com/cuda/cusparse/index.html}},\ \bibinfo {note}
  {[Online; accessed 2019-8-25]}\BibitemShut {NoStop}%
\bibitem [{cuB()}]{cuBLAS_2}%
  \BibitemOpen
  \href@noop {} {\bibinfo {title} {{cuBLAS}}},\ \bibinfo {howpublished}
  {\url{https://docs.nvidia.com/cuda/cublas/index.html}},\ \bibinfo {note}
  {[Online; accessed 2019-8-25]}\BibitemShut {NoStop}%
\bibitem [{\citenamefont {Behnel}\ \emph {et~al.}(2011)\citenamefont {Behnel},
  \citenamefont {Bradshaw}, \citenamefont {Citro}, \citenamefont {Dalcin},
  \citenamefont {Seljebotn},\ and\ \citenamefont {Smith}}]{Cython}%
  \BibitemOpen
  \bibfield  {author} {\bibinfo {author} {\bibfnamefont {S.}~\bibnamefont
  {Behnel}}, \bibinfo {author} {\bibfnamefont {R.}~\bibnamefont {Bradshaw}},
  \bibinfo {author} {\bibfnamefont {C.}~\bibnamefont {Citro}}, \bibinfo
  {author} {\bibfnamefont {L.}~\bibnamefont {Dalcin}}, \bibinfo {author}
  {\bibfnamefont {D.~S.}\ \bibnamefont {Seljebotn}},\ and\ \bibinfo {author}
  {\bibfnamefont {K.}~\bibnamefont {Smith}},\ }\bibfield  {title} {\bibinfo
  {title} {{Cython: The best of both worlds}},\ }\href@noop {} {\bibfield
  {journal} {\bibinfo  {journal} {Computing in Science \& Engineering}\
  }\textbf {\bibinfo {volume} {13}},\ \bibinfo {pages} {31} (\bibinfo {year}
  {2011})}\BibitemShut {NoStop}%
\bibitem [{\citenamefont {McGibbon}\ and\ \citenamefont
  {Zhao}()}]{CUDAwrapper}%
  \BibitemOpen
  \bibfield  {author} {\bibinfo {author} {\bibfnamefont {R.}~\bibnamefont
  {McGibbon}}\ and\ \bibinfo {author} {\bibfnamefont {Y.}~\bibnamefont
  {Zhao}},\ }\href@noop {} {\bibinfo {title} {npcuda-example}}\BibitemShut
  {NoStop}%
\bibitem [{\citenamefont {Skilling}(2004)}]{Skilling2004}%
  \BibitemOpen
  \bibfield  {author} {\bibinfo {author} {\bibfnamefont {J.}~\bibnamefont
  {Skilling}},\ }\bibfield  {title} {\bibinfo {title} {{Nested Sampling}},\
  }\href {https://doi.org/10.1063/1.1835238} {\bibfield  {journal} {\bibinfo
  {journal} {AIP Conference Proceedings}\ }\textbf {\bibinfo {volume} {735}},\
  \bibinfo {pages} {395} (\bibinfo {year} {2004})}\BibitemShut {NoStop}%
\bibitem [{\citenamefont {Skilling}(2006)}]{Skilling2006}%
  \BibitemOpen
  \bibfield  {author} {\bibinfo {author} {\bibfnamefont {J.}~\bibnamefont
  {Skilling}},\ }\bibfield  {title} {\bibinfo {title} {{Nested sampling for
  general Bayesian computation}},\ }\href {https://doi.org/10.1214/06-BA127}
  {\bibfield  {journal} {\bibinfo  {journal} {Bayesian Anal.}\ }\textbf
  {\bibinfo {volume} {1}},\ \bibinfo {pages} {833} (\bibinfo {year}
  {2006})}\BibitemShut {NoStop}%
\bibitem [{\citenamefont {Metropolis}\ and\ \citenamefont
  {Ulam}(1949)}]{Metropolis1949}%
  \BibitemOpen
  \bibfield  {author} {\bibinfo {author} {\bibfnamefont {N.}~\bibnamefont
  {Metropolis}}\ and\ \bibinfo {author} {\bibfnamefont {S.}~\bibnamefont
  {Ulam}},\ }\bibfield  {title} {\bibinfo {title} {{The Monte Carlo Method}},\
  }\href {https://doi.org/10.1080/01621459.1949.10483310} {\bibfield  {journal}
  {\bibinfo  {journal} {Journal of the American Statistical Association}\
  }\textbf {\bibinfo {volume} {44}},\ \bibinfo {pages} {335} (\bibinfo {year}
  {1949})}\BibitemShut {NoStop}%
\bibitem [{\citenamefont {Metropolis}\ \emph {et~al.}(1953)\citenamefont
  {Metropolis}, \citenamefont {Rosenbluth}, \citenamefont {Rosenbluth},
  \citenamefont {Teller},\ and\ \citenamefont {Teller}}]{Metropolis1953}%
  \BibitemOpen
  \bibfield  {author} {\bibinfo {author} {\bibfnamefont {N.}~\bibnamefont
  {Metropolis}}, \bibinfo {author} {\bibfnamefont {A.~W.}\ \bibnamefont
  {Rosenbluth}}, \bibinfo {author} {\bibfnamefont {M.~N.}\ \bibnamefont
  {Rosenbluth}}, \bibinfo {author} {\bibfnamefont {A.~H.}\ \bibnamefont
  {Teller}},\ and\ \bibinfo {author} {\bibfnamefont {E.}~\bibnamefont
  {Teller}},\ }\bibfield  {title} {\bibinfo {title} {{Equation of State
  Calculations by Fast Computing Machines}},\ }\href
  {https://doi.org/10.1063/1.1699114} {\bibfield  {journal} {\bibinfo
  {journal} {The Journal of Chemical Physics}\ }\textbf {\bibinfo {volume}
  {21}},\ \bibinfo {pages} {1087} (\bibinfo {year} {1953})}\BibitemShut
  {NoStop}%
\bibitem [{\citenamefont {Hastings}(1970)}]{Hastings1970}%
  \BibitemOpen
  \bibfield  {author} {\bibinfo {author} {\bibfnamefont {W.~K.}\ \bibnamefont
  {Hastings}},\ }\bibfield  {title} {\bibinfo {title} {{Monte Carlo sampling
  methods using Markov chains and their applications}},\ }\href
  {https://doi.org/10.1093/biomet/57.1.97} {\bibfield  {journal} {\bibinfo
  {journal} {Biometrika}\ }\textbf {\bibinfo {volume} {57}},\ \bibinfo {pages}
  {97} (\bibinfo {year} {1970})}\BibitemShut {NoStop}%
\bibitem [{\citenamefont {Foreman-Mackey}\ \emph {et~al.}(2013)\citenamefont
  {Foreman-Mackey}, \citenamefont {Hogg}, \citenamefont {Lang},\ and\
  \citenamefont {Goodman}}]{emcee}%
  \BibitemOpen
  \bibfield  {author} {\bibinfo {author} {\bibfnamefont {D.}~\bibnamefont
  {Foreman-Mackey}}, \bibinfo {author} {\bibfnamefont {D.~W.}\ \bibnamefont
  {Hogg}}, \bibinfo {author} {\bibfnamefont {D.}~\bibnamefont {Lang}},\ and\
  \bibinfo {author} {\bibfnamefont {J.}~\bibnamefont {Goodman}},\ }\bibfield
  {title} {\bibinfo {title} {{emcee: The MCMC Hammer}},\ }\href
  {https://doi.org/10.1086/670067} {\bibfield  {journal} {\bibinfo  {journal}
  {Publications of the Astronomical Society of the Pacific}\ }\textbf {\bibinfo
  {volume} {125}},\ \bibinfo {pages} {306} (\bibinfo {year} {2013})},\ \Eprint
  {https://arxiv.org/abs/1202.3665} {arXiv:1202.3665 [astro-ph.IM]}
  \BibitemShut {NoStop}%
\bibitem [{\citenamefont {Vousden}\ \emph {et~al.}(2016)\citenamefont
  {Vousden}, \citenamefont {Farr},\ and\ \citenamefont {Mandel}}]{Vousden2016}%
  \BibitemOpen
  \bibfield  {author} {\bibinfo {author} {\bibfnamefont {W.}~\bibnamefont
  {Vousden}}, \bibinfo {author} {\bibfnamefont {W.}~\bibnamefont {Farr}},\ and\
  \bibinfo {author} {\bibfnamefont {I.}~\bibnamefont {Mandel}},\ }\bibfield
  {title} {\bibinfo {title} {{Dynamic temperature selection for parallel
  tempering in Markov chain Monte Carlo simulations}},\ }\href
  {https://doi.org/10.1093/mnras/stv2422} {\bibfield  {journal} {\bibinfo
  {journal} {\mnras}\ }\textbf {\bibinfo {volume} {455}},\ \bibinfo {pages}
  {1919} (\bibinfo {year} {2016})},\ \Eprint {https://arxiv.org/abs/1501.05823}
  {arXiv:1501.05823 [astro-ph.IM]} \BibitemShut {NoStop}%
\bibitem [{\citenamefont {Goodman}\ and\ \citenamefont
  {Weare}(2010)}]{Goodman2010}%
  \BibitemOpen
  \bibfield  {author} {\bibinfo {author} {\bibfnamefont {J.}~\bibnamefont
  {Goodman}}\ and\ \bibinfo {author} {\bibfnamefont {J.}~\bibnamefont
  {Weare}},\ }\bibfield  {title} {\bibinfo {title} {{Ensemble samplers with
  affine invariance}},\ }\href {https://doi.org/10.2140/camcos.2010.5.65}
  {\bibfield  {journal} {\bibinfo  {journal} {Communications in Applied
  Mathematics and Computational Science, Vol.$\sim$5, No.$\sim$1,
  p.$\sim$65-80, 2010}\ }\textbf {\bibinfo {volume} {5}},\ \bibinfo {pages}
  {65} (\bibinfo {year} {2010})}\BibitemShut {NoStop}%
\bibitem [{\citenamefont {Swendsen}\ and\ \citenamefont
  {Wang}(1986)}]{Swendsen1986}%
  \BibitemOpen
  \bibfield  {author} {\bibinfo {author} {\bibfnamefont {R.~H.}\ \bibnamefont
  {Swendsen}}\ and\ \bibinfo {author} {\bibfnamefont {J.-S.}\ \bibnamefont
  {Wang}},\ }\bibfield  {title} {\bibinfo {title} {{Replica Monte Carlo
  simulation of spin glasses}},\ }\href
  {https://doi.org/10.1103/PhysRevLett.57.2607} {\bibfield  {journal} {\bibinfo
   {journal} {\prl}\ }\textbf {\bibinfo {volume} {57}},\ \bibinfo {pages}
  {2607} (\bibinfo {year} {1986})}\BibitemShut {NoStop}%
\bibitem [{\citenamefont {Earl}\ and\ \citenamefont {Deem}(2005)}]{Earl2005}%
  \BibitemOpen
  \bibfield  {author} {\bibinfo {author} {\bibfnamefont {D.~J.}\ \bibnamefont
  {Earl}}\ and\ \bibinfo {author} {\bibfnamefont {M.~W.}\ \bibnamefont
  {Deem}},\ }\bibfield  {title} {\bibinfo {title} {{Parallel tempering: Theory,
  applications, and new perspectives}},\ }\href
  {https://doi.org/10.1039/B509983H} {\bibfield  {journal} {\bibinfo  {journal}
  {Physical Chemistry Chemical Physics (Incorporating Faraday Transactions)}\
  }\textbf {\bibinfo {volume} {7}},\ \bibinfo {pages} {3910} (\bibinfo {year}
  {2005})},\ \Eprint {https://arxiv.org/abs/physics/0508111}
  {arXiv:physics/0508111 [physics.comp-ph]} \BibitemShut {NoStop}%
\bibitem [{\citenamefont {Sokal}(1997)}]{Sokal1997}%
  \BibitemOpen
  \bibfield  {author} {\bibinfo {author} {\bibfnamefont {A.}~\bibnamefont
  {Sokal}},\ }\bibinfo {title} {{Monte Carlo Methods in Statistical Mechanics:
  Foundations and New Algorithms}},\ in\ \href
  {https://doi.org/10.1007/978-1-4899-0319-8_6} {\emph {\bibinfo {booktitle}
  {Functional Integration: Basics and Applications}}},\ \bibinfo {editor}
  {edited by\ \bibinfo {editor} {\bibfnamefont {C.}~\bibnamefont
  {DeWitt-Morette}}, \bibinfo {editor} {\bibfnamefont {P.}~\bibnamefont
  {Cartier}},\ and\ \bibinfo {editor} {\bibfnamefont {A.}~\bibnamefont
  {Folacci}}}\ (\bibinfo  {publisher} {Springer US},\ \bibinfo {address}
  {Boston, MA},\ \bibinfo {year} {1997})\ pp.\ \bibinfo {pages}
  {131--192}\BibitemShut {NoStop}%
\bibitem [{\citenamefont {Foreman-Mackey}(2018)}]{autocorrelation_2}%
  \BibitemOpen
  \bibfield  {author} {\bibinfo {author} {\bibfnamefont {D.}~\bibnamefont
  {Foreman-Mackey}},\ }\href@noop {} {\bibinfo {title} {Autocorrelation time
  estimation}},\ \bibinfo {howpublished}
  {\href{https://dfm.io/posts/autocorr/}{https://dfm.io/posts/autocorr/}}
  (\bibinfo {year} {2018})\BibitemShut {NoStop}%
\bibitem [{\citenamefont {Pratten}\ \emph {et~al.}(2020)\citenamefont
  {Pratten}, \citenamefont {Husa}, \citenamefont {Garcia-Quiros}, \citenamefont
  {Colleoni}, \citenamefont {Ramos-Buades}, \citenamefont {Estelles},\ and\
  \citenamefont {Jaume}}]{Pratten2020PhenomX}%
  \BibitemOpen
  \bibfield  {author} {\bibinfo {author} {\bibfnamefont {G.}~\bibnamefont
  {Pratten}}, \bibinfo {author} {\bibfnamefont {S.}~\bibnamefont {Husa}},
  \bibinfo {author} {\bibfnamefont {C.}~\bibnamefont {Garcia-Quiros}}, \bibinfo
  {author} {\bibfnamefont {M.}~\bibnamefont {Colleoni}}, \bibinfo {author}
  {\bibfnamefont {A.}~\bibnamefont {Ramos-Buades}}, \bibinfo {author}
  {\bibfnamefont {H.}~\bibnamefont {Estelles}},\ and\ \bibinfo {author}
  {\bibfnamefont {R.}~\bibnamefont {Jaume}},\ }\bibfield  {title} {\bibinfo
  {title} {{Setting the cornerstone for the IMRPhenomX family of models for
  gravitational waves from compact binaries: The dominant harmonic for
  non-precessing quasi-circular black holes}},\ }\href@noop {} {\bibfield
  {journal} {\bibinfo  {journal} {arXiv e-prints}\ ,\ \bibinfo {pages}
  {arXiv:2001.11412}} (\bibinfo {year} {2020})},\ \Eprint
  {https://arxiv.org/abs/2001.11412} {arXiv:2001.11412 [gr-qc]} \BibitemShut
  {NoStop}%
\bibitem [{\citenamefont {Garc\'ia-Quir{\'{o}}s}\ \emph
  {et~al.}(2020)\citenamefont {Garc\'ia-Quir{\'{o}}s}, \citenamefont
  {Colleoni}, \citenamefont {Husa}, \citenamefont {Estell{\'{e}}s},
  \citenamefont {Pratten}, \citenamefont {Ramos-Buades}, \citenamefont
  {Mateu-Lucena},\ and\ \citenamefont {Jaume}}]{GarciaQuiros2020PhenomXHM}%
  \BibitemOpen
  \bibfield  {author} {\bibinfo {author} {\bibfnamefont {C.}~\bibnamefont
  {Garc\'ia-Quir{\'{o}}s}}, \bibinfo {author} {\bibfnamefont {M.}~\bibnamefont
  {Colleoni}}, \bibinfo {author} {\bibfnamefont {S.}~\bibnamefont {Husa}},
  \bibinfo {author} {\bibfnamefont {H.}~\bibnamefont {Estell{\'{e}}s}},
  \bibinfo {author} {\bibfnamefont {G.}~\bibnamefont {Pratten}}, \bibinfo
  {author} {\bibfnamefont {A.}~\bibnamefont {Ramos-Buades}}, \bibinfo {author}
  {\bibfnamefont {M.}~\bibnamefont {Mateu-Lucena}},\ and\ \bibinfo {author}
  {\bibfnamefont {R.}~\bibnamefont {Jaume}},\ }\bibfield  {title} {\bibinfo
  {title} {{IMRPhenomXHM: A multi-mode frequency-domain model for the
  gravitational wave signal from non-precessing black-hole binaries}},\
  }\href@noop {} {\bibfield  {journal} {\bibinfo  {journal} {arXiv e-prints}\
  ,\ \bibinfo {pages} {arXiv:2001.10914}} (\bibinfo {year} {2020})},\ \Eprint
  {https://arxiv.org/abs/2001.10914} {arXiv:2001.10914 [gr-qc]} \BibitemShut
  {NoStop}%
\bibitem [{\citenamefont {{Pratten}}\ \emph {et~al.}(2020)\citenamefont
  {{Pratten}}, \citenamefont {{Garc{\'\i}a-Quir{\'o}s}}, \citenamefont
  {{Colleoni}}, \citenamefont {{Ramos-Buades}}, \citenamefont {{Estell{\'e}s}},
  \citenamefont {{Mateu-Lucena}}, \citenamefont {{Jaume}}, \citenamefont
  {{Haney}}, \citenamefont {{Keitel}}, \citenamefont {{Thompson}},\ and\
  \citenamefont {{Husa}}}]{Pratten2020IMRPhenomXPHM}%
  \BibitemOpen
  \bibfield  {author} {\bibinfo {author} {\bibfnamefont {G.}~\bibnamefont
  {{Pratten}}}, \bibinfo {author} {\bibfnamefont {C.}~\bibnamefont
  {{Garc{\'\i}a-Quir{\'o}s}}}, \bibinfo {author} {\bibfnamefont
  {M.}~\bibnamefont {{Colleoni}}}, \bibinfo {author} {\bibfnamefont
  {A.}~\bibnamefont {{Ramos-Buades}}}, \bibinfo {author} {\bibfnamefont
  {H.}~\bibnamefont {{Estell{\'e}s}}}, \bibinfo {author} {\bibfnamefont
  {M.}~\bibnamefont {{Mateu-Lucena}}}, \bibinfo {author} {\bibfnamefont
  {R.}~\bibnamefont {{Jaume}}}, \bibinfo {author} {\bibfnamefont
  {M.}~\bibnamefont {{Haney}}}, \bibinfo {author} {\bibfnamefont
  {D.}~\bibnamefont {{Keitel}}}, \bibinfo {author} {\bibfnamefont {J.~E.}\
  \bibnamefont {{Thompson}}},\ and\ \bibinfo {author} {\bibfnamefont
  {S.}~\bibnamefont {{Husa}}},\ }\bibfield  {title} {\bibinfo {title} {{Let's
  twist again: computationally efficient models for the dominant and
  sub-dominant harmonic modes of precessing binary black holes}},\ }\href@noop
  {} {\bibfield  {journal} {\bibinfo  {journal} {arXiv e-prints}\ ,\ \bibinfo
  {eid} {arXiv:2004.06503}} (\bibinfo {year} {2020})},\ \Eprint
  {https://arxiv.org/abs/2004.06503} {arXiv:2004.06503 [gr-qc]} \BibitemShut
  {NoStop}%
\bibitem [{\citenamefont {{Astropy Collaboration}}\ \emph
  {et~al.}(2013)\citenamefont {{Astropy Collaboration}}, \citenamefont
  {Robitaille}, \citenamefont {Tollerud}, \citenamefont {Greenfield},
  \citenamefont {Droettboom}, \citenamefont {Bray}, \citenamefont {Aldcroft},
  \citenamefont {Davis}, \citenamefont {Ginsburg}, \citenamefont
  {Price-Whelan}, \citenamefont {Kerzendorf}, \citenamefont {Conley},
  \citenamefont {Crighton}, \citenamefont {Barbary}, \citenamefont {Muna},
  \citenamefont {Ferguson}, \citenamefont {Grollier}, \citenamefont {Parikh},
  \citenamefont {Nair}, \citenamefont {Unther}, \citenamefont {Deil},
  \citenamefont {Woillez}, \citenamefont {Conseil}, \citenamefont {Kramer},
  \citenamefont {Turner}, \citenamefont {Singer}, \citenamefont {Fox},
  \citenamefont {Weaver}, \citenamefont {Zabalza}, \citenamefont {Edwards},
  \citenamefont {{Azalee Bostroem}}, \citenamefont {Burke}, \citenamefont
  {Casey}, \citenamefont {Crawford}, \citenamefont {Dencheva}, \citenamefont
  {Ely}, \citenamefont {Jenness}, \citenamefont {Labrie}, \citenamefont {Lim},
  \citenamefont {Pierfederici}, \citenamefont {Pontzen}, \citenamefont {Ptak},
  \citenamefont {Refsdal}, \citenamefont {Servillat},\ and\ \citenamefont
  {Streicher}}]{Astropy}%
  \BibitemOpen
  \bibfield  {author} {\bibinfo {author} {\bibnamefont {{Astropy
  Collaboration}}}, \bibinfo {author} {\bibfnamefont {T.}~\bibnamefont
  {Robitaille}}, \bibinfo {author} {\bibfnamefont {E.}~\bibnamefont
  {Tollerud}}, \bibinfo {author} {\bibfnamefont {P.}~\bibnamefont
  {Greenfield}}, \bibinfo {author} {\bibfnamefont {M.}~\bibnamefont
  {Droettboom}}, \bibinfo {author} {\bibfnamefont {E.}~\bibnamefont {Bray}},
  \bibinfo {author} {\bibfnamefont {T.}~\bibnamefont {Aldcroft}}, \bibinfo
  {author} {\bibfnamefont {M.}~\bibnamefont {Davis}}, \bibinfo {author}
  {\bibfnamefont {A.}~\bibnamefont {Ginsburg}}, \bibinfo {author}
  {\bibfnamefont {A.}~\bibnamefont {Price-Whelan}}, \bibinfo {author}
  {\bibfnamefont {W.}~\bibnamefont {Kerzendorf}}, \bibinfo {author}
  {\bibfnamefont {A.}~\bibnamefont {Conley}}, \bibinfo {author} {\bibfnamefont
  {N.}~\bibnamefont {Crighton}}, \bibinfo {author} {\bibfnamefont
  {K.}~\bibnamefont {Barbary}}, \bibinfo {author} {\bibfnamefont
  {D.}~\bibnamefont {Muna}}, \bibinfo {author} {\bibfnamefont {H.}~\bibnamefont
  {Ferguson}}, \bibinfo {author} {\bibfnamefont {F.}~\bibnamefont {Grollier}},
  \bibinfo {author} {\bibfnamefont {M.}~\bibnamefont {Parikh}}, \bibinfo
  {author} {\bibfnamefont {P.}~\bibnamefont {Nair}}, \bibinfo {author}
  {\bibfnamefont {H.}~\bibnamefont {Unther}}, \bibinfo {author} {\bibfnamefont
  {C.}~\bibnamefont {Deil}}, \bibinfo {author} {\bibfnamefont {J.}~\bibnamefont
  {Woillez}}, \bibinfo {author} {\bibfnamefont {S.}~\bibnamefont {Conseil}},
  \bibinfo {author} {\bibfnamefont {R.}~\bibnamefont {Kramer}}, \bibinfo
  {author} {\bibfnamefont {J.}~\bibnamefont {Turner}}, \bibinfo {author}
  {\bibfnamefont {L.}~\bibnamefont {Singer}}, \bibinfo {author} {\bibfnamefont
  {R.}~\bibnamefont {Fox}}, \bibinfo {author} {\bibfnamefont {B.}~\bibnamefont
  {Weaver}}, \bibinfo {author} {\bibfnamefont {V.}~\bibnamefont {Zabalza}},
  \bibinfo {author} {\bibfnamefont {Z.}~\bibnamefont {Edwards}}, \bibinfo
  {author} {\bibfnamefont {K.}~\bibnamefont {{Azalee Bostroem}}}, \bibinfo
  {author} {\bibfnamefont {D.}~\bibnamefont {Burke}}, \bibinfo {author}
  {\bibfnamefont {A.}~\bibnamefont {Casey}}, \bibinfo {author} {\bibfnamefont
  {S.}~\bibnamefont {Crawford}}, \bibinfo {author} {\bibfnamefont
  {N.}~\bibnamefont {Dencheva}}, \bibinfo {author} {\bibfnamefont
  {J.}~\bibnamefont {Ely}}, \bibinfo {author} {\bibfnamefont {T.}~\bibnamefont
  {Jenness}}, \bibinfo {author} {\bibfnamefont {K.}~\bibnamefont {Labrie}},
  \bibinfo {author} {\bibfnamefont {P.}~\bibnamefont {Lim}}, \bibinfo {author}
  {\bibfnamefont {F.}~\bibnamefont {Pierfederici}}, \bibinfo {author}
  {\bibfnamefont {A.}~\bibnamefont {Pontzen}}, \bibinfo {author} {\bibfnamefont
  {A.}~\bibnamefont {Ptak}}, \bibinfo {author} {\bibfnamefont {B.}~\bibnamefont
  {Refsdal}}, \bibinfo {author} {\bibfnamefont {M.}~\bibnamefont {Servillat}},\
  and\ \bibinfo {author} {\bibfnamefont {O.}~\bibnamefont {Streicher}},\
  }\bibfield  {title} {\bibinfo {title} {{Astropy: A community Python package
  for astronomy}},\ }\href {https://doi.org/10.1051/0004-6361/201322068}
  {\bibfield  {journal} {\bibinfo  {journal} {\aap}\ }\textbf {\bibinfo
  {volume} {558}},\ \bibinfo {pages} {A33} (\bibinfo {year} {2013})},\ \Eprint
  {https://arxiv.org/abs/1307.6212} {arXiv:1307.6212 [astro-ph.IM]}
  \BibitemShut {NoStop}%
\bibitem [{\citenamefont {van~der Walt}\ \emph {et~al.}(2011)\citenamefont
  {van~der Walt}, \citenamefont {Colbert},\ and\ \citenamefont
  {Varoquaux}}]{Numpy}%
  \BibitemOpen
  \bibfield  {author} {\bibinfo {author} {\bibfnamefont {S.}~\bibnamefont
  {van~der Walt}}, \bibinfo {author} {\bibfnamefont {S.~C.}\ \bibnamefont
  {Colbert}},\ and\ \bibinfo {author} {\bibfnamefont {G.}~\bibnamefont
  {Varoquaux}},\ }\bibfield  {title} {\bibinfo {title} {{The NumPy Array: A
  Structure for Efficient Numerical Computation}},\ }\href
  {https://doi.org/10.1109/MCSE.2011.37} {\bibfield  {journal} {\bibinfo
  {journal} {Computing In Science \& Engineering}\ }\textbf {\bibinfo {volume}
  {13}},\ \bibinfo {pages} {22} (\bibinfo {year} {2011})}\BibitemShut {NoStop}%
\bibitem [{\citenamefont {Hunter}(2007)}]{Matplotlib}%
  \BibitemOpen
  \bibfield  {author} {\bibinfo {author} {\bibfnamefont {J.~D.}\ \bibnamefont
  {Hunter}},\ }\bibfield  {title} {\bibinfo {title} {{Matplotlib: A 2D graphics
  environment}},\ }\href {https://doi.org/10.1109/MCSE.2007.55} {\bibfield
  {journal} {\bibinfo  {journal} {Computing In Science \& Engineering}\
  }\textbf {\bibinfo {volume} {9}},\ \bibinfo {pages} {90} (\bibinfo {year}
  {2007})}\BibitemShut {NoStop}%
\end{thebibliography}
\end{document}